\documentclass[a4paper,11pt]{article}
\usepackage{jheppub} % for details on the use of the package, please see the JINST-author-manual
\usepackage{lineno}
\usepackage{dsfont}
\usepackage{calc}
\usepackage{tikz}
\usetikzlibrary{decorations.markings}
\usetikzlibrary{decorations.pathmorphing}
\usetikzlibrary{decorations.pathreplacing}
\usepackage{mathrsfs}
\usepackage{bm}

\newcommand{\normal}[1]{{\hspace{-0.1ex}:\hspace{-0.6ex} #1 \hspace{-0.6ex}:\hspace{0.2ex}}}
\newcommand{\s}{\hspace{0.2ex}}
\newcommand{\barred}[1]{\hspace{0.2ex}\overline{\hspace{-0.2ex} #1}}

\newcommand{\llangle}{\langle \hspace{-2pt}\langle}
\newcommand{\rrangle}{\rangle\hspace{-2pt}\rangle}

\title{\boldmath The 2d Free Boson Minkowski CFT with Asymptotic Charges}

\author{Nicholas Agia and Daniel L.~Jafferis}
\affiliation{Center for the Fundamental Laws of Nature, Harvard University,
Cambridge, MA, USA}
\emailAdd{nagia@g.harvard.edu}
\emailAdd{jafferis@g.harvard.edu}

\abstract{We explicitly construct the states of the 2d free boson on the infinite line with specified asymptotic charges. The Minkowski CFT states derive from (the analytic continuation of) the shrinking limit of Euclidean angular quantization, wherein the superselection sector for fixed asymptotic charges corresponds to specific endpoint operators for the angular quantization on the sphere. When applied to string theory, we obtain a new class of fundamental strings that are neither open nor closed and which we dub ``stretched strings''. We describe constant-time snapshots of the worldsheets of the different types of stretched strings, and as a special case we find a heuristic picture of Maldacena's ``long strings'' in the $c=1$ string theory.}

\begin{document}

\maketitle
\flushbottom

\section{Introduction}\label{intro}

The 2d free boson is one of the simplest quantum field theories, so it is perhaps surprising that there is still more to learn about it. Similarly, a great deal is known about conformal field theories in states on a sphere due to the famous state/operator correspondence. Far less is understood about CFT states in Minkowski space, where the spectrum is continuous and gapless. In this paper, we describe the 2d free boson CFT in Minkowski space by taking the ``shrinking limit'' of the angular quantization introduced in \cite{Agia22}. This procedure constructs a wide class of states with nontrivial asymptotic conditions at spatial infinity, which here corresponds to injecting momentum or winding into the system. While we focus on the free boson CFT for simplicity and concreteness, the same methodology allows one to study asymptotic charges in any CFT with a WZW-like sector. A generalization of these ideas should also be able to describe higher-dimensional Minkowski CFT states with arbitrary asymptotic conditions.

One main line of motivation to understand asymptotic conditions in CFTs and QFTs more generally stems from the prevalence of such considerations in applications to holography. Our more immediate motivation, however, is describing string theory states obtained from angular quantization between operators with Euclidean time winding. It was argued in \cite{Jafferis21} why such a quantization scheme is required when there are operators with nonzero time winding present. This paper is thus a precursor for future work showing that this new class of string provides a string-theoretic description of the microstates of the cigar black hole. Along the way, we shall show that these strings from angular quantization also account for Maldacena's ``long strings'' in the $c=1$ string theory \cite{Maldacena05}. While this paper is concerned only with CFT, we shall already find a heuristic picture reminiscent of the latter.

The organization of this paper is as follows. In Section \ref{AQ review}, we review the basics of angular quantization together with the explicit results for the noncompact and compact free bosons. In Section \ref{shrinking limit}, we provide the asymptotic conditions that the states which survive the shrinking limit of angular quantization must obey. In Section \ref{zero net charge}, we explicitly perform the shrinking limit to obtain the free boson Minkowski CFT states for the special case in which the sum of the asymptotic charges vanishes. Section \ref{nonzero net charge} tackles the general case of nonvanishing net asymptotic charge. Finally, Section \ref{prelude} contains a short nod to string theory where we describe the Lorentzian worldsheets of the different basic types of ``stretched strings'' defined by angular quantization, of which the long folded string of Maldacena is a special case. To this end, Figure \ref{pure momentum special} represents the most important finding of this paper.

\section{Review of Angular Quantization}\label{AQ review}

Angular quantization in 2d CFT is a technique developed in \cite{Agia22} to decompose sphere correlators as a trace over states living on a meridian connecting two antipodal points. At each of the two endpoint poles lies a local operator. In order to obtain a Hilbert space of states with an associated evolution operator, the endpoint operators must be regularized by excising holes around them on which suitable boundary conditions are placed. These boundary conditions must have the property that shrinking the excised holes down to a point reproduces the original operator, up to a constant. If one such hole has radius $\varepsilon$ on the sphere with stereographic coordinate $z$, then a boundary condition $B_i$ placed at $|z| = \varepsilon$ which shrinks to an operator $\mathcal{O}_i(0)$ at the origin may be thought of as a state $|B_i\rrangle_{\varepsilon}$ in radial quantization with the property
\begin{equation}
|B_i\rrangle_{\varepsilon} \ \stackrel{\varepsilon\rightarrow 0}{\longrightarrow} \ b_i i^{-s_i}\varepsilon^{\Delta_i}\mathcal{O}_i(\varepsilon).
\end{equation}
Here, $\Delta_i$ is the scaling dimension of the operator $\mathcal{O}_i$, $b_i$ is a numerical constant determined by the specific boundary condition chosen and the phase $i^{-s_i}$ where $s_i$ is the spin of the operator $\mathcal{O}_i$ was factored out of $b_i$ for convenience; moreover, placing the operator at $z = \varepsilon$ as $\varepsilon\rightarrow 0$ gives a precise meaning to the normalization and phase of $b_i$. Likewise, there is an operator $\mathcal{O}_j(\infty)$ at infinity, which is to be thought of as the origin of the other stereographic patch with transition function $z' = \frac{1}{z}$. The other boundary condition $B_j$ placed at $|z| = \frac{1}{\varepsilon}$ may then be thought of as the BPZ conjugate state ${}_{\varepsilon}\llangle B_j|$ with the property
\begin{equation}
{}_{\varepsilon}\llangle B_j| \ \stackrel{\varepsilon\rightarrow 0}{\longrightarrow} \ b_j i^{s_j}\varepsilon^{-\Delta_j}\mathcal{O}_j(\tfrac{1}{\varepsilon}).
\end{equation}
In the regulated set-up, all sphere correlators can then be sliced along constant-angle rays, defining the angular quantization Hilbert space $\mathcal{H}_{ij}^{B_i B_j}(\varepsilon)$ with a Hamiltonian $H_{\text{R}}^{B_i B_j}(\varepsilon)$ which generates counterclockwise rotations. By construction, the ``thermal'' traces over the angular quantization Hilbert space on the annulus are proportional to the original sphere correlators in the shrinking limit $\varepsilon\rightarrow 0$. That is,
\begin{multline}\label{AQ sphere}
\mathrm{Tr}_{\mathcal{H}_{ij}^{B_i B_j}}\left[\mathcal{O}_1(z_1,\barred{z}_1)\cdots\mathcal{O}_n(z_n,\barred{z}_n)e^{-2\pi H_{\text{R}}^{B_i B_j}(\varepsilon)}\right]_{S^2} 
\\ \stackrel{\varepsilon\rightarrow 0}{\longrightarrow} \ b_i b_j i^{s_j-s_i}\varepsilon^{\Delta_i-\Delta_j}\left\langle \mathcal{O}_i(\varepsilon)\mathcal{O}_1(z_1,\barred{z}_1)\cdots\mathcal{O}_n(z_n,\barred{z}_n)\mathcal{O}_j(\tfrac{1}{\varepsilon})\right\rangle_{S^2}.
\end{multline}
Of course, the sphere correlator appearing explicitly in \eqref{AQ sphere} decays as $\varepsilon^{2\Delta_j}$, so traces in the shrinking limit of angular quantization always become $\varepsilon^{\Delta_i+\Delta_j}$ times the finite physical quantity. The equality in \eqref{AQ sphere} for arbitrary bulk operator insertions encapsulates the required locality of the endpoint boundary conditions.

The Hamiltonian $H_{\text{R}}^{B_i B_j}$ is labeled with `R' because it becomes the Rindler Hamiltonian if the angular coordinate on the sphere is analytically continued to a real time coordinate. When originally defining angular quantization, it is preferable to work on the sphere because of the natural interpretation of the shrinking limit as the insertion of two endpoint local operators. Nevertheless, the physical interpretation of the angular quantization Hilbert space descends from the shrinking limit in the cylinder conformal frame. This transformation is effected as usual by $z = e^{i\tau + \sigma}$, where $\tau$ is the Euclidean time coordinate on the cylinder and $\sigma$ is the spatial coordinate. At finite regulator, the cylinder has length $2L$, where $L = -\ln\varepsilon$, and the relation \eqref{AQ sphere} in this frame becomes\footnote{Note that the thermal trace in angular quantization in a given conformal frame is defined by the partition function in that frame, so it picks up the same Weyl anomaly that the correlator does. In this case, $Z_{\text{flat plane}} = \varepsilon^{c/6}Z_{\text{flat cylinder}}$, where $c$ is the central charge.}
\begin{multline}\label{AQ cylinder}
\mathrm{Tr}_{\mathcal{H}_{ij}^{B_i B_j}}\left[\mathcal{O}_1(\tau_1,\sigma_1)\cdots\mathcal{O}_n(\tau_n,\sigma_n)e^{-2\pi H_{\text{R}}^{B_i B_j}(L)}\right]_{S^1\times\mathds{R}}
\\ \stackrel{L\rightarrow\infty}{\longrightarrow} \ b_i b_j \left\langle \mathcal{O}_i(0,-L)\mathcal{O}_1(\tau_1,\sigma_1)\cdots\mathcal{O}_n(\tau_n,\sigma_n)\mathcal{O}_j(0,L)\right\rangle_{S^1\times\mathds{R}}.
\end{multline}
The states $|\psi\rangle \in \mathcal{H}_{ij}^{B_i B_j}(L)$ live on the spatial line parametrized by $\sigma$, with $-L \leqslant \sigma \leqslant L$, obeying boundary condition $B_i$ at $\sigma = -L$ and obeying boundary condition $B_j$ at $\sigma = L$. The time coordinate $\tau$ in \eqref{AQ cylinder} was compactified with periodicity $\beta = 2\pi$, but now we may consider the theory on the semi-infinite strip with $\tau$ decompactified for which $\mathcal{H}_{ij}^{B_i B_j}(L)$ and $H_{\text{R}}^{B_i B_j}(L)$ are unchanged. Finally, the shrinking limit is obtained by taking $L\rightarrow\infty$. In this limit, the vector space $\mathcal{H}_{ij}^{B_i B_j}(L)$ becomes the angular quantization vector space $\mathcal{H}_{ij}$; the only reason we have indicated possible $L$-dependence is that the limit $\mathcal{H}_{ij}$ may be a quotient of $\mathcal{H}_{ij}^{B_i B_j}(L)$ if the inner product develops a null subspace. Moreover, the angular quantization state space $\mathcal{H}_{ij}$ should only depend on the original endpoint operators $\mathcal{O}_i$ and $\mathcal{O}_j$, not on which specific boundary conditions $B_i$ and $B_j$ are chosen (as long as they are local and shrink to the operators $\mathcal{O}_i$ and $\mathcal{O}_j$, respectively). This independence of boundary condition is not obvious, and below we shall demonstrate the equivalence of Neumann-class and Dirichlet-class boundary conditions for the free boson.

The angular quantization Hilbert space $\mathcal{H}_{ij}$ describes a state space of the CFT on Minkowski space, i.e.~the states live on $\mathds{R}$ and are evolved in real time via the Hamiltonian $H_{ij} = \lim_{L\rightarrow\infty}[H_{\text{R}}^{B_i B_j}(L) - a(L)]$, where $a(L)$ is a possible $c$-number offset to render the eigenvalues of $H_{ij}$ finite. In particular, the Minkowski spectrum of states for any CFT is continuous and gapless, and there is no state/operator correspondence. Instead, the angular quantization associated to $\mathcal{H}_{ij}$ imparts asymptotic conditions on any state $|\psi\rangle$ which describe its behavior as $\sigma\rightarrow -\infty$ (from the $L\rightarrow\infty$ limit of $B_i$) and its behavior as $\sigma\rightarrow \infty$ (from the $L\rightarrow\infty$ limit of $B_j$). In this sense, angular quantization and hence Minkowski CFT itself possesses an asymptotics/operator correspondence, meaning that to each \emph{pair} of local operators in the CFT corresponds a state space $\mathcal{H}_{ij}$ with asymptotic conditions determined by that pair. As such, the full Hilbert space of the Minkowski CFT should be regarded as
\begin{equation}
\mathcal{H} = \bigoplus_{i,j}\mathcal{H}_{ij},
\end{equation}
where each pair of primaries labels a superselection sector of the theory (because no finite-energy operation can change the asymptotic conditions). If the CFT possesses a global symmetry, and if the endpoint operators $\mathcal{O}_i$ and $\mathcal{O}_j$ are charged under that symmetry, then the asymptotic conditions associated to the states in $\mathcal{H}_{ij}$ can be thought of as injecting those charges at the boundary of space. It is in this sense that angular quantization with nontrivial endpoint operators constructs the Minkowski space CFT with specified asymptotic charges. As the example considered in this paper, the shrinking limit of the finite-regulator angular quantization descriptions reviewed below describes the free boson Minkowski CFT with asymptotic momentum and winding charges. 

\subsection{Noncompact Free Boson with Neumann-Class Boundary Conditions}

Consider a noncompact free boson $X$ whose endpoint local operators are the exponential primaries $\mathcal{O}_i = \normal{e^{ik_1 X}}$ and $\mathcal{O}_j = \normal{e^{ik_2 X}}$. The Neumann-class boundary conditions shrinking to these operators fix the normal derivative of $X$ at the boundaries. Specifically, it was proven in \cite{Agia22} that the boundary conditions
\begin{align}
\left(z\partial X + \barred{z}\barred{\partial}X\right)\Big|_{|z|=\varepsilon} & = -ik_1
\\ \left(z\partial X + \barred{z}\barred{\partial}X\right)\Big|_{|z|=\frac{1}{\varepsilon}} & = +ik_2
\end{align}
on the two-holed sphere have the desired property. On the regulated cylinder with boundaries at $\sigma = \pm L$, these conditions read
\begin{align}
\partial_{\sigma}X\Big|_{\sigma=-L} & = -ik_1
\\ \partial_{\sigma}X\Big|_{\sigma=+L} & = +ik_2.
\end{align}
The finite regulator $\mathcal{H}_{k_1 k_2}^{\text{N}}(L)$ and $H_{\text{R}}^{\text{N}}(L)$ are then obtained by canonical quantization. Specifically, the mode expansion of the free boson on the semi-infinite strip is\footnote{We use a slightly different convention for the constant mode here compared to that used in \cite{Agia22}, namely $x_{\text{here}} = x_{\text{there}} - \frac{i(k_1+k_2)L}{12}$.}
\begin{multline}\label{noncompact N X}
X(\tau,\sigma) = x - \frac{\pi i p}{L}\tau - \frac{i(k_1-k_2)}{2}\sigma - \frac{i(k_1+k_2)}{4L}(\tau^2-\sigma^2)
\\ + i\sqrt{2}\sum_{\substack{\ell\in\mathds{Z} \\ \ell \neq 0}}\frac{\alpha_{\ell}}{\ell}e^{-\frac{\pi\ell\tau}{2L}}\cos\left(\frac{\pi\ell(\sigma+L)}{2L}\right),
\end{multline}
where the nonvanishing commutators are $[x,p] = i$ and $[\alpha_{\ell},\alpha_{\ell'}] = n\delta_{\ell,-\ell'}$. Then, the Rindler Hamiltonian associated to the Neumann-class boundary conditions is given by
\begin{equation}\label{noncompact N H}
H_{\text{R}}^{\text{N}}(L) = \frac{\pi p^2}{2L} - \frac{i(k_1+k_2)}{2\pi}x + \frac{\pi}{2L}\sum_{\ell=1}^{\infty}\alpha_{-\ell}\alpha_{\ell} + \frac{(k_1-k_2)^2}{8\pi}L + \frac{(k_1+k_2)^2}{12\pi}L - \frac{\pi}{48L}.
\end{equation}
Note that the constant mode $x$ appears explicitly in the Hamiltonian so that the cylinder evolution operator $e^{-2\pi H_{\text{R}}^{\text{N}}}$ contains the factor $e^{i(k_1+k_2)x}$ one expects from the endpoint operators $\normal{e^{ik_1 X}}$ and $\normal{e^{ik_2 X}}$. The finite-regulator angular quantization Hilbert space $\mathcal{H}_{k_1 k_2}^{\text{N}}$ is then built up from the oscillator vacuum $|p;\{0\}\rangle_{k_1 k_2}$, which is an eigenstate of the center-of-mass momentum operator $p$, by acting with all possible combinations of creation operators $\alpha_{-\ell}$, $\ell > 0$. That is, the Fock basis of $\mathcal{H}_{k_1 k_2}^{\text{N}}$ consists of the states
\begin{equation}
|p;\{N_{\ell}\}\rangle_{k_1 k_2} \equiv \left(\prod_{\ell=1}^{\infty}\frac{1}{\sqrt{\ell^{N_{\ell}}N_{\ell}!}}\alpha_{-\ell}^{N_{\ell}}\right)|p;\{0\}\rangle_{k_1 k_2};
\end{equation}
we use the scripts on the states to denote explicitly the factor $\mathcal{H}_{k_1 k_2}$ in the total Hilbert space $\mathcal{H}$ in which they reside. Using \eqref{noncompact N X}, the mode expansion of an exponential operator on the semi-infinite strip is
\begin{align}
\notag \normal{e^{ikX(\tau,\sigma)}} & = \left[\frac{\pi}{L}e^{\frac{\pi\tau}{L}}\cos\left(\frac{\pi\sigma}{2L}\right)\right]^{k^2/2}e^{\frac{(k_1-k_2)k}{2}\sigma + \frac{(k_1+k_2)k}{4L}(\tau^2+\sigma^2)}e^{ikx}e^{\frac{\pi k\tau}{L}p}
\\ \notag & \hspace{15pt} \times \exp\left[\sqrt{2}k\sum_{\ell=1}^{\infty}\frac{\alpha_{-\ell}}{\ell}e^{\frac{\pi\ell\tau}{2L}}\cos\left(\frac{\pi\ell(\sigma+L)}{2L}\right)\right]
\\ & \hspace{15pt} \times \exp\left[-\sqrt{2}k\sum_{\ell'=1}^{\infty}\frac{\alpha_{\ell'}}{\ell'}e^{-\frac{\pi\ell'\tau}{2L}}\cos\left(\frac{\pi\ell'(\sigma+L)}{2L}\right)\right].
\end{align}
Using these expressions, the thermal one-point function of a bulk exponential primary $\mathcal{O}_{k_3}(\tau,\sigma)$ on the cylinder at finite regulator is computed to be
\begin{multline}
\mathrm{Tr}_{\mathcal{H}_{k_1 k_2}^{\text{N}}}\hspace{-2pt}\left[\mathcal{O}_{k_3}(\tau,\sigma)e^{-2\pi H_{\text{R}}^{\text{N}}(L)}\right]_{S^1\times\mathds{R}}
\\ = \frac{\delta(k_1+k_2+k_3)}{\sqrt{2} \eta(\frac{2iL}{\pi})}\frac{e^{-(k_1^2+k_2^2)L/2}}{e^{(k_1^2-k_2^2)\sigma/2}}\left[e^{\frac{L}{6}}\eta\left(\frac{2iL}{\pi}\right)\vartheta_4\hspace{-2pt}\left(\hspace{-2pt}-\frac{i\sigma}{\pi}\bigg|\frac{2iL}{\pi}\right)\right]^{k_3^2/2}.
\end{multline}
The term in brackets approaches unity as $L\rightarrow\infty$.

\subsection{Noncompact Free Boson with Dirichlet-Class Boundary Conditions}

We again consider the angular quantization of the noncompact free boson $X$ with the same endpoint operators $\normal{e^{ik_1 X(0)}}$ and $\normal{e^{ik_2 X(\infty)}}$, but now with Dirichlet-class boundary conditions which fix the tangential derivative of $X$ at the boundaries. The appropriate Dirichlet-class boundary conditions are
\begin{equation}
\left(z\partial X - \barred{z}\barred{\partial}X\right)\Big|_{|z|=\varepsilon} = \left(z\partial X - \barred{z}\barred{\partial}X\right)\Big|_{|z|=\frac{1}{\varepsilon}} = 0
\end{equation}
on the two-holed sphere, or equivalently 
\begin{equation}
\partial_{\tau}X\Big|_{\sigma = \pm L} = 0
\end{equation}
on the regulated cylinder or strip. Of course, these boundary conditions do not depend on $k_1$ or $k_2$ and so cannot by themselves shrink to the desired endpoint operators $\normal{e^{ik_1 X}}$ and $\normal{e^{ik_2 X}}$. Instead, the above Dirichlet-class boundary conditions are imposed by boundary Lagrange multipliers, and the parameters $k_1$ and $k_2$ appear in the equations of motions for these Lagrange multipliers; for the details, see \cite{Agia22}. The mode expansion of the free boson on the semi-infinite strip is\footnote{We have rescaled $x_{\text{s}}\rightarrow \frac{x_{\text{s}}}{L}$ and $p_{\text{s}}\rightarrow p_{\text{s}}L$ compared to the conventions of \cite{Agia22}.}
\begin{equation}\label{noncompact D X}
X(\tau,\sigma) = x + \left(\frac{x_{\text{s}}}{L} - \frac{i(k_1-k_2)}{2}\right)\sigma + \sqrt{2}\sum_{\substack{\ell\in\mathds{Z} \\ \ell \neq 0}}\frac{\alpha_{\ell}}{\ell}e^{-\frac{\pi\ell\tau}{2L}}\sin\left(\frac{\pi\ell(\sigma+L)}{2L}\right),
\end{equation}
where the nonvanishing mode commutators are $[x,p] = [x_{\text{s}},p_{\text{s}}] = i$ and $[\alpha_{\ell},\alpha_{\ell'}] = \ell\delta_{\ell,-\ell'}$. The momenta $p$ and $p_{\text{s}}$ canonically conjugate to $x$ and $x_{\text{s}}$, respectively, appear in the mode expansions of the boundary Lagrange multipliers; since the latter are not needed in anything that follows, we shall not write them here. Note that now $X$ has two zero-momentum-modes, the center-of-mass mode $x$ and also the `slope' mode $x_{\text{s}}$. The constant $-\frac{i(k_1-k_2)}{2}$ was separated out in the definition of $x_{\text{s}}$ in \eqref{noncompact D X} for convenience, essentially to mimic the analogous term appearing in \eqref{noncompact N X}. The Rindler Hamiltonian associated to the Dirichlet-class boundary conditions is given by
\begin{equation}\label{noncompact D H}
H_{\text{R}}^{\text{D}}(L) = -\frac{i(k_1+k_2)}{2\pi}x + \frac{1}{2\pi L}x_{\text{s}}^2 + \frac{\pi}{2L}\sum_{\ell=1}^{\infty}\alpha_{-\ell}\alpha_{\ell} + \frac{(k_1-k_2)^2}{8\pi}L - \frac{\pi}{48L}.
\end{equation}
The Fock basis of $\mathcal{H}_{k_1 k_2}^{\text{D}}$ again consists of the states $|p',x_{\text{s}};\{N_{\ell}\}\rangle_{k_1 k_2}$, which we take to be simultaneous eigenstates of the center-of-mass momentum $p$ and slope mode $x_{\text{s}}$. A peculiarity about the slope mode is that its eigenstates, despite forming a continuum, are normalized to $\langle x_{\text{s}}|x_{\text{s}}\rangle = 1$; this unusual property is due to a proper treatment of the boundary Lagrange multipliers, as described in \cite{Agia22}. Using \eqref{noncompact D X}, the mode expansion of an exponential operator on the semi-infinite strip is
\begin{multline}
\normal{e^{ikX(\tau,\sigma)}} = \left[\frac{\pi}{4L\cos(\frac{\pi\sigma}{2L})}\right]^{\frac{k^2}{2}}\hspace{-3pt}e^{\frac{(k_1-k_2)k}{2}\sigma}e^{ikx}e^{\frac{ik\sigma}{L}x_{\text{s}}}\exp\hspace{-2pt}\left[i\sqrt{2}k\sum_{\ell=1}^{\infty}\frac{\alpha_{-\ell}}{\ell}e^{\frac{\pi\ell\tau}{2L}}\sin\hspace{-2pt}\left(\hspace{-2pt}\frac{\pi\ell(\sigma\hspace{-2pt}+\hspace{-2pt}L)}{2L}\hspace{-2pt}\right)\right]
\\ \times \exp\left[i\sqrt{2}k\sum_{\ell'=1}^{\infty}\frac{\alpha_{\ell'}}{\ell'}e^{-\frac{\pi\ell'\tau}{2L}}\sin\hspace{-1pt}\left(\hspace{-1pt}\frac{\pi\ell'(\sigma\hspace{-1pt}+\hspace{-1pt}L)}{2L}\hspace{-1pt}\right)\right].
\end{multline}
Using these expressions, the thermal one-point function of a bulk exponential primary $\mathcal{O}_{k_3}(\tau,\sigma)$ on the cylinder at finite regulator is computed to be
\begin{equation}
\mathrm{Tr}_{\mathcal{H}_{k_1 k_2}^{\text{D}}}\left[\mathcal{O}_{k_3}(\tau,\sigma)e^{-2\pi H_{\text{R}}^{\text{D}}(L)}\right]_{S^1\times\mathds{R}} = \frac{\delta(k_1+k_2+k_3)}{4\sqrt{2}\pi^2\eta(\frac{2iL}{\pi})}\frac{e^{-(k_1^2+k_2^2)L/2}}{e^{(k_1^2-k_2^2)\sigma/2}}\left[\frac{e^{\frac{L}{2}}\eta^3(\frac{2iL}{\pi})}{\vartheta_4\left(-\frac{i\sigma}{\pi}\big|\frac{2iL}{\pi}\right)}\right]^{k_3^2/2}.
\end{equation}
The term in brackets again approaches unity as $L\rightarrow\infty$.

\subsection{Compact Free Boson}

Now consider a compact free boson $X$ of radius $R$, i.e.~$X \sim X + 2\pi R$. We take the endpoint operators to be exponential primaries $\mathcal{O}_{n_1,w_1}(0)$ and $\mathcal{O}_{n_2,w_2}(\infty)$, where $n$ and $w$ are integers denoting momentum and winding, respectively. For the excision regulator we shall employ Dirichlet-class boundary conditions, suppressing the label `$\text{D}$' since it is the only regulator for the compact boson used in this paper. The relevant Dirichlet-class boundary conditions are
\begin{align}
\left(z\partial X - \barred{z}\barred{\partial}X\right)\Big|_{|z| = \varepsilon} & = -iw_1 R
\\ \left(z\partial X - \barred{z}\barred{\partial}X\right)\Big|_{|z|=\frac{1}{\varepsilon}} & = +iw_2 R
\end{align}
on the two-holed sphere, or equivalently
\begin{align}
\partial_{\tau}X\Big|_{\sigma=-L} & = +w_1 R
\\ \partial_{\tau}X\Big|_{\sigma=+L} & = -w_2 R
\end{align}
on the regulated cylinder or strip. As with the noncompact boson, the momenta $n_1$ and $n_2$ of the endpoint operators do not appear in the boundary conditions; as before, the above relations are imposed via boundary Lagrange multipliers, and $n_1$ and $n_2$ appear in their equations of motion. The mode expansion of the compact boson on the semi-infinite strip is
\begin{multline}\label{compact X}
X(\tau,\sigma) = x + \left(\frac{x_{\text{s}}}{L} - \frac{i(n_1-n_2)}{2R}\right)\sigma + \frac{(w_1-w_2)R}{2}\tau - \frac{(w_1+w_2)R}{2L}\tau\sigma
\\ + \sqrt{2}\sum_{\substack{\ell\in\mathds{Z} \\ \ell \neq 0}}\frac{\alpha_{\ell}}{\ell}e^{-\frac{\pi\ell\tau}{2L}}\sin\left(\frac{\pi\ell(\sigma+L)}{2L}\right),
\end{multline}
where once again $[x,p] = [x_{\text{s}},p_{\text{s}}] = i$ and $[\alpha_{\ell},\alpha_{\ell'}] = \ell\delta_{\ell,-\ell'}$, with $x$ and $x_{\text{s}}$ being the two independent zero-momentum-modes. The associated Rindler Hamiltonian for the compact boson is
\begin{multline}\label{compact H}
H_{\text{R}}(L) = -\frac{i(n_1+n_2)}{2\pi R}x + \frac{i(w_1-w_2)R}{2}p - \frac{i(w_1+w_2)R}{2}p_{\text{s}} + \frac{1}{2\pi L}x_{\text{s}}^2 + \frac{\pi}{2L}\sum_{\ell=1}^{\infty}\alpha_{-\ell}\alpha_{\ell}
\\ + \frac{(n_1-n_2)^2}{8\pi R^2}L + \frac{(w_1+w_2)^2 R^2}{24\pi}L + \frac{(w_1-w_2)^2 R^2}{8\pi}L - \frac{\pi}{48L}.
\end{multline}
The Fock basis of the finite-regulator Hilbert space $\mathcal{H}_{n_1,w_1}^{n_2,w_2}(L)$ similarly consists of the states $|m,x_{\text{s}};\{N_{\ell}\}\rangle_{n_1,w_1}^{n_2,w_2}$, for which $p|m,x_{\text{s}};\{N_{\ell}\}\rangle_{n_1,w_1}^{n_2,w_2} = \frac{m}{R}|m,x_{\text{s}};\{N_{\ell}\}\rangle_{n_1,w_1}^{n_2,w_2}$ for any given $m \in \mathds{Z}$. As before, the slope mode eigenstates obey $\langle x_{\text{s}}|x_{\text{s}}\rangle = 1$. The mode expansion of a bulk exponential primary $\mathcal{O}_{n,w}$, heuristically of the form $\normal{e^{i(k_{\text{L}}X_{\text{L}}+k_{\text{R}}X_{\text{R}})}}$ with $k_{\text{L}} = \frac{n}{R} + wR$ and $k_{\text{R}} = \frac{n}{R} - wR$, on the semi-infinite strip is given precisely by
\begin{align}
\notag \mathcal{O}_{n,w}(\tau,\sigma) & = \left[\frac{\pi}{4L\cos(\frac{\pi\sigma}{2L})}\right]^{\frac{n^2}{2R^2}}\left[\frac{\pi}{L}\cos\left(\frac{\pi\sigma}{2L}\right)\right]^{\frac{w^2 R^2}{2}}e^{\frac{(w_1+w_2)wR^2}{12}L + \frac{\pi i}{2}nw - \frac{\pi i\sigma}{2L}nw + \frac{\pi\tau}{2L}w^2 R^2}
\\ \notag & \hspace{15pt} \times e^{[\frac{(n_1-n_2)n}{2R^2}+\frac{(w_1-w_2)wR^2}{2}]\sigma}e^{\frac{i[(n_1-n_2)w + n(w_1-w_2)]}{2}\tau}e^{\frac{(w_1+w_2)R}{4L}[wR(\tau^2-\sigma^2) - \frac{2in}{R}\tau\sigma]}
\\ \notag & \hspace{15pt} \times e^{\frac{in}{R}x}e^{-\pi i wR p}e^{\pi iwR p_{\text{s}}}e^{\frac{1}{L}(\frac{in}{R}\sigma-wR\tau)x_{\text{s}}}
\\ \notag & \hspace{15pt} \times \exp\left\{i\sqrt{2}\sum_{\ell=1}^{\infty}\frac{\alpha_{-\ell}}{\ell}e^{\frac{\pi\ell\tau}{2L}}\left[\frac{n}{R}\sin\left(\frac{\pi\ell(\sigma+L)}{2L}\right) + iwR\cos\left(\frac{\pi\ell(\sigma+L)}{2L}\right)\right]\right\}
\\ & \hspace{15pt} \times \exp\left\{i\sqrt{2}\sum_{\ell'=1}^{\infty}\frac{\alpha_{\ell'}}{\ell'}e^{-\frac{\pi\ell'\tau}{2L}}\left[\frac{n}{R}\sin\left(\frac{\pi\ell'(\sigma+L)}{2L}\right) - iwR\cos\left(\frac{\pi\ell'(\sigma+L)}{2L}\right)\right]\right\}.
\end{align}
Using these expressions, the thermal one-point function of a bulk exponential primary $\mathcal{O}_{n_3,w_3}(\tau,\sigma)$ on the cylinder at finite regulator is computed to be
\begin{multline}
\mathrm{Tr}_{\mathcal{H}_{n_1,w_1}^{n_2,w_2}}\left[\mathcal{O}_{n_3,w_3}(\tau,\sigma)e^{-2\pi H_{\text{R}}(L)}\right]_{S^1\times\mathds{R}} = \frac{R\delta_{n_1+n_2+n_3,0}\delta_{w_1+w_2+w_3,0}}{4\sqrt{2}\pi^2}
\\ \times \frac{(-1)^{n_3 w_1}e^{-(\Delta_1+\Delta_2)L}}{e^{i(n_1 w_1 - n_2 w_2)\tau}e^{(\Delta_1-\Delta_2)\sigma}}\frac{1}{\eta(\frac{2iL}{\pi})}\frac{[e^{\frac{L}{6}}\eta(\frac{2iL}{\pi})]^{\frac{1}{2}(\frac{3n_3^2}{R^2}+w_3^2 R^2)}}{\vartheta_4\big(\hspace{-2pt}-\hspace{-2pt}\frac{i\sigma}{\pi}\big|\frac{2iL}{\pi}\big){}^{\frac{1}{2}(\frac{n_3^2}{R^2}-w_3^2 R^2)}},
\end{multline}
where $\Delta_j = \frac{1}{2}(\frac{n_j^2}{R^2} + w_j^2 R^2)$.

\section{The Shrinking Limit: Asymptotic Conditions}\label{shrinking limit}

Now we would like to describe in detail the shrinking limit $L\rightarrow\infty$ for the free boson angular quantizations briefly reviewed above. This limit is not trivial, as many expressions have parts that blow up or decay as $L \rightarrow \infty$, which can obfuscate the meaningful finite quantities. Nevertheless, some aspects of this limit are straightforward, for instance the manner in which the discrete oscillator modes become a continuum in the infinite-volume limit, a familiar feature from introductory quantum mechanics. However, one of the most important aspects of angular quantization at finite regulator is the treatment of the boundary conditions, which in the shrinking limit transmute into the asymptotic conditions, so the behavior of the fields near spatial infinity will be of paramount importance.

Let us be clear about the end goal of the shrinking limit. Ultimately, we are constructing the CFT on $\mathds{R}^{1,1}$ as a collection of superselection sectors corresponding to specific asymptotic conditions. For each superselection sector, there is a vector space $\mathcal{H}_{ij}$ of states with corresponding Hamiltonian $H_{ij}$, and each state $|\psi\rangle_{ij} \in \mathcal{H}_{ij}$ obeys the asymptotic conditions specified by the original primaries $\mathcal{O}_i$ and $\mathcal{O}_j$. Moreover, for each pair of states in $\mathcal{H}_{ij}$ the norm of the Hamiltonian $H_{ij}$ differs by a finite amount. Finally, there must exist an inner product pairing on the total state vector space $\bigoplus_{i,j}\mathcal{H}_{ij}$. In short, we aim for an enumeration of finite-energy states together with a vector pairing thereof.

In the free boson examples studied here, the existence of a Fock construction for the angular quantization at finite regulator greatly simplifies the task of enumeration, allowing us to describe the resulting states concretely and explicitly. Likewise, that the CFT has a field description allows us to express the asymptotic conditions directly in terms of the field itself.

The question remains: what are the asymptotic conditions corresponding to the free boson exponential primary endpoint operators? As we have already mentioned, these asymptotic conditions cannot depend on the specific finite-regulator boundary conditions used, as long as they shrink to the desired operators. For instance, the Neumann-class and Dirichlet-class boundary conditions described in the previous section implement the boundary conditions quite differently (the former fixes the normal derivative while the latter fixes the tangential derivative), but they both must fix the same asymptotic conditions. Fortunately, these asymptotic conditions are simple to obtain from the same source which motivated the boundary conditions --- the OPE. Specifically, the singular parts of the OPEs of $\partial X$ and $\barred{\partial}X$ with an exponential primary $\normal{e^{ikX}}$ are
\begin{align}
\partial X(z)\normal{e^{ikX(0)}} & \sim -\frac{ik}{2z}\normal{e^{ikX(0)}}
\\ \barred{\partial}X(\barred{z})\normal{e^{ikX(0)}} & \sim -\frac{ik}{2\barred{z}}\normal{e^{ikX(0)}}.
\end{align}
The first OPE means that $\partial X(z)$ behaves, to leading order, like the function $-\frac{ik}{2z}$ when in the presence of an exponential operator $\normal{e^{ikX}}$ at the origin, and the second OPE means that $\barred{\partial}X(\barred{z})$ similarly behaves like $-\frac{ik}{2\barred{z}}$. As usual, these statements are operatorial, meaning they apply inside any correlator as long as no other operator is closer to the exponential operator in question. Thus, both $z\partial X(z)$ and $\barred{z}\partial X(z)$ behave like the constant $-\frac{ik}{2}$ near the operator $\normal{e^{ikX(0)}}$. In fact, these statements uniquely characterize the exponential primary operator, and hence they \emph{are} the asymptotic conditions. Previously, we chose to use $z\partial X(z) \pm \barred{z}\partial X(z)$ instead since these are the more natural linear combinations on the cylinder; we make this transition again here since the Minkowski space CFT is obtained upon analytic continuation of the $L\rightarrow\infty$ limit on the strip. Therefore, the asymptotic conditions in the sector $\mathcal{H}_{k_1 k_2}$ for the noncompact free boson are
\begin{align}
\label{noncompact asymptotics 1}\hspace{-130pt}\text{noncompact free boson:}\quad \qquad \partial_{\sigma}X(\tau,\sigma) & \ \stackrel{\sigma\rightarrow -\infty}{\longrightarrow} \ -ik_1
\\ \partial_{\tau}X(\tau,\sigma) & \ \stackrel{\sigma\rightarrow -\infty}{\longrightarrow} \ 0
\\ \partial_{\sigma}X(\tau,\sigma) & \ \stackrel{\sigma\rightarrow +\infty}{\longrightarrow} \ +ik_2
\\ \label{noncompact asymptotics 4} \partial_{\tau}X(\tau,\sigma) & \ \stackrel{\sigma\rightarrow +\infty}{\longrightarrow} \ 0,
\end{align}
for all Euclidean $\tau \in \mathds{R}$. These relations must hold in arbitrary correlators, but it suffices to check the asymptotic conditions by computing the expectation value $\langle X(\tau,\sigma)\rangle_{k_1 k_2}$ in any state in $\mathcal{H}_{k_1 k_2}$. Likewise, the asymptotic conditions in the sector $\mathcal{H}_{n_1,w_1}^{n_2,w_2}$ for the compact free boson are
\begin{align}
\label{compact asymptotics 1}\hspace{-105pt}\text{compact free boson:}\quad \qquad \partial_{\sigma}X(\tau,\sigma) & \ \stackrel{\sigma\rightarrow -\infty}{\longrightarrow} \ -\frac{in_1}{R}
\\ \partial_{\tau}X(\tau,\sigma) & \ \stackrel{\sigma\rightarrow -\infty}{\longrightarrow} \ +w_1 R
\\ \partial_{\sigma}X(\tau,\sigma) & \ \stackrel{\sigma\rightarrow +\infty}{\longrightarrow} \ +\frac{in_2}{R}
\\ \label{compact asymptotics 4} \partial_{\tau}X(\tau,\sigma) & \ \stackrel{\sigma\rightarrow +\infty}{\longrightarrow} \ -w_2 R,
\end{align}
again for all Euclidean $\tau \in \mathds{R}$.

Several comments are in order. The first is that the asymptotic conditions \eqref{noncompact asymptotics 1}-\eqref{noncompact asymptotics 4} and \eqref{compact asymptotics 1}-\eqref{compact asymptotics 4} fix both the behavior of the spatial derivative \emph{and} that of the temporal derivative of the free boson. In contrast, the boundary conditions used in the finite-regulator angular quantization fixed either the spatial derivative \emph{or} that of the temporal derivative. The reason is because the temporal derivative of $X$ is always related to its conjugate momentum, and placing such conditions on both fields simultaneously would violate the classical Poisson algebra, leading to an inconsistent canonical quantization. It is thus nontrivial to ensure all asymptotic conditions are obeyed despite starting from states which initially satisfy half of them and only at one point.

The second comment is that the asymptotic conditions are far simpler in the special case that the net asymptotic charge vanishes, i.e.~for $k_1+k_2 = 0$ in the noncompact case and for $n_1+n_2=w_1+w_2=0$ in the compact case. While this scenario does not look much different than the generic one in \eqref{noncompact asymptotics 1}-\eqref{noncompact asymptotics 4} or \eqref{compact asymptotics 1}-\eqref{compact asymptotics 4}, the simplification arises because these asymptotic conditions can be satisfied by letting the expectation values of $\partial_{\sigma}X(\tau,\sigma)$ and $\partial_{\tau}X(\tau,\sigma)$ be constants for all $\tau$ and $\sigma$. An exactly linear configuration automatically solves the classical equation of motion, and hence the asymptotic conditions can be obeyed without any contribution from the oscillators. On the other hand, when the sum of the asymptotic charges does not vanish, it is impossible to satisfy the asymptotic conditions without contributions from the oscillators. In fact, any finite oscillator contribution cannot change the asymptotics, so the nonzero net charge case requires infinitely-excited oscillator states. Due to this sharp difference, the net vanishing and net nonzero charge scenarios are treated separately in the following two sections.

Lastly, the third comment is that \eqref{noncompact asymptotics 1}-\eqref{noncompact asymptotics 4} and \eqref{compact asymptotics 1}-\eqref{compact asymptotics 4} apply only for real $\tau$, i.e.~in Euclidean signature. Indeed, the asymptotic conditions in angular quantization arise because the shrinking limit in Euclidean signature is always conformally equivalent to the shrinking of holes on the sphere into local endpoint operators. Obviously there is no analog of shrinking a noncompact Lorentzian time direction to a point and recovering a local operator. Instead, the Euclidean data is enough to completely specify the states of the theory, from which we may simply derive the asymptotic behavior in the Lorentzian theory by performing the analytic continuation. For this reason, all expressions below will be evaluated for arbitrary complex $\tau$; computing $\langle X(\tau,\sigma)\rangle_{ij}$ and restricting to real $\tau$ will verify the asymptotic conditions, and subsequently restricting to pure imaginary $\tau$ will reveal the Lorentzian consequences of injecting asymptotic charges.

\section{Special Case: Zero Net Charge}\label{zero net charge}

As discussed in Section \ref{shrinking limit}, the case wherein the net asymptotic charge vanishes (namely $k_1 + k_2 = 0$ and $n_1 + n_2 = w_1 + w_2 = 0$ for the noncompact and compact boson, respectively) has the simplest shrinking limit because the oscillators merely need to have vanishing contributions in the asymptotic limit. We shall first treat the oscillator sectors separately before including the zero-momentum-modes.

We already know that the oscillator ground state is a part of the angular quantization Hilbert space when the net asymptotic charge vanishes because $\langle \{0\}|\alpha_{\ell}|\{0\}\rangle = 0$. It is simple to see that the entire Fock space built on top of $|\{0\}\rangle$ is also part of this space by showing that the total contribution to $\langle X(\tau,\sigma)\rangle$ from the sum over $\alpha_{\ell}$ vanishes for $\sigma\rightarrow \pm\infty$ as $L\rightarrow\infty$. Specifically, consider the oscillator state $|\{N_{\ell}\}\rangle$. From 
\begin{align}
\langle\{N_{\ell'}-\mathrm{sgn}(\ell)\delta_{|\ell|,\ell'}\}|\alpha_{\ell}|\{N_{\ell'}\}\rangle & = \frac{\langle\{0\}|\alpha_{|\ell|}^{N_{|\ell|}-\mathrm{sgn}(\ell)}\alpha_{\ell}\alpha_{-|\ell|}^{N_{|\ell|}}|\{0\}\rangle}{\sqrt{|\ell|^{2N_{|\ell|}-\mathrm{sgn}(\ell)}(N_{|\ell|}-\mathrm{sgn}(\ell))!N_{|\ell|}!}}
\\ & = \sqrt{|\ell|(N_{|\ell|}+\theta_{\ell})},
\end{align}
where $\theta_{\ell}$ is unity if $\ell > 0$ and vanishes otherwise, the oscillator contribution to the matrix element of $X(\tau,\sigma)$ is
\begin{multline}
\langle \{N_{\ell'}-\mathrm{sgn}(\ell)\delta_{|\ell|,\ell'}\}|X_{\text{osc}}(\tau,\sigma)|\{N_{\ell'}\}\rangle
\\ = \sqrt{2}\sum_{\ell=1}^{\infty}\sqrt{\frac{N_{\ell}}{\ell}}\left[\sqrt{1+\frac{1}{N_{\ell}}}e^{-\frac{\pi\ell\tau}{2L}} \mp e^{\frac{\pi\ell\tau}{2L}}\right]\genfrac{\{}{\}}{0pt}{0}{i\cos}{\sin}\left(\frac{\pi\ell(\sigma+L)}{2L}\right),
\end{multline}
where the top line is for Neumann-class boundary conditions and the bottom line is for Dirichlet-class boundary conditions. The surviving momentum states in the shrinking limit, constructed explicitly in the subsection below, are those for which the mode number $\ell$ is scaled with $L$ as $L\rightarrow\infty$ with $\frac{\ell}{L}$ held finite. For such states, the summand is proportional to $\frac{1}{\sqrt{\ell}}$, and there is no possible enhancement from having infinitely many terms in the sum because only finitely many of the occupation numbers $N_{\ell}$ can be nonzero by definition of a Fock space, so it may seem like the above contribution vanishes. However, the sum over $\ell$ then also contributes a factor of $L$ when being turned into an integral, so it seems like the above matrix element is infinite in the shrinking limit. This infinity is simply due to the fact that the basis oscillator momentum states are not themselves normalizable. In forming normalizable sums over these states, the above matrix element necessarily vanishes as $\sigma\rightarrow \pm\infty$ since the wavepacket has no support there. Finally, there are the terms from mode numbers $\ell$ which do not scale with $L$ as $L\rightarrow\infty$. All such terms certainly remain finite, but they are also independent of position since the dependence on the latter occurs only in the form $\frac{\ell\tau}{L}$ and $\frac{\ell\sigma}{L}$, both of which vanish in the shrinking limit. These contributions from finite mode numbers are hence constants, which do not alter the derivative asymptotics; in fact, we never have to consider states involving finite mode numbers in the shrinking limit because their effects can always be absorbed into a redefinition of the constant mode of $X$. Therefore, we have proven that the simple Fock space structure of the oscillators persists after the shrinking limit in the case of vanishing net asymptotic charge.

\subsection{Oscillator States}

This subsection details the continuum limit of momentum modes for a free field in a box and hence may be skipped without loss of continuity; it is included merely for completeness. The pure oscillator sectors of the free theories discussed in this paper are all of the same form, consisting of quantized momentum solutions of the Klein-Gordon equation with either pure Neumann or pure Dirichlet boundary conditions. In this way, $L$ acts as an infrared regulator by replacing the infinite spatial volume of Minkowski space $\mathds{R}^{1,1}$ with the finite volume $2L$. In the shrinking limit, the discrete oscillator states $|\{N_{\ell}\}\rangle$ thus become the continuum of momentum states in the usual way. However, even though $|\{N_{\ell}\}\rangle$ is the most convenient oscillator basis to use at finite regulator, it is a rather inconvenient basis in which to consider the shrinking limit. For fixed $\ell$, the state $|N_{\ell} = 1\rangle$ is a superposition of waves with spatial momenta $\pm \frac{\pi \ell}{2L}$ on the strip, so the finite-momentum states are obtained by writing the mode number as $\ell = \frac{qL}{\pi}$ and taking the limit $L\rightarrow\infty$ while keeping $q$ finite; the resulting continuum state is a standing-wave superposition involving both momentum $q$ and momentum $-q$. Let us note that states with multiple finite-momentum excitations in the shrinking limit arise from scaling multiple mode numbers to infinity in this way, and as such it is highly atypical for two modes to be scaled to land on the same finite momentum $q$. To reflect the phase space suppression of such coincident-momenta states, it is best to change bases from $|\{N_{\ell}\}\rangle$, where $N_{\ell}$ is any natural number for each $\ell$, to a basis consisting of a finite number of modes excited with no assumption about different mode numbers being related. That is, we change to a new basis of $j$-particle states for any $j = 0,1,2,\dotsc$, spanned by states of the form $|\{\ell_1,\ell_2,\dotsc,\ell_j\}\rangle$ defined by
\begin{equation}
|\{\ell_1,\ell_2,\dotsc,\ell_j\}\rangle \equiv \left(\prod_{i=1}^j\frac{i^{\ell_i}}{\sqrt{\ell_i}}\alpha_{-\ell_i}\right)|\{0\}\rangle,
\end{equation}
where $|\{0\}\rangle$ is the oscillator vacuum; the choice of phase $i^{\ell_i}$ will be convenient in the shrinking limit below. These states are orthonormalized when all of the $\ell_i$ are distinct but are missing the symmetry factors when some of the mode numbers coincide. Nevertheless, this lack of normality allows for the simple completeness relation
\begin{equation}
\mathds{1} = \sum_{j=0}^{\infty}\frac{1}{j!}\sum_{\ell_1=1}^{\infty}\cdots\sum_{\ell_j=1}^{\infty}\left|\{\ell_1,\dotsc,\ell_j\}\right\rangle\left\langle \{\ell_1,\dotsc,\ell_j\}\right|,
\end{equation}
where all terms in the inner sums with distinct $\{\ell_i\}$ are overcounted by $j!$, and the terms with some of the mode numbers coincident are less overcounted but compensated by their larger norms.

Start with a fixed large $L$ and parametrize the shrinking limit by writing $L(\lambda)$ where $L(0) = L$ and $L(1) = \infty$, for example. Then we consider sequences of states by writing each mode number as
\begin{equation}
\ell(\lambda) = \frac{q\big(L(\lambda)\big)}{\pi}L(\lambda),
\end{equation}
where $q(\infty) = q$ is the desired oscillator momentum in the continuum limit; the subleading order $\frac{1}{L}$ correction to $q(L)$ is important because this is what actually forces $\ell(\lambda)$ to be an integer as a function of $\lambda$. There are of course infinitely many ways to choose $q(L)$, but the one-particle states in the shrinking limit are equivalence classes thereof that form a complete Dirac-orthonormal basis of $L^2(\mathds{R})$, so we must also be careful not to overcount. For generic $q$ and $L$, $\frac{qL}{\pi}$ is irrational, so two seemingly easy options are to define the subleading piece in $q(L)$ so that $\ell(\lambda)$ is either the floor or the ceiling of $\frac{qL}{\pi}$. However, it is not obvious that these two limits actually lead to independent states because they are not orthonormal. A far more natural choice is to define the subleading piece in $q(L)$ so that $\ell(\lambda)$ is either always even or always odd, because the oscillators in the Neumann or Dirichlet mode expansions, for example in \eqref{noncompact N X} and \eqref{noncompact D X} respectively, involve either $\cos(\frac{\pi \ell \sigma}{2L} + \frac{\pi \ell}{2})$ or $\sin(\frac{\pi\ell\sigma}{2L} + \frac{\pi \ell}{2})$. For any of the Neumann and Dirichlet oscillator expansions on the very wide strip, we may then rewrite
\begin{align}
i\sqrt{2}\sum_{\substack{\ell\in\mathds{Z} \\ \ell\neq 0}}\frac{\alpha_{\ell}}{\ell}e^{-\frac{\pi\ell\tau}{2L}}\hspace{-2pt}\cos\left(\hspace{-2pt}\frac{\pi\ell(\sigma\hspace{-2pt}+\hspace{-2pt}L)}{2L}\hspace{-2pt}\right) & = \frac{i}{\sqrt{2}}\sum_{\substack{\ell\in\mathds{Z} \\ \ell\neq 0}}\hspace{-2pt}\left(\frac{\alpha_{\ell}^{\text{L}}}{\ell}e^{\frac{\pi i\ell}{L}(\sigma+i\tau)} + \frac{\alpha_{\ell}^{\text{R}}}{\ell}e^{-\frac{\pi i\ell}{L}(\sigma-i\tau)}\right)
\\ \sqrt{2}\sum_{\substack{\ell\in\mathds{Z} \\ \ell\neq 0}}\frac{\alpha_{\ell}}{\ell}e^{-\frac{\pi\ell\tau}{2L}}\hspace{-2pt}\sin\hspace{-2pt}\left(\hspace{-2pt}\frac{\pi\ell(\sigma\hspace{-2pt}+\hspace{-2pt}L)}{2L}\hspace{-2pt}\right) & = \frac{i}{\sqrt{2}}\sum_{\substack{\ell\in\mathds{Z} \\ \ell\neq 0}}\hspace{-2pt}\left(-\frac{\alpha_{\ell}^{\text{L}}}{\ell}e^{\frac{\pi i\ell}{L}(\sigma+i\tau)} + \frac{\alpha_{\ell}^{\text{R}}}{\ell}e^{-\frac{\pi i\ell}{L}(\sigma-i\tau)}\right),
\end{align}
where we have defined the strange-looking operators
\begin{align}
\alpha_{\ell}^{\text{L}} & \equiv \frac{(-1)^{\ell}}{2}\left[\alpha_{2\ell} - i\left(\frac{\ell e^{-\frac{\mathrm{sgn}(\ell)\pi i}{2L}(\sigma+i\tau)}}{|\ell|-1/2}\right)\alpha_{2\ell-\mathrm{sgn}(\ell)}\right]
\\ \alpha_{\ell}^{\text{R}} & \equiv \frac{(-1)^{\ell}}{2}\left[\alpha_{2\ell} + i\left(\frac{\ell e^{\frac{\mathrm{sgn}(\ell)\pi i}{2L}(\sigma-i\tau)}}{|\ell|-1/2}\right)\alpha_{2\ell-\mathrm{sgn}(\ell)}\right],
\end{align}
where $\mathrm{sgn}(\ell) \equiv \frac{\ell}{|\ell|}$ and we suppress the $\tau$ and $\sigma$ dependence in the notation. These operators satisfy
\begin{align}\label{LL oscillator finite commutator}
[\alpha_{\ell}^{\text{L}},\alpha_{\ell'}^{\text{L}}] & = [\alpha_{\ell}^{\text{R}},\alpha_{\ell'}^{\text{R}}] = \ell\left(1 + \frac{1}{2(2|\ell|-1)}\right)\delta_{\ell,-\ell'}
\\ \label{LR oscillator finite commutator} [\alpha_{\ell}^{\text{L}},\alpha_{\ell'}^{\text{R}}] & = \frac{\ell}{2}\left(1 - \frac{e^{-\frac{\mathrm{sgn}(\ell)\pi i}{L}\sigma}}{1-1/2|\ell|}\right)\delta_{\ell,-\ell'}.
\end{align}
Moreover, the oscillator part of any of the Hamiltonians \eqref{noncompact N H}, \eqref{noncompact D H} or \eqref{compact H} is universal, and is rewritten as
\begin{multline}
\frac{\pi}{2L}\sum_{\ell=1}^{\infty}\alpha_{-\ell}\alpha_{\ell} = \frac{\pi}{2L}\sum_{\ell=1}^{\infty}\left\{\left[1 + \left(1-\frac{1}{2\ell}\right)^{2}\right]\sec^2\left(\frac{\pi\sigma}{2L}\right)\left[\alpha_{-\ell}^{\text{L}}\alpha_{\ell}^{\text{L}} + \alpha_{-\ell}^{\text{R}}\alpha_{\ell}^{\text{R}}\right] \right. 
\\ \left. + \left[\frac{1\hspace{-2pt}-\hspace{-2pt}\frac{1}{4\ell}}{\ell}\sec^2\hspace{-2pt}\left(\frac{\pi\sigma}{2L}\right)-2\tan^2\hspace{-2pt}\left(\frac{\pi\sigma}{2L}\right)\right]\hspace{-2pt}\left[\alpha_{-\ell}^{\text{L}}\alpha_{\ell}^{\text{R}} + \alpha_{-\ell}^{\text{R}}\alpha_{\ell}^{\text{L}}\right] - 2i\tan\hspace{-2pt}\left(\frac{\pi\sigma}{2L}\right)\hspace{-3pt}\left[\alpha_{-\ell}^{\text{L}}\alpha_{\ell}^{\text{R}} - \alpha_{-\ell}^{\text{R}}\alpha_{\ell}^{\text{L}}\right]\hspace{-2pt}\right\},
\end{multline}
which is of course independent of $\tau$ and $\sigma$. For any fixed $\tau$ and $\sigma$, the operators  $\alpha_{-\ell}^{\text{L}}$ and $\alpha_{-\ell}^{\text{R}}$ create a valid one-particle basis, though such a description at finite regulator is complicated because these are obviously not eigenstates. Now write $\ell = \frac{q(L)}{\pi}L$ with $q(L) > 0$ understood as above, and define
\begin{alignat}{2}
\sqrt{\frac{q}{2\pi}}\s\s a(-q) & \equiv \lim_{L\rightarrow\infty} \alpha_{\frac{q(L)}{\pi}L}^{\text{L}} \qquad\quad & \sqrt{\frac{q}{2\pi}}\s\s a(-q)^{\dagger} & \equiv \lim_{L\rightarrow\infty} \alpha_{-\frac{q(L)}{\pi}L}^{\text{L}}
\\ \sqrt{\frac{q}{2\pi}}\s\s a(q) & \equiv \lim_{L\rightarrow\infty} \alpha_{\frac{q(L)}{\pi}L}^{\text{R}} \qquad\quad & \sqrt{\frac{q}{2\pi}}\s\s a(q)^{\dagger} & \equiv \lim_{L\rightarrow\infty} \alpha_{-\frac{q(L)}{\pi}L}^{\text{R}}.
\end{alignat}
From the distributional limit
\begin{equation}
2L\delta_{\ell_1,\ell_2} \ \stackrel{L\rightarrow \infty}{\longrightarrow} \ 2\pi\delta(q_1 - q_2),
\end{equation}
we immediately see that the limits of the finite-regulator commutators \eqref{LL oscillator finite commutator} and \eqref{LR oscillator finite commutator} are
\begin{align}
[a(q_1),a(q_2)^{\dagger}] & = 2\pi \delta(q_1-q_2)
\\ [a(q_1),a(q_2)] & = [a(q_1)^{\dagger}, a(q_2)^{\dagger}] = 0
\end{align}
for all $q_1,q_2 \in \mathds{R}$. Moreover, the shrinking limits of any of the Neumann and Dirichlet oscillator expansions on the very wide strip are
\begin{align}
i\sqrt{2}\hspace{-1pt}\sum_{\substack{\ell\in\mathds{Z} \\ \ell\neq 0}}\hspace{-2pt}\frac{\alpha_{\ell}}{\ell}e^{-\frac{\pi\ell\tau}{2L}}\hspace{-2pt}\cos\hspace{-2pt}\left(\hspace{-2pt}\frac{\pi \ell(\sigma\hspace{-2pt}+\hspace{-2pt}L)}{2L}\hspace{-2pt}\right) & \stackrel{L\rightarrow\infty}{\longrightarrow} \frac{i}{\sqrt{2}}\hspace{-2pt}\int_{-\infty}^{\infty}\hspace{-2pt}\frac{dq}{\sqrt{2\pi|q|}}\hspace{-3pt}\left[\hspace{-1pt}a(q)e^{\hspace{-0.5pt}-\hspace{-0.5pt}|q|\tau - iq\sigma} \hspace{-3pt}-\hspace{-2pt} a(q)^{\dagger}e^{|q|\tau + iq\sigma}\hspace{-1pt}\right]
\\ \sqrt{2}\hspace{-1pt}\sum_{\substack{\ell\in\mathds{Z} \\ \ell\neq 0}}\hspace{-2pt}\frac{\alpha_{\ell}}{\ell}e^{-\frac{\pi\ell\tau}{2L}}\hspace{-2pt}\sin\hspace{-2pt}\left(\hspace{-2pt}\frac{\pi \ell(\sigma\hspace{-2pt}+\hspace{-2pt}L)}{2L}\hspace{-2pt}\right) & \stackrel{L\rightarrow\infty}{\longrightarrow} \frac{i}{\sqrt{2}}\hspace{-2pt}\int_{-\infty}^{\infty}\hspace{-3pt}\frac{dq}{\sqrt{2\pi|q|}}\mathrm{sgn}(q)\hspace{-3pt}\left[\hspace{-1pt}a(q)e^{-|q|\tau - iq\sigma} \hspace{-3pt}-\hspace{-2pt} a(q)^{\dagger}e^{|q|\tau + iq\sigma}\hspace{-1pt}\right]\hspace{-2pt}.
\end{align}
Therefore, $a(q)^{\dagger}$ indeed creates a properly normalized plane-wave wavefunction $e^{iq\sigma}$ for any real value of the momentum $q$, so that the set of $|\{q\}\rangle = a(q)^{\dagger}|\{0\}\rangle$ manifestly forms a complete Dirac-orthonormal basis of the one-particle Hilbert space $L^2(\mathds{R})$, as required. At finite regulator, the left- and right-moving oscillators $\alpha_{\ell}^{\text{L}}$ and $\alpha_{\ell}^{\text{R}}$ may be expanded in terms of $a(\pm q)$, receiving both corrections to the overall normalization due to \eqref{LL oscillator finite commutator} as well as couplings between $a(q)$ and $a(-q)$ due to \eqref{LR oscillator finite commutator}. The latter is of course expected, because the Neumann or Dirichlet boundary conditions on a long but finite line couple the left- and right-moving modes due to reflection at the boundary. For example, taking the finite-regulator basis to consist of $\alpha^{\text{L}}_{\ell}$ and $\alpha^{\text{R}}_{\ell}$ evaluated at $\tau = \sigma = 0$, the subleading expansions read
\begin{align}
\alpha_{\ell>0}^{\text{L}} & = \sqrt{\frac{q}{2\pi}}a(-q) - \frac{1}{16L}\sqrt{\frac{2\pi}{q}}\left[a(q) - a(-q)\right] + \mathcal{O}\left(\frac{1}{L^2}\right)
\\ \alpha_{\ell>0}^{\text{R}} & = \sqrt{\frac{q}{2\pi}}a(q) + \frac{1}{16L}\sqrt{\frac{2\pi}{q}}\left[a(q) - a(-q)\right] + \mathcal{O}\left(\frac{1}{L^2}\right),
\end{align}
together with their conjugates. The leading contribution to the universal oscillator part of the Rindler Hamiltonians as well as its order $\frac{1}{L}$ correction is then given by
\begin{align}
\frac{\pi}{2L}\sum_{\ell=1}^{\infty}\alpha_{-\ell}\alpha_{\ell} & \stackrel{L\rightarrow\infty}{\longrightarrow} \int_{-\infty}^{\infty}\frac{dq}{2\pi}|q|a(q)^{\dagger}a(q) - \frac{\pi}{4L}\int_{-\infty}^{\infty}\frac{dq}{2\pi}\hspace{-2pt}\left[a(q)^{\dagger}a(q) \hspace{-1pt}-\hspace{-1pt} a(-q)^{\dagger}a(q)\right] \hspace{-1pt}+\hspace{-1pt} \mathcal{O}\left(\frac{1}{L^2}\right).
\end{align}
The one-particle states with finite $q$ thus all have finite energy relative to the oscillator vacuum and form normalizable states via superposition in the usual way.

Having detailed the shrinking limit of the one-particle states, it is now simple to analyze the multiparticle states $|\{q_1,\dotsc,q_j\}\rangle \equiv a(q_1)^{\dagger}\cdots a(q_j)^{\dagger}|\{0\}\rangle$, where each $q_i \in \mathds{R}\setminus\{0\}$, normalized as
\begin{equation}
\left\langle \{q_1',q_2',\dotsc,q_{j'}'\}\big|\{q_1,q_2,\dotsc,q_j\}\right\rangle = \delta_{j'j}(2\pi)^j\sum_{\sigma \in S_j}\prod_{i=1}^j \delta(q'_i - q_{\sigma(i)}),
\end{equation}
where $S_j$ is the group of permutations on $j$ elements; these states obey the completeness relation
\begin{equation}
\mathds{1} = \sum_{j=0}^{\infty}\frac{1}{j!}\int_{-\infty}^{\infty}\frac{dq_1}{2\pi}\cdots\int_{-\infty}^{\infty}\frac{dq_j}{2\pi}\left|\{q_1,\dotsc,q_j\}\right\rangle\left\langle\{q_1,\dotsc,q_j\}\right|.
\end{equation}
We know that the shrinking limits of the oscillators are $(-1)^{\ell}\alpha_{-2\ell} \rightarrow \sqrt{\frac{q}{2\pi}}[a(q)^{\dagger} + a(-q)^{\dagger}]$ and $-i(-1)^{\ell}\alpha_{-2\ell+1} \rightarrow \sqrt{\frac{q}{2\pi}}[a(q)^{\dagger} - a(-q)^{\dagger}]$, where $q$ is the limit of $\frac{\pi\ell}{L}$. The shrinking limit of a finite-regulator oscillator state, scaling each $\ell_i = \frac{q_i L}{\pi}$ with finite $q_i > 0$ and choosing each $\ell_i$ to be either even or odd throughout the limit, is then
\begin{equation}
\left|\{\ell_1,\ell_2,\dotsc,\ell_j\}\right\rangle \ \stackrel{L\rightarrow\infty}{\longrightarrow} \ \frac{1}{(2L)^{j/2}}\sum_{\vec{s}\in\mathds{Z}_2^j}(-1)^{s_1+s_2+\dotsc+s_j}\left|\{(-1)^{s_1}q_1,(-1)^{s_2}q_2,\dotsc,(-1)^{s_j}q_j\}\right\rangle,
\end{equation}
where we sum over the $2^j$ choices of signs in the momenta. The unconventional phase in the definition of $|\{\ell_1,\dotsc,\ell_j\}\rangle$ was chosen to cancel the factors of $(-1)^{\ell}$ and $-i(-1)^{\ell}$ in changing basis from circular polarization to linear polarization. The states $|\{q_1,\dotsc,q_j\}\rangle$ with some $\{q_i\}$ coincident have divergent normalizations (whose regularized form is $2L$ for each coincidence) but are correspondingly higher codimension in phase space. 

Finally, we can confirm we have properly constructed the normalizable Hilbert space. The generic oscillator state in the finite-regulator Hilbert space is
\begin{equation}
|\psi_J(L)\rangle = \sum_{j=0}^{J}\sum_{\ell_1,\dotsc,\ell_j=1}^{\infty} f_j(\{\ell_1,\dotsc,\ell_j\},L)\left|\{\ell_1,\dotsc,\ell_j\}\right\rangle,
\end{equation}
which has finite norm for all $f_j(\{\ell\},L) \in L^2(\mathds{N}^j)$; by definition this Hilbert space is the (closure of the) direct sum of the $j$-particle subspaces, so the states consist of arbitrarily but finitely many particles. Consider a $j$-particle wavefunction $f_j(\{\ell\},L)$ that is sharply peaked at some $\{\ell_0\}$ and has compact support of width $\sigma$, where $\mathrm{Vol}[B^j(1)]\sigma^j \simeq \mu[\mathrm{supp} f_j(\{\ell\},L)]$; a basis of $L^2(\mathds{N}^j)$ may be formed out of such functions. For this $j$-particle state to remain a nontrivial $j$-particle state in the shrinking limit, we must take $\{\ell_0\} = \{\frac{q_0 L}{\pi}\}$ as well as $\sigma = L\Delta q$ for finite and positive $q_0$ and $\Delta q$; we need not consider the case where only a subset of the mode numbers is scaled with $L$, since they only affect the dressings of the finite-energy states at finite IR regulator. To preserve normalizability, $f_j(\{\ell\},L)$ must scale as order $1/(2L)^{j/2}$, so define $\widetilde{f}_j(\{q\},L) \equiv (2L)^{j/2}f_j(\{\frac{qL}{\pi}\},L)$ which is peaked at $\{|q|\} = \{q_0\}$, has width $\Delta q$ where $\mathrm{Vol}[B^j(1)](\Delta q)^j \simeq \mu[\mathrm{supp}\widetilde{f}_j(\{q\},L)]$ and scales as order $(2L)^0$. Hence,
\begin{equation}
\widetilde{f}_j(\{q\}) \equiv \lim_{L\rightarrow\infty} \widetilde{f}_j(\{q\},L) \in L^2\hspace{-2pt}\left(\mathds{R}^j\right).
\end{equation}
The shrinking limit of these states is therefore
\begin{align}
\notag \sum_{j=0}^J & \sum_{\ell_1,\dotsc,\ell_j=1}^{\infty} f_j(\{\ell_1,\dotsc,\ell_j\},L)\left|\{\ell_1,\dotsc,\ell_j\}\right\rangle 
\\ & \hspace{-15pt}\stackrel{L\rightarrow\infty}{\longrightarrow} \sum_{j=0}^{J}\sum_{\vec{s}\in\mathds{Z}_2^j}\hspace{-3pt}\left(\hspace{-2pt}(2L)^j\hspace{-3pt}\int_{-\infty}^{\infty}\hspace{-2pt}\frac{dq_1}{2\pi}\cdots\hspace{-2pt}\int_{-\infty}^{\infty}\hspace{-2pt}\frac{dq_j}{2\pi}\right)\hspace{-3pt}\left(\frac{\widetilde{f}_j(\{(-1)^s k\})}{(2L)^{j/2}}\right)\hspace{-3pt}\left(\frac{(-1)^{\sum s}}{(2L)^{j/2}}\left|\{(-1)^s k\}\right\rangle\right)
\\ & = \sum_{j=0}^J\sum_{\vec{s}\in\mathds{Z}_2^j}\hspace{-2pt}(-1)^{\sum s}\hspace{-5pt}\int_{-\infty}^{\infty}\hspace{-3pt}\frac{dq_1}{2\pi}\cdots\hspace{-2pt}\int_{-\infty}^{\infty}\hspace{-3pt}\frac{dq_j}{2\pi}\widetilde{f}_j(\{(-1)^{s}k\})\left|\{(-1)^{s_1}k_1,\dotsc,(-1)^{s_j}k_j\}\right\rangle\hspace{-1pt},
\end{align}
which is indeed a normalizable state in the continuum theory. Note that this result also holds in the generalized basis sense for $f_j(\{\ell\}) = \delta_{\{\ell\},\{\ell_0\}}$ which does not follow the assumptions (it has $\sigma = 0$ and the limit of $(2L)^{j/2}\delta_{\{\ell\},\{\ell_0\}}$ is not in Lebesgue space); its limit produces $|\{\ell_0\}\rangle \rightarrow \sum_{\vec{s}\in\mathds{Z}_2^j}\frac{(-1)^{\sum s}}{(2L)^{j/2}}|\{(-1)^s q_0\}\rangle$ that we concluded above. Thus we have constructed the complete oscillator sector of the angular quantization Hilbert spaces of finite-energy normalizable states for the free boson with zero net charge. The Neumann and Dirichlet oscillators are manifestly equivalent in the shrinking limit, as the latter differ only by a phase for the left-moving states compared to the former.

\subsection{Noncompact Boson on \texorpdfstring{$\bm{\mathds{R}}$}{R} and the Equivalence of Neumann and Dirichlet}

Now we incorporate the shrinking limit of the non-oscillator sectors. The state space must consist of all (possibly Dirac-)normalizable states of finite energy. The guiding principle we use is that of the Reeh-Schlieder theorem \cite{Reeh61}, which asserts that any state may be arbitrarily-well-approximated by acting on a ``vacuum'' state by the algebra of local operators in a finite (Lorentzian) spacetime region. Since we have already treated the oscillators, consider the finite-regulator state $|p;\{0\}\rangle_{k,-k}$ for the Neumann-class noncompact boson angular quantization, where $p$ is a fixed constant not scaled with $L$. From Section \ref{AQ review}, the actions of the free boson, the Hamiltonian and an exponential primary on it are given by
\begin{align}
X(\tau,\sigma)|p;\{0\}\rangle_{k,-k} & = \left[x - \frac{\pi ip}{L}\tau - ik\sigma\right]|p;\{0\}\rangle_{k,-k} + \text{oscillators}
\\ H_{\text{R}}^{\text{N}}(L)|p;\{0\}\rangle_{k,-k} & = \left[\frac{\pi p^2}{2L} + \frac{k^2}{2\pi}L - \frac{\pi}{48L}\right]|p;\{0\}\rangle_{k,-k}
\\ \notag \mathcal{O}_{k'}(\tau,\sigma)|p;\{0\}\rangle_{k,-k} & = \left[\frac{\pi}{L}e^{\frac{\pi\tau}{L}}\cos\left(\frac{\pi\sigma}{2L}\right)\right]^{k'^2/2}e^{kk'\sigma}e^{\frac{\pi k'\tau}{L}p}
\\ & \hspace{15pt}\times \exp\hspace{-2pt}\left[\sqrt{2}k'\sum_{\ell=1}^{\infty}\frac{\alpha_{-\ell}}{\ell}e^{\frac{\pi\ell\tau}{2L}}\cos\hspace{-2pt}\left(\hspace{-2pt}\frac{\pi\ell(\sigma\hspace{-2pt}+\hspace{-2pt}L)}{2L}\hspace{-2pt}\right)\right]\hspace{-2pt}|p+k';\{0\}\rangle_{k,-k}.
\end{align}
The first expression above shows that the expectation values of the free boson derivatives in the oscillator vacuum satisfy
\begin{align}
\langle \partial_{\sigma}X(\tau,\sigma)\rangle_{k,-k} & \ \stackrel{L\rightarrow\infty}{\longrightarrow} \ -ik
\\ \langle \partial_{\tau}X(\tau,\sigma)\rangle_{k,-k} & \ \stackrel{L\rightarrow\infty}{\longrightarrow} \ 0,
\end{align}
showing that the asymptotic conditions \eqref{noncompact asymptotics 1}-\eqref{noncompact asymptotics 4} are indeed obeyed. Note that the state in which we are computing expectation values is a properly normalizable state obtained by smearing $|p;\{0\}\rangle_{k,-k}$ in $p$. To show that the exponential operator above is normalizable, we compute
\begin{align}
\notag &{}_{-k,k}\langle\{0\};p|\mathcal{O}_{k''}(\tau_2,\sigma_2)\mathcal{O}_{k'}(\tau_1,\sigma_1)|p;\{0\}\rangle_{-k,k} 
\\ \notag & \hspace{30pt} = 2\pi\delta(k'+k'')\left[\frac{\pi^2}{L^2}e^{\frac{\pi(\tau_1+\tau_2)}{L}}\cos\left(\frac{\pi\sigma_1}{2L}\right)\cos\left(\frac{\pi\sigma_2}{2L}\right)\right]^{k'^2/2}e^{\frac{\pi}{L}[(\tau_1-\tau_2)p-k'\tau_2]}e^{kk'(\sigma_1-\sigma_2)}
\\ & \hspace{45pt} \times \exp\left[2k'^2\sum_{\ell=1}^{\infty}\frac{1}{\ell}e^{\frac{\pi\ell(\tau_1-\tau_2)}{2L}}\cos\left(\frac{\pi\ell(\sigma_1+L)}{2L}\right)\cos\left(\frac{\pi\ell(\sigma_2+L)}{2L}\right)\right].
\end{align}
It is simple to evaluate the sum remaining in the last line, and taking its $L\rightarrow\infty$ limit gives
\begin{multline}
\sum_{\ell=1}^{\infty}\frac{1}{\ell}e^{\frac{\pi\ell\tau_{12}}{2L}}\hspace{-2pt}\cos\hspace{-2pt}\left(\hspace{-3pt}\frac{\pi\ell(\sigma_1\hspace{-2pt}+\hspace{-2pt}L)}{2L}\hspace{-3pt}\right)\hspace{-2pt}\cos\hspace{-2pt}\left(\hspace{-3pt}\frac{\pi\ell(\sigma_2\hspace{-2pt}+\hspace{-2pt}L)}{2L}\hspace{-3pt}\right) \stackrel{L\rightarrow\infty}{\longrightarrow} -\frac{1}{8}\hspace{-1pt}\ln\hspace{-3pt}\left[\hspace{-1pt}\frac{\pi^4[(|\tau_{12}|^2\hspace{-3pt}+\hspace{-2pt}\sigma_{12}^2)^2\hspace{-3pt}-\hspace{-2pt}(2\sigma_{12}\mathrm{Im}\s\tau_{12})^2]}{L^4}\hspace{-1pt}\right]
\\ + \frac{i}{4}\hspace{-2pt}\left[\tan^{-1}\hspace{-2pt}\left(\hspace{-2pt}\frac{\mathrm{Re}\s\tau_{12}}{\sigma_{12}\hspace{-2pt}+\hspace{-2pt}\mathrm{Im}\s\tau_{12}}\hspace{-2pt}\right) - \tan^{-1}\hspace{-2pt}\left(\hspace{-2pt}\frac{\mathrm{Re}\s\tau_{12}}{\sigma_{12}\hspace{-2pt}-\hspace{-2pt}\mathrm{Im}\s\tau_{12}}\hspace{-2pt}\right)\right] + \frac{\pi i}{8}[\mathrm{sgn}(\sigma_{12}+\mathrm{Im}\s\tau_{12}) - \mathrm{sgn}(\sigma_{12}-\mathrm{Im}\s\tau_{12})],
\end{multline}
where $\tau_{ij} \equiv \tau_i - \tau_j$ and $\sigma_{ij} \equiv \sigma_i - \sigma_j$. The shrinking limit of the above matrix element is then
\begin{multline}\label{noncompact N O norm}
\hspace{-10pt} {}_{-k,k}\hspace{-1pt}\langle\{\hspace{-1pt}0\hspace{-1pt}\}\hspace{-1pt};\hspace{-1pt}p|\mathcal{O}_{k''}\hspace{-1pt}(\hspace{-1pt}\tau_2,\hspace{-1pt}\sigma_2\hspace{-1pt})\mathcal{O}_{k'}\hspace{-1pt}(\hspace{-1pt}\tau_1,\hspace{-1pt}\sigma_1\hspace{-1pt})|p;\hspace{-1pt}\{\hspace{-1pt}0\hspace{-1pt}\}\hspace{-1pt}\rangle_{\hspace{-1pt}-k,k} \hspace{1pt} \stackrel{L\rightarrow\infty}{\longrightarrow} \hspace{1pt} \frac{2\pi\delta(k'\hspace{-2pt}+\hspace{-2pt}k'')e^{kk'\sigma_{12}}}{[(|\tau_{12}|^2+\sigma_{12}^2)^2-(2\sigma_{12}\mathrm{Im}\s\tau_{12})^2]^{k'^2/4}}
\\ \times \hspace{-2pt}\exp\hspace{-2pt}\left\{\hspace{-2pt}\frac{ik'^2}{2}\hspace{-2pt}\left[\tan^{-1}\hspace{-4pt}\left(\hspace{-3pt}\frac{\mathrm{Re}\s\tau_{12}}{\sigma_{12}\hspace{-2pt}+\hspace{-2pt}\mathrm{Im}\s\tau_{12}}\hspace{-3pt}\right) \hspace{-3pt} - \hspace{-2pt} \tan^{-1}\hspace{-4pt}\left(\hspace{-3pt}\frac{\mathrm{Re}\s\tau_{12}}{\sigma_{12}\hspace{-2pt}-\hspace{-2pt}\mathrm{Im}\s\tau_{12}}\hspace{-3pt}\right) \hspace{-2pt}+\hspace{-2pt} \pi\theta(|\mathrm{Im}\s\tau_{12}|\hspace{-2pt}-\hspace{-2pt}|\sigma_{12}|)\mathrm{sgn}(\mathrm{Im}\s\tau_{12})\hspace{-2pt}\right]\hspace{-2pt}\right\},
\end{multline}
and hence the exponential operator is normalizable. We conclude that the states $|p;\{0\}\rangle_{k,-k}$ for all $p \in \mathds{R}$ must be included because any one of them is obtained from any other by applying a normalizable local operator. Therefore, the Hilbert space $\mathcal{H}_{k,-k}$ for the noncompact boson on $\mathds{R}$ has the basis $|p;\{q_1,\dotsc,q_j\}\rangle_{k,-k}$ for $p,q_i\in\mathds{R}$ and $j \in \mathds{N}$, and the mode expansion and Hamiltonian are
\begin{align}
\label{noncompact shrinking N X} X(\tau,\sigma) & = x - ik\sigma + \frac{i}{\sqrt{2}}\int_{-\infty}^{\infty}\frac{dq}{\sqrt{2\pi|q|}}\left[a(q)e^{-|q|\tau - iq\sigma} - a(q)^{\dagger}e^{|q|\tau + iq\sigma}\right]
\\ \label{noncompact shrinking N H} H_{k,-k} & = \lim_{L\rightarrow\infty}\left[H_{\text{R}}^{\text{N}}(L) - \frac{k^2}{2\pi}L\right] = \int_{-\infty}^{\infty}\frac{dq}{2\pi}|q|a(q)^{\dagger}a(q),
\end{align}
where $[x,p] = i$ and $[a(q),a(q')^{\dagger}] = 2\pi\delta(q-q')$. The center-of-mass $x$ is now a true zero-mode. The inner product is that computed from the commutation relations among the mode operators.

Now we obtain the shrinking limit for the Dirichlet-class noncompact boson angular quantization, demonstrating the equivalence of the Neumann-class and Dirichlet-class boundary conditions. So consider the finite-regulator state $|p,x_{\text{s}};\{0\}\rangle_{k,-k}$ in $\mathcal{H}_{k,-k}^{\text{D}}$. From Section \ref{AQ review}, the actions of the free boson, the Hamiltonian and an exponential primary on it are given by
\begin{align}
X(\tau,\sigma)|p,x_{\text{s}};\{0\}\rangle_{k,-k} & = \left[x + \left(\frac{x_{\text{s}}}{L} - ik\right)\sigma\right]|p,x_{\text{s}};\{0\}\rangle_{k,-k} + \text{oscillators}
\\ H_{\text{R}}^{\text{D}}(L)|p,x_{\text{s}};\{0\}\rangle & = \left[\frac{1}{2\pi L}x_{\text{s}}^2 + \frac{k^2}{2\pi}L - \frac{\pi}{48L}\right]|p,x_{\text{s}};\{0\}\rangle
\\ \notag \mathcal{O}_{k'}\hspace{-1pt}(\hspace{-1pt}\tau,\sigma\hspace{-1pt})|p,x_{\text{s}};\{0\}\rangle_{k,-k} & = \left[\frac{\pi}{4L\cos(\frac{\pi\sigma}{2L})}\right]^{\frac{k'^2}{2}}e^{kk'\sigma}e^{\frac{ik'\sigma}{L}x_{\text{s}}}
\\ & \hspace{10pt}\times \hspace{-2pt}\exp\hspace{-3pt}\left[i\sqrt{2}k'\hspace{-2pt}\sum_{\ell=1}^{\infty}\hspace{-2pt}\frac{\alpha_{-\ell}\hspace{-2pt}}{\ell}\hspace{-1pt}e^{\frac{\pi\ell\tau\hspace{-2pt}}{2L}}\hspace{-2pt}\sin\hspace{-2pt}\left(\hspace{-3pt}\frac{\pi\ell(\sigma\hspace{-3pt}+\hspace{-3pt}L)}{2L}\hspace{-3pt}\right)\hspace{-2pt}\right]\hspace{-3pt}|p\hspace{-2pt}+\hspace{-2pt}k'\hspace{-2pt},\hspace{-1pt}x_{\text{s}};\hspace{-1pt}\{0\}\rangle_{k,-k}.
\end{align}
In order for the energy between states to remain finite, the difference in values of $x_{\text{s}}$ cannot scale faster than $\sqrt{L}$. It is also required that any one value of $\frac{x_{\text{s}}}{L}$ vanish as $L\rightarrow\infty$, so that the free boson again satisfies the asymptotic conditions
\begin{align}
\langle \partial_{\sigma}X(\tau,\sigma)\rangle_{k,-k} & \ \stackrel{L\rightarrow\infty}{\longrightarrow} \ -ik
\\ \langle \partial_{\tau}X(\tau,\sigma)\rangle_{k,-k} & \ \stackrel{L\rightarrow\infty}{\longrightarrow} \ 0.
\end{align}
For the normalizability of the exponential operators, we compute
\begin{multline}
{}_{-k,k}\langle p,\hspace{-1pt}x_{\text{s}};\hspace{-1pt}\{0\}|\mathcal{O}_{k''}\hspace{-1pt}(\hspace{-1pt}\tau_2,\hspace{-1pt}\sigma_2\hspace{-1pt})\mathcal{O}_{k'}\hspace{-1pt}(\hspace{-1pt}\tau_1,\hspace{-1pt}\sigma_1\hspace{-1pt})|p,\hspace{-1pt}x_{\text{s}};\hspace{-1pt}\{0\}\rangle_{k,-k} = 2\pi\delta(k'+k'')\frac{e^{kk'(\sigma_1-\sigma_2)}e^{\frac{ik'}{L}(\sigma_1-\sigma_2)x_{\text{s}}}}{[\hspace{-1pt}\frac{16L^2}{\pi^2}\hspace{-2pt}\cos(\hspace{-1pt}\frac{\pi\sigma_1}{2L}\hspace{-1pt})\hspace{-2pt}\cos(\hspace{-1pt}\frac{\pi\sigma_2}{2L}\hspace{-1pt})]^{\frac{k'^2}{2}}}
\\ \times \exp\left[2k'^2\sum_{\ell=1}^{\infty}\frac{1}{\ell}e^{\frac{\pi\ell(\tau_1-\tau_2)}{2L}}\sin\left(\frac{\pi\ell(\sigma_1+L)}{2L}\right)\sin\left(\frac{\pi\ell(\sigma_2+L)}{2L}\right)\right].
\end{multline}
Once again, the remaining sum is easy to compute, and its limit as $L\rightarrow\infty$ is
\begin{multline}
\sum_{\ell=1}^{\infty}\frac{e^{\frac{\pi\ell\tau_{12}}{2L}}}{\ell}\sin\hspace{-3pt}\left(\hspace{-3pt}\frac{\pi\ell(\sigma_1\hspace{-3pt}+\hspace{-3pt}L)}{2L}\hspace{-3pt}\right)\hspace{-2pt}\sin\hspace{-3pt}\left(\hspace{-3pt}\frac{\pi\ell(\sigma_2\hspace{-3pt}+\hspace{-3pt}L)}{2L}\hspace{-3pt}\right) \ \stackrel{L\rightarrow\infty}{\longrightarrow} \ \frac{1}{8}\ln\left[\frac{(\frac{4L}{\pi})^4}{(|\tau_{12}|^2+\sigma_{12}^2)^2-(2\sigma_{12}\mathrm{Im}\s\tau_{12})^2}\right]
\\ + \frac{i}{4}\hspace{-3pt}\left[\tan^{-1}\hspace{-4pt}\left(\hspace{-3pt}\frac{\mathrm{Re}\tau_{12}}{\sigma_{12}\hspace{-3pt}+\hspace{-3pt}\mathrm{Im}\tau_{12}}\hspace{-3pt}\right) \hspace{-3pt}-\hspace{-3pt} \tan^{-1}\hspace{-4pt}\left(\hspace{-3pt}\frac{\mathrm{Re}\tau_{12}}{\sigma_{12}\hspace{-3pt}-\hspace{-3pt}\mathrm{Im}\tau_{12}}\hspace{-3pt}\right) \hspace{-3pt}+\hspace{-3pt} \frac{\pi}{2}\mathrm{sgn}(\sigma_{12}\hspace{-3pt}+\hspace{-3pt}\mathrm{Im}\tau_{12}) \hspace{-3pt}-\hspace{-3pt} \frac{\pi}{2}\mathrm{sgn}(\sigma_{12}\hspace{-3pt}-\hspace{-3pt}\mathrm{Im}\tau_{12})\right]\hspace{-2pt}.
\end{multline}
Hence, we have computed
\begin{multline}\label{noncompact D O norm}
\hspace{-10pt}{}_{-k,k}\langle p,\hspace{-1pt}x_{\text{s}};\hspace{-1pt}\{0\}|\mathcal{O}_{k''}\hspace{-1pt}(\hspace{-1pt}\tau_2,\hspace{-1pt}\sigma_2\hspace{-1pt})\mathcal{O}_{k'}\hspace{-1pt}(\hspace{-1pt}\tau_1,\hspace{-1pt}\sigma_1\hspace{-1pt})|p,\hspace{-1pt}x_{\text{s}};\hspace{-1pt}\{0\}\rangle_{k,-k} \hspace{1pt}\stackrel{L\rightarrow\infty}{\longrightarrow}\hspace{1pt} \frac{2\pi\delta(k'+k'')e^{kk'\sigma_{12}}}{[(|\tau_{12}|^2\hspace{-2pt}+\hspace{-2pt}\sigma_{12}^2)^2\hspace{-2pt}-\hspace{-2pt}(2\sigma_{12}\mathrm{Im}\s\tau_{12})^2]^{\frac{k'^2}{4}}}
\\ \times \hspace{-2pt}\exp\hspace{-2pt}\left\{\hspace{-2pt}\frac{ik'^2}{2}\hspace{-3pt}\left[\tan^{-1}\hspace{-4pt}\left(\hspace{-3pt}\frac{\mathrm{Re}\s\tau_{12}}{\sigma_{12}\hspace{-2pt}+\hspace{-2pt}\mathrm{Im}\s\tau_{12}}\hspace{-3pt}\right) \hspace{-3pt}-\hspace{-2pt} \tan^{-1}\hspace{-4pt}\left(\hspace{-3pt}\frac{\mathrm{Re}\s\tau_{12}}{\sigma_{12}\hspace{-2pt}-\hspace{-2pt}\mathrm{Im}\s\tau_{12}}\hspace{-3pt}\right) \hspace{-2pt}+\hspace{-2pt} \pi\theta(|\mathrm{Im}\tau_{12}|\hspace{-2pt}-\hspace{-2pt}|\sigma_{12}|)\mathrm{sgn}(\mathrm{Im}\tau_{12})\right]\hspace{-2pt}\right\}\hspace{-2pt},
\end{multline}
which again proves that the exponential operators are normalizable. Now, however, we see that neither the mode operator $x_{\text{s}}$ nor its conjugate $p_{\text{s}}$ appear anywhere in the shrinking limit of the algebra of local normalizable operators. Hence, there is no finite-energy process by which the eigenvalue of $x_{\text{s}}$ in a state $|p,x_{\text{s}};\{0\}\rangle$ can change. As such, the eigenvalue of $x_{\text{s}}$ effectively `freezes' and no longer labels states in the $L\rightarrow\infty$ limit. More properly, the $(x_{\text{s}},p_{\text{s}})$-sector of $\mathcal{H}_{k,-k}^{\text{D}}(L)$ decouples in this limit, by which we should quotient. Therefore, the shrinking limit of the Dirichlet-class noncompact free boson angular quantization consists of a Hilbert space $\mathcal{H}_{k,-k}$ spanned by states of the form $|p,\{q_1,\dotsc,q_j\}\rangle_{k,-k}$, where $p,q_i\in\mathds{R}$ and $j \in \mathds{N}$, for which the mode expansions of the free boson and the Hamiltonian are
\begin{align}
\label{noncompact shrinking D X} X(\tau,\sigma) & = x - ik\sigma + \frac{i}{\sqrt{2}}\int_{-\infty}^{\infty}\frac{dq}{\sqrt{2\pi|q|}}\mathrm{sgn}(q)\left[a(q)e^{-|q|\tau-iq\sigma} - a(q)^{\dagger}e^{|q|\tau+iq\sigma}\right]
\\ \label{noncompact shrinking D H} H_{k,-k} & = \lim_{L\rightarrow\infty}\left[H_{\text{R}}^{\text{D}}(L) - \frac{k^2}{2\pi}L\right] = \int_{-\infty}^{\infty}\frac{dq}{2\pi}|q|a(q)^{\dagger}a(q),
\end{align}
where $[x,p] = i$ and $[a(q),a(q')^{\dagger}] = 2\pi\delta(q-q')$. The sole zero-mode subspace is the $(x,p)$-sector, and the inner product follows from the commutation relations among the mode operators as usual.

From the above constructions, it is almost obvious that the shrinking limits of the Neumann-class and the Dirichlet-class approaches are equivalent. Both obtain a Fock space structure with basis states of the form $|p;\{q_1,\dotsc,q_j\}\rangle_{k,-k}$ together with the mode operators $x$, $p$ and $a(q)$, $q \in \mathds{R}\setminus\{0\}$. The equivalence is then proven by exhibiting an isometry between these Hilbert spaces which maps one Hamiltonian to the other. Momentarily scripting the operators with `$\text{N}$' or `$\text{D}$' based on the regulator used, it is immediately clear from \eqref{noncompact shrinking N X} and \eqref{noncompact shrinking N H} as well as \eqref{noncompact shrinking D X} and \eqref{noncompact shrinking D H} that the desired isometry is effected by
\begin{align}
|p,\{0\}\rangle_{k,-k}^{\text{N}} & \ \longmapsto \ |p,\{0\}\rangle_{k,-k}^{\text{D}}
\\ x_{\text{N}} & \ \longmapsto \ x_{\text{D}}
\\ p_{\text{N}} & \ \longmapsto \ p_{\text{D}}
\\ a_{\text{N}}(q) & \ \longmapsto \ \mathrm{sgn}(q)a_{\text{D}}(q).
\end{align}
This isometry merely flips the sign of the left-moving excitations relative to the right-moving ones, which of course is exactly what one expects --- the difference between a wave reflected from a Neumann boundary and a Dirichlet boundary is precisely a phase shift by $\pi$. There is but one subtle point which is worthy of addressing. The above isometry manifestly maps $X(\tau,\sigma)$ and $H_{k,-k}$ between their respective expressions, but the exponential operators require more care. We found that the exponentials in the two approaches have the mode expansions
\begin{align}
\mathcal{O}_{k'}^{\text{N}}(\tau,\sigma) & = e^{kk'\sigma}e^{ik'x}\mathscr{V}^{\text{N}}_{k'}[a_{\text{N}}(q)](\tau,\sigma)
\\ \mathcal{O}_{k'}^{\text{D}}(\tau,\sigma) & = e^{kk'\sigma}e^{ik'x}\mathscr{V}^{\text{D}}_{k'}[a_{\text{D}}(q)](\tau,\sigma)
\end{align}
where
\begin{align}
\mathscr{V}^{\text{N}}_{k'} & \hspace{-2pt}\equiv\hspace{-2pt} \lim_{L\rightarrow\infty}\hspace{-4pt}\left\{\hspace{-4pt}\left(\hspace{-1.5pt}\frac{\pi}{L}\hspace{-1.5pt}\right)^{\hspace{-4pt}\frac{k'^2}{2}}\hspace{-6pt}\exp\hspace{-4pt}\left[\hspace{-2pt}\sqrt{\hspace{-1pt}2}k'\hspace{-2pt}\sum_{\ell=1}^{\infty}\hspace{-3pt}\frac{\alpha_{\hspace{-0.5pt}-\hspace{-0.5pt}\ell}}{\ell}\hspace{-1pt}e^{\hspace{-1pt}\frac{\pi\ell\tau\hspace{-2pt}}{2L}}\hspace{-2pt}\cos\hspace{-3pt}\left(\hspace{-3pt}\frac{\pi\ell(\hspace{-1pt}\sigma\hspace{-3pt}+\hspace{-3pt}L\hspace{-1pt})}{2L}\hspace{-3pt}\right)\hspace{-3pt}\right]\hspace{-3pt}\exp\hspace{-4pt}\left[\hspace{-2pt}-\hspace{-1pt}\sqrt{\hspace{-1pt}2}k'\hspace{-2pt}\sum_{\ell'=1}^{\infty}\hspace{-3pt}\frac{\alpha_{\ell'}\hspace{-2pt}}{\ell'}\hspace{-1pt}e^{\hspace{-1pt}-\hspace{-1pt}\frac{\pi\ell'\hspace{-1pt}\tau\hspace{-2pt}}{2L}}\hspace{-2pt}\cos\hspace{-3pt}\left(\hspace{-3pt}\frac{\pi\ell'\hspace{-1pt}(\hspace{-1pt}\sigma\hspace{-3pt}+\hspace{-3pt}L\hspace{-1pt})}{2L}\hspace{-3pt}\right)\hspace{-3pt}\right]\hspace{-3pt}\right\}
\\ \mathscr{V}^{\text{D}}_{k'} & \hspace{-2pt}\equiv\hspace{-2pt} \lim_{L\rightarrow\infty}\hspace{-4pt}\left\{\hspace{-4pt}\left(\hspace{-1.5pt}\frac{\pi}{4L}\hspace{-1.5pt}\right)^{\hspace{-4pt}\frac{k'^2}{2}}\hspace{-6pt}\exp\hspace{-4pt}\left[\hspace{-1pt}i\hspace{-1pt}\sqrt{\hspace{-1pt}2}k'\hspace{-2pt}\sum_{\ell=1}^{\infty}\hspace{-3pt}\frac{\alpha_{\hspace{-0.5pt}-\hspace{-0.5pt}\ell}}{\ell}\hspace{-1pt}e^{\hspace{-1pt}\frac{\pi\ell\tau\hspace{-2pt}}{2L}}\hspace{-2pt}\sin\hspace{-3pt}\left(\hspace{-3pt}\frac{\pi\ell(\hspace{-1pt}\sigma\hspace{-3pt}+\hspace{-3pt}L\hspace{-1pt})}{2L}\hspace{-3pt}\right)\hspace{-4pt}\right]\hspace{-3pt}\exp\hspace{-4pt}\left[\hspace{-1pt}i\hspace{-1pt}\sqrt{\hspace{-1pt}2}k'\hspace{-2pt}\sum_{\ell'=1}^{\infty}\hspace{-3pt}\frac{\alpha_{\ell'}\hspace{-2pt}}{\ell'}\hspace{-1pt}e^{\hspace{-1pt}-\hspace{-1pt}\frac{\pi\ell'\hspace{-1pt}\tau\hspace{-2pt}}{2L}}\hspace{-2pt}\sin\hspace{-3pt}\left(\hspace{-3pt}\frac{\pi\ell'\hspace{-1pt}(\hspace{-1pt}\sigma\hspace{-3pt}+\hspace{-3pt}L\hspace{-1pt})}{2L}\hspace{-3pt}\right)\hspace{-4pt}\right]\hspace{-3pt}\right\}.
\end{align}
For the oscillator exponentials $\mathscr{V}_{k'}$, there is no simple way to express the shrinking limit that is independent of $L$. Indeed, the operators $\mathscr{V}_{k'}(\tau,\sigma)$ have finite norm; the diverging norms of the individual exponentials are canceled by the explicit factors of $\frac{1}{L^{k'^2/2}}$. In these expressions, the limit $L\rightarrow\infty$ is understood to mean that all matrix elements are to be computed by using the finite-regulator sums and only afterward taking the $L\rightarrow\infty$ of the $c$-number result. Such a procedure is required not only to make sense of the $L^{k'^2/2}$ factors in the denominator but also to resolve a curiosity in the isometry between the Neumann-class and Dirichlet-class approaches. If one were blindly to apply the leading shrinking limits of the sums in the above expressions for $\mathscr{V}_{k'}^{\text{N}}$ and $\mathscr{V}_{k'}^{\text{D}}$, then one would find that the map $a_{\text{N}}(q) \mapsto \mathrm{sgn}(q)a_{\text{D}}(q)$ indeed maps the exponentials to each other, since after all they simply descend from \eqref{noncompact shrinking N X} and \eqref{noncompact shrinking D X}. However, the prefactors involved are $(\frac{\pi}{L})^{k'^2/2}$ and $(\frac{\pi}{4L})^{k'^2/2}$, and so one might be tempted to conclude $\mathscr{V}_{k'}^{\text{N}}[\mathrm{sgn}(q)a_{\text{D}}(q)] \stackrel{?}{=} 2^{k'^2}\mathscr{V}_{k'}^{\text{D}}[a_{\text{D}}(q)]$. Nevertheless, it is in fact true that
\begin{equation}\label{oscillator equality}
\mathscr{V}_{k'}^{\text{N}}[\mathrm{sgn}(q)a_{\text{D}}(q)] = \mathscr{V}_{k'}^{\text{D}}[a_{\text{D}}(q)]!
\end{equation}
This perhaps surprising result emphasizes the need to perform the oscillator sums before taking the limit $L\rightarrow\infty$. The easiest way to verify \eqref{oscillator equality} is by noting that the exponential norms computed in \eqref{noncompact N O norm} and \eqref{noncompact D O norm} are manifestly equal. It is simple to see that the discrete sums must be performed first whenever there are summations which must be commuted past each other, where the corresponding integrals would have a divergence; the sums may of course be performed asymptotically in the limit $L\rightarrow\infty$ since only the contributions diverging and constant in $L$ are required. Then, the magical factor of $2$ which makes all Neumann-class and Dirichlet-class computations equal is due to 
\begin{multline}
\sum_{\ell=1}^{\infty}\frac{1}{\ell}e^{\frac{\pi\ell\tau}{2L}}\hspace{-2pt}\left[\cos\hspace{-2pt}\left(\hspace{-2pt}\frac{\pi\ell(\sigma_1\hspace{-2pt}+\hspace{-2pt}L)}{2L}\hspace{-2pt}\right)\cos\hspace{-2pt}\left(\hspace{-2pt}\frac{\pi\ell(\sigma_2\hspace{-2pt}+\hspace{-2pt}L)}{2L}\hspace{-2pt}\right) - \sin\hspace{-2pt}\left(\hspace{-2pt}\frac{\pi\ell(\sigma_1\hspace{-2pt}+\hspace{-2pt}L)}{2L}\hspace{-2pt}\right)\sin\hspace{-2pt}\left(\hspace{-2pt}\frac{\pi\ell(\sigma_2\hspace{-2pt}+\hspace{-2pt}L)}{2L}\hspace{-2pt}\right)\right]
\\ = \sum_{\ell=1}^{\infty}\frac{(-1)^{\ell}}{\ell}e^{\frac{\pi\ell\tau}{2L}}\cos\hspace{-2pt}\left(\hspace{-2pt}\frac{\pi\ell(\sigma_1\hspace{-2pt}+\hspace{-2pt}\sigma_2)}{2L}\hspace{-2pt}\right) \ \stackrel{L\rightarrow\infty}{\longrightarrow} \ \sum_{\ell=1}^{\infty}\frac{(-1)^{\ell}}{\ell} = -\ln 2.
\end{multline}

\subsection{Compact Boson on \texorpdfstring{$\bm{\mathds{R}}$}{R}}

Having presented the shrinking limit of angular quantization in great detail for the noncompact boson, we may easily adapt the results to the compact boson of radius $R$, which we reviewed in Section \ref{AQ review} exclusively for Dirichlet-class boundary conditions. We are still working in the special case where the total asymptotic charge vanishes, i.e.~here we construct $\mathcal{H}_{n_1,w_1}^{n_2,w_2}$ for $(n_1,w_1) = (n,w)$ and $(n_2,w_2) = (-n,-w)$. The operators of interest acting on a finite-regulator oscillator vacuum $|m,x_{\text{s}};\{0\}\rangle_{n,w}^{-n,-w}$ (for which we temporarily omit the $n$ and $w$ labels) are
\begin{align}
X(\tau,\sigma)|m,x_{\text{s}};\{0\}\rangle & = \left[x + \left(\frac{x_{\text{s}}}{L} - \frac{in}{R}\right)\sigma + wR\tau\right]|m,x_{\text{s}};\{0\}\rangle + \text{oscillators}
\\ H_{\text{R}}(L)|m,x_{\text{s}};\{0\}\rangle & = \left[imw + \frac{1}{2\pi L}x_{\text{s}}^2 + \frac{n^2}{2\pi R^2}L + \frac{w^2 R^2}{2\pi}L - \frac{\pi}{48L}\right]|m,x_{\text{s}};\{0\}\rangle
\\ \notag \mathcal{O}_{n',w'}(\tau,\sigma)|m,x_{\text{s}};\{0\}\rangle & = \left[\frac{\pi}{4L\cos(\frac{\pi\sigma}{2L})}\right]^{\frac{n'^2}{2R^2}}\left[\frac{\pi}{L}\cos\left(\frac{\pi\sigma}{2L}\right)\right]^{\frac{w'^2 R^2}{2}}e^{\frac{\pi i}{2}n'w'-\frac{\pi i\sigma}{2L}n'w'+\frac{\pi\tau}{2L}w'^2 R^2}
\\ \notag & \hspace{15pt} \times e^{[\frac{nn'}{R^2}+ww'R^2]\sigma}e^{i[nw'+n'w]\tau}e^{-\pi i mw'}e^{\frac{1}{L}(\frac{in'}{R}\sigma-w'R\tau)x_{\text{s}}}
\\ & \hspace{-100pt}\times \hspace{-2pt}\exp\hspace{-3pt}\left\{\hspace{-2pt}i\sqrt{\hspace{-1pt}2}\hspace{-1pt}\sum_{\ell=1}^{\infty}\hspace{-3pt}\frac{\alpha_{\hspace{-0.5pt}-\hspace{-0.5pt}\ell}\hspace{-2pt}}{\ell}e^{\hspace{-2pt}\frac{\pi\ell\tau'\hspace{-2pt}}{2L}}\hspace{-4pt}\left[\hspace{-2pt}\frac{n'\hspace{-1pt}}{R}\hspace{-2pt}\sin\hspace{-3pt}\left(\hspace{-3pt}\frac{\pi\ell(\hspace{-1pt}\sigma\hspace{-3pt}+\hspace{-3pt}L\hspace{-1pt})}{2L}\hspace{-3pt}\right) \hspace{-3pt}+\hspace{-3pt} iw'\hspace{-2pt}R\hspace{-2pt}\cos\hspace{-3pt}\left(\hspace{-3pt}\frac{\pi\ell(\hspace{-1pt}\sigma\hspace{-3pt}+\hspace{-3pt}L\hspace{-1pt})}{2L}\hspace{-3pt}\right)\hspace{-3pt}\right]\hspace{-3pt}\right\}\hspace{-3pt}|m\hspace{-3pt}+\hspace{-3pt}n'\hspace{-2pt},x_{\text{s}}\hspace{-3pt}-\hspace{-3pt}\pi w'\hspace{-2pt}R;\hspace{-1pt}\{\hspace{-1pt}0\hspace{-1pt}\}\rangle.
\end{align}
As before, finite energy difference and the asymptotic conditions require that $\frac{x_{\text{s}}}{L} \rightarrow 0$ and $\frac{\Delta x_{\text{s}}}{\sqrt{L}} \rightarrow 0$ for any fixed sloped eigenvalues $x_{\text{s}}$ as $L\rightarrow\infty$. Then, the expectation values of the free boson derivatives are
\begin{align}
\left\langle \partial_{\sigma}X(\tau,\sigma)\right\rangle & \ \stackrel{L\rightarrow\infty}{\longrightarrow} \ -\frac{in}{R}
\\ \left\langle \partial_{\tau}X(\tau,\sigma)\right\rangle & \ \stackrel{L\rightarrow\infty}{\longrightarrow} \ wR,
\end{align}
showing that the compact boson asymptotic conditions \eqref{compact asymptotics 1}-\eqref{compact asymptotics 4} are indeed satisfied in this case. For normalizability of the exponential operators, we compute
\begin{multline}
\langle m,x_{\text{s}};\{0\}|\mathcal{O}_{n'',w''}(\tau_2,\sigma_2)\mathcal{O}_{n',w'}(\tau_1,\sigma_1)|m,x_{\text{s}};\{0\}\rangle 
\\ = \delta_{n',-n''}\delta_{w',-w''}\left[\frac{\pi^2}{16L^2\cos(\frac{\pi\sigma_1}{2L})\cos(\frac{\pi\sigma_2}{2L})}\right]^{\frac{n'^2}{2R^2}}\left[\frac{\pi^2}{L^2}\cos\left(\frac{\pi\sigma_1}{2L}\right)\cos\left(\frac{\pi\sigma_2}{2L}\right)\right]^{\frac{w'^2 R^2}{2}}
\\ \times e^{\hspace{-1pt}-\hspace{-1pt}\frac{\pi i(\hspace{-1pt}\sigma_1\hspace{-1pt}+\hspace{-1pt}\sigma_2\hspace{-1pt})}{2L}\hspace{-1pt}n'\hspace{-1pt}w'\hspace{-1pt}+\frac{\pi(\hspace{-1pt}\tau_1\hspace{-1pt}+\hspace{-1pt}\tau_2\hspace{-1pt})}{2L}\hspace{-1pt}w'^2 \hspace{-2pt}R^2}e^{[\hspace{-1pt}\frac{nn'\hspace{-2pt}}{R^2}\hspace{-1pt}+\hspace{-1pt}ww'\hspace{-2pt}R^2]\sigma_{12}}e^{i[nw'\hspace{-2pt}+n'\hspace{-1pt}w]\tau_{12}}e^{\frac{1}{L}\hspace{-1pt}(\hspace{-1pt}\frac{in'\hspace{-2pt}}{R}\sigma_{12} - w'\hspace{-2pt}R\tau_{12}\hspace{-1pt})x_{\text{s}}}e^{\frac{\pi}{L}\hspace{-1pt}(\hspace{-1pt}\frac{in'\hspace{-2pt}}{R}\sigma_2-w'\hspace{-2pt}R\tau_2\hspace{-1pt})w'\hspace{-2pt}R}
\\ \times \exp\hspace{-2pt}\left\{\hspace{-2pt}2\hspace{-2pt}\sum_{\ell=1}^{\infty}\hspace{-3pt}\frac{e^{\hspace{-1pt}\frac{\pi\ell\tau_{12}}{2L}}\hspace{-3pt}}{\ell}\hspace{-3pt}\left[\hspace{-3pt}\frac{n'\hspace{-1pt}}{R}\hspace{-3pt}\sin\hspace{-4pt}\left(\hspace{-4pt}\frac{\pi\hspace{-0.5pt}\ell\hspace{-0.5pt}(\hspace{-1pt}\sigma_1\hspace{-3pt}+\hspace{-3pt}L\hspace{-1pt})}{2L}\hspace{-4pt}\right) \hspace{-4pt}+\hspace{-3pt} iw'\hspace{-2pt}R\hspace{-2pt}\cos\hspace{-4pt}\left(\hspace{-4pt}\frac{\pi\hspace{-0.5pt}\ell\hspace{-0.5pt}(\hspace{-1pt}\sigma_1\hspace{-3pt}+\hspace{-3pt}L\hspace{-1pt})}{2L}\hspace{-4pt}\right)\hspace{-4pt}\right]\hspace{-5pt}\left[\hspace{-3pt}\frac{n'\hspace{-1pt}}{R}\hspace{-3pt}\sin\hspace{-4pt}\left(\hspace{-4pt}\frac{\pi\hspace{-0.5pt}\ell\hspace{-0.5pt}(\hspace{-1pt}\sigma_2\hspace{-3pt}+\hspace{-3pt}L\hspace{-1pt})}{2L}\hspace{-4pt}\right) \hspace{-4pt}-\hspace{-3pt} iw'\hspace{-2pt}R\hspace{-2pt}\cos\hspace{-4pt}\left(\hspace{-4pt}\frac{\pi\hspace{-0.5pt}\ell\hspace{-0.5pt}(\hspace{-1pt}\sigma_2\hspace{-3pt}+\hspace{-3pt}L\hspace{-1pt})}{2L}\hspace{-4pt}\right)\hspace{-4pt}\right]\hspace{-4pt}\right\}\hspace{-2pt}.\hspace{-5pt}
\end{multline}
As before, the remaining sum is straightforward to evaluate, and taking the $L\rightarrow\infty$ limit results in
\begin{multline}
\langle m,x_{\text{s}};\{0\}|\mathcal{O}_{n'',w''}(\tau_2,\sigma_2)\mathcal{O}_{n',w'}(\tau_1,\sigma_1)|m,x_{\text{s}};\{0\}\rangle 
\\ \stackrel{L\rightarrow\infty}{\longrightarrow} \ \delta_{n',-n''}\delta_{w',-w''}\frac{(-1)^{n'w'}e^{[\frac{nn'}{R^2}+ww'R^2]\sigma_{12}}e^{i[nw'+n'w]\tau_{12}}}{[(|\tau_{12}|^2+\sigma_{12}^2)^2-(2\sigma_{12}\mathrm{Im}\s\tau_{12})^2]^{\Delta_{n',w'}}}\left(\frac{\sigma_{12}-i\tau_{12}}{\sigma_{12}+i\tau_{12}}\right)^{n'w'}
\\ \times \exp\left\{2i\Delta_{n',w'}\hspace{-3pt}\left[\tan^{-1}\hspace{-4pt}\left(\hspace{-3pt}\frac{\mathrm{Re}\tau_{12}}{\sigma_{12}\hspace{-3pt}+\hspace{-3pt}\mathrm{Im}\tau_{12}}\hspace{-3pt}\right) \hspace{-3pt}-\hspace{-3pt} \tan^{-1}\hspace{-4pt}\left(\hspace{-3pt}\frac{\mathrm{Re}\tau_{12}}{\sigma_{12}\hspace{-3pt}-\hspace{-3pt}\mathrm{Im}\tau_{12}}\hspace{-3pt}\right) \hspace{-3pt}+\hspace{-3pt} \pi\theta(|\mathrm{Im}\s\tau_{12}|-|\sigma_{12}|)\mathrm{sgn}(\mathrm{Im}\s\tau_{12})\right]\right\},
\end{multline}
where $\Delta_{n',w'} = \frac{1}{4}(\frac{n'^2}{R^2} + w'^2 R^2)$, and hence these operators are indeed normalizable. As in the noncompact case, the operator $x_{\text{s}}$ does not appear in the mode expansion of any local normalizable operator in the shrinking limit. However, in contrast with the noncompact case, its conjugate momentum $p_{\text{s}}$ does appear in the mode expansions of the exponential operators. The entire effect of the slope mode is thus that acting with $\mathcal{O}_{n',w'}$ shifts the state $|x_{\text{s}}\rangle$ to $|x_{\text{s}}-\pi w'R\rangle$. As a result, the $(x_{\text{s}},p_{\text{s}})$-sector for the compact boson angular quantization does not get eliminated completely in the shrinking limit, but rather $p_{\text{s}}$ becomes compactified (at radius $\frac{1}{\pi R}$), rendering the spectrum of $x_{\text{s}}$ discrete. As before, shifting the entire spectrum of $x_{\text{s}}$ by a constant does nothing, so we might as well set one of its eigenvalues to be $x_{\text{s}} = 0$, followed by the operator rescaling $x_{\text{s}} \mapsto \pi R x_{\text{s}}$ and $p_{\text{s}} \mapsto \frac{p_{\text{s}}}{\pi R}$. Then, the oscillator vacuum states in the shrinking limit are written $|m,m_{\text{s}};\{0\}\rangle$, where $m,m_{\text{s}} \in \mathds{Z}$. The mode expansion of an exponential operator then becomes
\begin{equation}
\mathcal{O}_{n',w'}(\tau,\sigma) = e^{\frac{\pi i}{2}n'w'}e^{[\frac{nn'}{R^2}+ww'R^2]\sigma}e^{i[nw'+n'w]\tau}e^{\frac{in'}{R}x}e^{-\pi iw'Rp}e^{iw'p_{\text{s}}}\mathscr{V}_{n',w'}[a(q)](\tau,\sigma),
\end{equation}
where
\begin{multline}
\hspace{-5pt}\mathscr{V}_{n',w'} \hspace{-2pt}\equiv\hspace{-2pt} \lim_{L\rightarrow\infty} \hspace{-3pt}\Bigg[\hspace{-3pt}\left(\hspace{-1pt}\frac{\pi}{4L}\hspace{-1pt}\right)^{\hspace{-2pt}\frac{n'^2}{2R^2}}\hspace{-3pt}\left(\hspace{-1pt}\frac{\pi}{L}\hspace{-1pt}\right)^{\hspace{-3pt}\frac{w'^2 R^2}{2}}\hspace{-5pt}\exp\hspace{-2pt}\left\{\hspace{-2pt}i\sqrt{2}\hspace{-1pt}\sum_{\ell=1}^{\infty}\hspace{-3pt}\frac{\alpha_{\hspace{-0.5pt}-\hspace{-0.5pt}\ell}\hspace{-2pt}}{\ell}e^{\hspace{-1pt}\frac{\pi\ell\tau}{2L}}\hspace{-4pt}\left[\hspace{-2pt}\frac{n'\hspace{-2pt}}{R}\hspace{-1pt}\sin\hspace{-3pt}\left(\frac{\pi\ell(\hspace{-1pt}\sigma\hspace{-3pt}+\hspace{-3pt}L\hspace{-1pt})}{2L}\hspace{-3pt}\right)\hspace{-3pt}+\hspace{-3pt}iw'\hspace{-2pt}R\hspace{-1pt}\cos\hspace{-3pt}\left(\hspace{-3pt}\frac{\pi\ell(\hspace{-1pt}\sigma\hspace{-3pt}+\hspace{-3pt}L\hspace{-1pt})}{2L}\hspace{-3pt}\right)\hspace{-2pt}\right]\hspace{-2pt}\right\}
\\ \times \exp\left\{\hspace{-2pt}i\sqrt{2}\hspace{-1pt}\sum_{\ell'=1}^{\infty}\hspace{-3pt}\frac{\alpha_{\ell'}\hspace{-2pt}}{\ell'}e^{\hspace{-1pt}-\hspace{-1pt}\frac{\pi\ell'\tau}{2L}}\hspace{-4pt}\left[\hspace{-2pt}\frac{n'\hspace{-2pt}}{R}\hspace{-1pt}\sin\hspace{-3pt}\left(\frac{\pi\ell'(\hspace{-1pt}\sigma\hspace{-3pt}+\hspace{-3pt}L\hspace{-1pt})}{2L}\hspace{-3pt}\right)\hspace{-3pt}-\hspace{-3pt}iw'\hspace{-2pt}R\hspace{-1pt}\cos\hspace{-3pt}\left(\hspace{-3pt}\frac{\pi\ell'(\hspace{-1pt}\sigma\hspace{-3pt}+\hspace{-3pt}L\hspace{-1pt})}{2L}\hspace{-3pt}\right)\hspace{-2pt}\right]\hspace{-2pt}\right\}\Bigg].
\end{multline}
Acting with $\mathcal{O}_{n',w'}$ on a zero-mode state $|m,m_{\text{s}}\rangle$ thus yields the state $|m+n',m_{\text{s}}-w'\rangle$. Therefore, the Hilbert space $\mathcal{H}_{n,w}^{-n,-w}$ for the compact boson on $\mathds{R}$ has the basis states $|m,m_{\text{s}};\{q_1,\dotsc,q_j\}\rangle_{n,w}^{-n,-w}$ for $m,m_{\text{s}}\in\mathds{Z}$, $q_i \in \mathds{R}\setminus 0$ and $j \in \mathds{N}$, and the mode expansion and Hamiltonian are
\begin{align}
X(\tau,\sigma) & = x \hspace{-2pt}-\hspace{-2pt} \frac{in}{R}\sigma \hspace{-2pt}+\hspace{-2pt} wR\tau \hspace{-2pt}+\hspace{-2pt} \frac{i}{\sqrt{2}}\int_{-\infty}^{\infty}\hspace{-2pt}\frac{dq}{\sqrt{2\pi|q|}}\mathrm{sgn}(q)\hspace{-2pt}\left[a(q)e^{-|q|\tau - iq\sigma} \hspace{-3pt}-\hspace{-2pt} a(q)^{\dagger}e^{|q|\tau + iq\sigma}\right]
\\ H_{n,w}^{-n,-w} & = \lim_{L\rightarrow\infty}\left[H_{\text{R}}(L) - \frac{n^2}{2\pi R^2}L - \frac{w^2 R^2}{2\pi}L\right] = iwRp + \int_{-\infty}^{\infty}\frac{dq}{2\pi}|q|a(q)^{\dagger}a(q),
\end{align}
where $[x,p] = i$ and $[a(q),a(q')^{\dagger}] = 2\pi\delta(q-q')$; the integer $m_{\text{s}}$ is the eigenvalue of $x_{\text{s}}$, and its conjugate momentum $p_{\text{s}}$ is compactified at unit radius. The compact boson CFT on Minkowski space thus has two zero-modes; the first is from the obvious shift symmetry, while the second does not affect the value of $X$ at all but whose entire role is to enforce winding conservation of correlators.

\section{General Case: Nonzero Net Charge}\label{nonzero net charge}

In Section \ref{zero net charge}, we constructed the noncompact and compact free boson Minkowski CFTs in the special case that the net asymptotic charge vanishes. There, we found a simple structure to each result, essentially given by a free oscillator Fock space built on top of a classical solution to the equations of motion. We must now turn to the general case of arbitrary endpoint exponential operators in angular quantization. The corresponding vector spaces $\mathcal{H}_{k_1 k_2}$ and $\mathcal{H}_{n_1,w_1}^{n_2,w_2}$ describe states with unbalanced asymptotic conditions which cannot be satisfied by any classical saddle, which is the origin of the greater technical difficulty. At finite regulator, the boundary conditions on the strip are solved in \eqref{noncompact N X} and \eqref{compact X} by a quadratic term whose derivatives at $\sigma = \pm L$ are asymmetric. However, in the shrinking limit $L\rightarrow\infty$, all such quadratic terms vanish due to the explicit factor of $\frac{1}{L}$, making it impossible to then satisfy the asymptotic conditions \eqref{noncompact asymptotics 1}-\eqref{noncompact asymptotics 4} or \eqref{compact asymptotics 1}-\eqref{compact asymptotics 4} without help from the oscillators. In the noncompact Dirichlet-class approach, the finite regulator mode expansion \eqref{noncompact D X} does not contain a quadratic term, but it also does not obey the asymptotic conditions anywhere. Fortunately, the zero-momentum-mode sectors remain similar in form to those of Section \ref{zero net charge}, so we are tasked with finding a suitable replacement for the ``oscillator vacuum'' on which to construct all other states.

The first point we must make is that the putative state spaces $\mathcal{H}_{k_1 k_2}$ and $\mathcal{H}_{n_1,w_1}^{n_2,w_2}$ would not be Hilbert spaces. Specifically, an inner product would not be defined between two states in $\mathcal{H}_{k_1 k_2}$ but rather would be a pairing between an element of $\mathcal{H}_{k_1 k_2}$ and an element of $\mathcal{H}_{-k_1,-k_2}$, or likewise between $\mathcal{H}_{n_1,w_1}^{n_2,w_2}$ and $\mathcal{H}_{-n_1,-w_1}^{-n_2,-w_2}$. To demonstrate why, we recall that angular quantization is defined so that it reproduces all Euclidean correlators on the sphere. In particular, at finite regulator in the cylinder frame, this requirement is nothing other than the Cardy condition but applied to non-conformally-invariant boundary states and conditions. The cylinder equivalence of matrix elements between boundary states and thermal traces bestows upon angular quantization a notion of Hermitian conjugation that is inherited from that of radial quantization. 
\begin{figure}
\centering
\begin{tikzpicture}
\draw (-5,-2.5) -- (-5,2.5);
\draw (-2,-2.5) -- (-2,2.5);
\draw (2,-2.5) -- (2,2.5);
\draw (5,-2.5) -- (5,2.5);
\draw[blue,line width=0.75pt,postaction=decorate,decoration={markings,mark = at position 0.525 with {\arrow{stealth}}}] (-3.5,-2.9) -- (-3.5,2.1);
\draw[line width=1.2pt] (-5,2.5) arc (-180:180:1.5 and 0.4);
\draw[line width=1.2pt] (-5,-2.5) arc (-180:0:1.5 and 0.4);
\draw[line width=1.2pt,dashed] (-5,-2.5) arc (180:0:1.5 and 0.4);
\draw[blue,line width=0.75pt,postaction=decorate,decoration={markings,mark = at position 0.5 with {\arrow{stealth}}}] (3.5,2.1) -- (3.5,-2.9);
\draw[line width=1.2pt] (2,2.5) arc (-180:180:1.5 and 0.4);
\draw[line width=1.2pt] (2,-2.5) arc (-180:0:1.5 and 0.4);
\draw[line width=1.2pt,dashed] (2,-2.5) arc (180:0:1.5 and 0.4);
\draw[red,thick,-stealth] ({-3.5-1.5*sin(10)},{0-0.4*cos(10)}) arc (-100:-155:1.5 and 0.4) node[midway,below]{$H_{ij}$};
\draw[red,thick,-stealth] ({3.5+1.5*sin(10)},{0-0.4*cos(10)}) arc (-80:-25:1.5 and 0.4) node[midway,below]{$H_{\bar{\imath}\bar{\jmath}}$};
\node[scale=2] (arrow) at (0,0) {$\Longrightarrow$};
\node[scale=0.9] (conjugate) at (0,-0.4) {conjugate};
\node[scale=1] (1) at (-3.5,-3.25) {$|B_i\rrangle$};
\node[scale=1] (2) at (-3.5,2.5) {$\llangle B_j|$};
\node[scale=1] (3) at (3.5,-3.25) {$|\barred{B_j}\rrangle$};
\node[scale=1] (4) at (3.5,2.5) {$\llangle \barred{B_i}|$};
\end{tikzpicture}
\caption{The Hermitian conjugate of a state in angular quantization belongs to the sector with conjugated endpoint operators. The blue line represents a state in angular quantization, with the arrow indicating orientation. The red arrow shows Euclidean time evolution by the associated Rindler Hamiltonian.}
\label{Hermitian conjugation}
\end{figure}
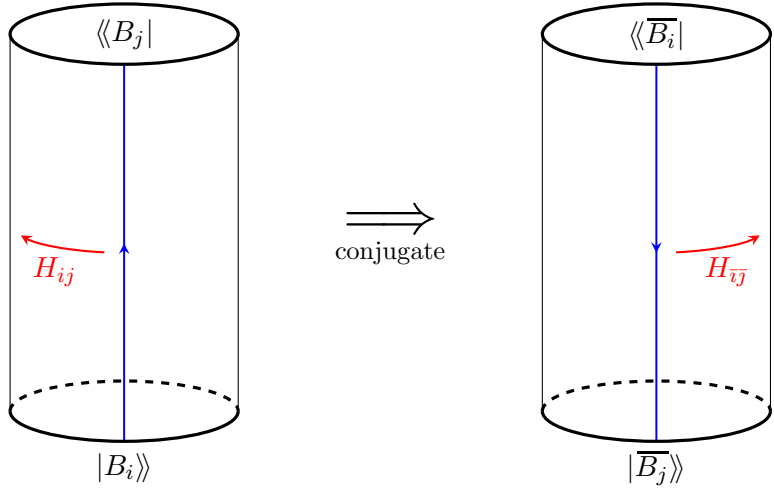
At finite regulator, the cylinder thermal trace over the angular quantization space $\mathcal{H}_{ij}$ is equal to the cylinder matrix element between boundary states $|B_i\rrangle$ and $|B_j\rrangle$ in radial quantization, as shown on the left in Figure \ref{Hermitian conjugation}. Then, applying radial quantization Hermitian conjugation inverts the cylinder, with the two boundary conditions being both interchanged and replaced by their respective BPZ conjugates $|\barred{B_i}\rrangle$ and $|\barred{B_j}\rrangle$, as shown on the right in Figure \ref{Hermitian conjugation}. The conjugated cylinder matrix element then equals the trace over the Hermitian-conjugated angular quantization space, each of whose states obeys the boundary condition $\barred{B_i}$ at the extreme left of the spatial slice and the boundary condition $\barred{B_j}$ at the extreme right. The boundary conditions $\barred{B_i}$ and $\barred{B_j}$ on the sphere shrink to the BPZ-conjugate operators $\barred{\mathcal{O}_i}(0)$ and $\barred{\mathcal{O}_j}(\infty)$, and hence the Hermitian conjugate of a state in the $\mathcal{H}_{ij}$ sector of angular quantization is a bra in the $\mathcal{H}_{\bar{\imath}\bar{\jmath}}$ sector. In the examples studied here, the endpoint operators are exponential operators $\mathcal{O}_{k} = \normal{e^{ikX}}$ whose conjugates are $\barred{\mathcal{O}_k} = \normal{e^{-ikX}} = \mathcal{O}_{-k}$. Therefore, the inner product pairing on the Minkowski free boson CFT would be between a state in $\mathcal{H}_{k_1 k_2}$ and a state in $\mathcal{H}_{-k_1,-k_2}$, or analogously between $\mathcal{H}_{n_1,w_1}^{n_2,w_2}$ and $\mathcal{H}_{-n_1,-w_1}^{-n_2,-w_2}$. Of course, there would be a canonical isomorphism between $\mathcal{H}_{k_1 k_2}$ and $\mathcal{H}_{-k_1,-k_2}$ simply given by negating all appearances of the external momenta. 

The second, and far more important, point we must make is that the putative state spaces $\mathcal{H}_{k_1 k_2}$ and $\mathcal{H}_{n_1,w_1}^{n_2,w_2}$ do not exist, at least in the familiar sense. When one speaks of a vector space of states, its elements are pure states of the theory, and local operators map each vector in the space to another vector. In the preceding Section \ref{zero net charge}, we were able to construct such vector spaces of pure states directly from the shrinking limit of the finite-regulator Hilbert spaces. In the present case, on the other hand, it is simple to see that none of the finitely-excited pure states satisfy the asymptotic conditions in the shrinking limit. Indeed, from the expressions given in Section \ref{AQ review}, the shrinking limit of the free boson expectation values in the oscillator vacua $|p;\{0\}\rangle$ or $|p,x_{\text{s}};\{0\}\rangle$ are
\begin{align}\label{noncompact osc vac}
\text{noncompact:} \qquad \langle X(\tau,\sigma)\rangle_{\text{osc.vac.}} & = \langle x\rangle - \frac{i(k_1-k_2)}{2}\sigma
\\ \label{compact osc vac} \text{compact:} \qquad \langle X(\tau,\sigma)\rangle_{\text{osc.vac.}} & = \langle x\rangle - \frac{i(n_1-n_2)}{2R}\sigma + \frac{(w_1-w_2)R}{2}\tau,
\end{align}
from which it is obvious that the asymptotic conditions \eqref{noncompact asymptotics 1}-\eqref{noncompact asymptotics 4} or \eqref{compact asymptotics 1}-\eqref{compact asymptotics 4} are not obeyed. Then, since we already showed that any finite oscillator excitation does not alter the $\sigma\rightarrow \pm\infty$ behavior of expectation values, we conclude that none of the states $|p;\{N_{\ell}\}\rangle$ or $|p,x_{\text{s}};\{N_{\ell}\}\rangle$ satisfy the asymptotic conditions. Of course, this failure is closely related to the fact that the angular quantization bare thermal traces on the cylinder vanish. The only nonvanishing traces are those for which bulk operators are inserted which provide the compensating momenta for momentum conservation to be satisfied. It is only the insertion of the correct bulk operators which yields finite answers, so in the logic of angular quantization one would expect that the insertion of such operators should be the key in obtaining the correct asymptotic behavior as well. 

Just as with a boundary condition, the meaning of an asymptotic condition lies in the behavior of correlators of local operators as the position of one of the operators approaches some limit point, here spatial infinity. Thus, the quantity $\langle X(\tau,\sigma)\rangle$ which must obey the asymptotic conditions must really be defined as
\begin{equation}\label{expectation value def}
\langle X(\tau,\sigma)\rangle \equiv \frac{\langle X(\tau,\sigma)\mathcal{O}_1(\tau_1,\sigma_1)\cdots\mathcal{O}_n(\tau_n,\sigma_n)\rangle}{\langle\mathcal{O}_1(\tau_1,\sigma_1)\cdots \mathcal{O}_n(\tau_n,\sigma_n)\rangle},
\end{equation}
where $\langle\mathcal{O}_1(\tau_1,\sigma_1)\cdots \mathcal{O}_n(\tau_n,\sigma_n)\rangle$ is any nonvanishing correlator of operators whose positions are held finite. Of course, the quantity \eqref{expectation value def} depends on all these other operators being inserted, which we suppress in the notation. The point of an asymptotic condition is that the quantity $\langle X(\tau,\sigma)\rangle$ in the limit $\sigma \rightarrow \pm\infty$ is completely independent of the other operator insertions in \eqref{expectation value def} and moreover equals a specific function of $\tau$ and $\sigma$. Now, as $|\sigma|$ grows large, the $n$-point OPE converges, which we use to replace $\mathcal{O}_1(\tau_1,\sigma_1)\cdots\mathcal{O}_n(\tau_n,\sigma_n)$ with a sum of local operators at some fixed coordinate $(\tau_0,\sigma_0)$. By linearity, the asymptotic condition holds generally if and only if it holds for each local operator in this sum separately. Thus we need only consider
\begin{equation}
\langle X(\tau,\sigma)\rangle = \frac{\langle X(\tau,\sigma)\mathcal{O}_{k}(\tau_0,\sigma_0)\rangle}{\langle\mathcal{O}_k(\tau_0,\sigma_0)\rangle},
\end{equation}
where $\mathcal{O}_k$ is an exponential primary, with $k = -(k_1+k_2)$ for the noncompact case and $k = (n,w) = -(n_1+n_2,w_1+w_2)$ for the compact case, or one of its descendants.

This discussion of asymptotic conditions involves only correlation functions of local operators. It is thus far more useful to use the more general definition of `state' --- a state is a linear map from the algebra of operators to the complex numbers. We take a state to be normalized so that it maps the identity operator to unity; many sources also require that the map be Hermitian, but we somewhat relax this condition here. This definition of a state $\rho$ is the one preferred in algebraic quantum field theory, which generalizes the map $\mathcal{O} \rightarrow \mathrm{Tr}[\mathcal{O}\rho]$ to a von Neumann algebra of operators (for which a trace does not exist if it is of Type III). For a review of the role of von Neumann algebras in quantum field theory, see for instance \cite{Witten23} and references therein.

We may now explicitly construct these states by using the equality of correlators with the shrinking limit of traces taken in angular quantization. We know we need only consider one local operator $\mathcal{O}_k(\tau_0,\sigma_0)$; first we consider the exponential primary, and in the following subsection we take into account all descendants by acting with arbitrary combinations of creation and annihilation operators. Note that the exponential operators $\mathcal{O}$ can all be split into a zero-momentum mode piece $\mathcal{O}^{(0)}$ times an oscillator creation exponential and an oscillator annihilation exponential. Then, using cyclicity of the trace, the nonvanishing cylinder traces in angular quantization take the form
\begin{multline}\label{trace splitting}
\mathrm{Tr}_{\mathcal{H}}\left[\mathcal{O}_k(\tau_0,\sigma_0) e^{-2\pi H_{\text{R}}}\right]_{S^1\times\mathds{R}} 
\\ = \mathrm{Tr}_{\mathcal{H}}\left[e^{\sum_{\ell=1}^{\infty}\frac{\alpha_{\ell}}{\ell}f_{\ell}^{-k}(-\tau_0,\sigma_0)^*}\mathcal{O}_{k}^{(0)}(\tau_0,\sigma_0) e^{-2\pi H_{\text{R}}}e^{\sum_{\ell'=1}^{\infty}\frac{\alpha_{-\ell'}}{\ell'}f_{\ell'}^{k}( \tau_0,\sigma_0)}\right]_{S^1\times\mathds{R}},
\end{multline}
where
\begin{align}\label{f noncompact N}
\text{noncompact (N):} \qquad f_{\ell}^k(\tau,\sigma) & = \sqrt{2}ke^{\frac{\pi\ell\tau}{2L}}\cos\left(\frac{\pi\ell(\sigma+L)}{2L}\right)
\\ \label{f noncompact D} \text{noncompact (D):} \qquad f_{\ell}^k(\tau,\sigma) & = i\sqrt{2}ke^{\frac{\pi\ell\tau}{2L}}\sin\left(\frac{\pi\ell(\sigma+L)}{2L}\right)
\\ \label{f compact} \text{compact:} \qquad f_{\ell}^{n,w}(\tau,\sigma) & = i\sqrt{2}e^{\frac{\pi\ell\tau}{2L}}\hspace{-2pt}\left[\hspace{-1pt}\frac{n}{R}\sin\hspace{-2pt}\left(\hspace{-2pt}\frac{\pi\ell(\sigma\hspace{-2pt}+\hspace{-2pt}L)}{2L}\hspace{-2pt}\right) \hspace{-2pt}+\hspace{-2pt} iwR\cos\hspace{-2pt}\left(\hspace{-2pt}\frac{\pi\ell(\sigma\hspace{-2pt}+\hspace{-2pt}L)}{2L}\hspace{-2pt}\right)\hspace{-1pt}\right].
\end{align}
Moreover, the asymptotic conditions dictate the behavior of the trace \eqref{trace splitting} with an extra $X(\tau,\sigma)$ inserted, $\mathrm{Tr}_{\mathcal{H}}[\mathcal{O}_k(\tau_0,\sigma_0)X(\tau,\sigma)e^{-2\pi H_{\text{R}}}]_{S^1\times\mathds{R}}$, as $\sigma\rightarrow \pm\infty$. Then, we define a state $\rho(\tau_0,\sigma_0)$ such that
\begin{equation}
\langle X(\tau,\sigma)\rangle(\tau_0,\sigma_0) = \mathrm{Tr}\left[\rho(\tau_0,\sigma_0)X(\tau,\sigma)\right].
\end{equation}
From the explicit expressions for the operators involved, we construct this state as
\begin{equation}
\rho(\tau_0,\sigma_0) = \rho^{(0)}\otimes \bigotimes_{\ell=1}^{\infty}\rho_{\ell}^{\text{osc}}(\tau_0,\sigma_0),
\end{equation}
where the zero-momentum-mode part of the state is
\begin{align}
\text{noncompact:} \qquad \rho^{(0)}_p & = |p\rangle\langle p|
\\ \text{compact:} \qquad \rho^{(0)}_{m,m_{\text{s}}} & = |m,m_{\text{s}}\rangle\langle m,m_{\text{s}}|,
\end{align}
and $\rho_{\ell}^{\text{osc}}$ is the density matrix
\begin{equation}\label{osc state general}
\rho^{\text{osc}}_{\ell}(\tau_0,\sigma_0) = \frac{1}{\mathcal{N}_{\ell}}\sum_{N_{\ell}=0}^{\infty}e^{-\frac{\pi^2}{2L}\alpha_{-\ell}\alpha_{\ell}}e^{\frac{\alpha_{-\ell}}{\ell}f_{\ell}^k(\tau_0,\sigma_0)}|N_{\ell}\rangle\langle N_{\ell}|e^{\frac{\alpha_{\ell}}{\ell}f_{\ell}^{-k}(-\tau_0,\sigma_0)^*}e^{-\frac{\pi^2}{2L}\alpha_{-\ell}\alpha_{\ell}}.
\end{equation}
This definition is very roughly a Poisson distribution for the system to be in the state $e^{\frac{\alpha_{-\ell}}{\ell}f_{\ell}^k(\tau_0,\sigma_0)}|N_{\ell}\rangle$, but this characterization should not be taken literally since these states are not orthogonal. Here, the normalization constant $\mathcal{N}_{\ell}$ is determined by the condition $\text{Tr}[\rho^{\text{osc}}_{\ell}(\tau_0,\sigma_0)] = 1$, namely
\begin{align}
\mathcal{N}_{\ell} & = \sum_{N_{\ell}=0}^{\infty}\frac{1}{\ell^{N_{\ell}}N_{\ell}!}\sum_{m=0}^{\infty}\frac{e^{-\frac{\pi^2\ell\hspace{-1pt}}{L}(\hspace{-1pt}N_{\ell}+m\hspace{-1pt})}}{m!^2}\hspace{-3pt}\left(\hspace{-2pt}\frac{1}{\ell^2}f_{\ell}^{-k}\hspace{-1pt}(\hspace{-1pt}-\tau_0,\sigma_0\hspace{-1pt})^* \hspace{-2pt}f_{\ell}^k\hspace{-1pt}(\hspace{-1pt}\tau_0,\sigma_0\hspace{-1pt})\hspace{-2pt}\right)^{\hspace{-3pt}m}\hspace{-4pt}\langle 0|\alpha_{\ell}^{N_{\ell}+m}\alpha_{-\ell}^{N_{\ell}+m}|0\rangle 
%\\ & = e^{\frac{1}{\ell}e^{-\frac{\pi^2\ell}{L}}f_{\ell}^{-k}(-\tau_0,\sigma_0)^* f_{\ell}^k(\tau_0,\sigma_0)}\sum_{N_{\ell}=0}^{\infty}e^{-\frac{\pi^2\ell}{L}N_{\ell}}L_{N_{\ell}}\left(-\frac{1}{\ell}e^{-\frac{\pi^2\ell}{L}}f_{\ell}^{-k}(-\tau_0,\sigma_0)^*f_{\ell}^k(\tau_0,\sigma_0)\right)
\\ & = \frac{1}{1-e^{-\frac{\pi^2\ell}{L}}}\exp\left[\frac{e^{-\frac{\pi^2\ell}{L}}}{\ell(1-e^{-\frac{\pi^2\ell}{L}})}f_{\ell}^{-k}(-\tau_0,\sigma_0)^* f_{\ell}^k(\tau_0,\sigma_0)\right].
\end{align}
Several comments are in order before proceeding. The definition \eqref{osc state general} may seem strange because its matrix eigenvalues are not all positive, whereas probabilities should be positive; equivalently, the density operator is not generally Hermitian. However, there is no problem here because $e^{\frac{\alpha_{-\ell}}{\ell}f_{\ell}^k(\tau_0,\sigma_0)}|N_{\ell}\rangle$ and $(\langle N_{\ell}|e^{\frac{\alpha_{\ell}}{\ell}f_{\ell}^{-k}(-\tau_0,\sigma_0)^*})^{\dagger}$ actually belong to different finite-regulator vector spaces, namely $\mathcal{H}_{k_1 k_2}$ or $\mathcal{H}_{n_1,w_1}^{n_2,w_2}$ for the former but $\mathcal{H}_{-k_1,-k_2}$ or $\mathcal{H}_{-n_1,-w_1}^{-n_2,-w_2}$ for the latter. Thus, the matrix elements need not satisfy a positivity condition except in the special case that the external momenta are pure imaginary, for which $\rho^{\text{osc}}_{\ell}$ is manifestly Hermitian, and it is simple to check that all the eigenvalues are indeed positive. Next, note that the original trace from which we defined this state satisfies the appropriate asymptotic conditions on the Euclidean cylinder, but the definition \eqref{osc state general} is supposed to be a valid state in the full angular quantization and so should not care whether or not $\tau$ is compact. That is, \eqref{osc state general} will also satisfy the asymptotic conditions on the whole Euclidean strip (i.e.~for any $\tau\in\mathds{R}$) because the conditions \eqref{noncompact asymptotics 1}-\eqref{noncompact asymptotics 4} or \eqref{compact asymptotics 1}-\eqref{compact asymptotics 4} do not depend on time. Nevertheless, the definition \eqref{osc state general} always retains the connection to the original cylinder due to the factors $e^{-\frac{\pi^2}{2L}\alpha_{-\ell}\alpha_{\ell}}$ which arose from the evolution operator $e^{-\beta H_{\text{R}}}$ with $\beta = 2\pi$. Thus, all states in this superselection sector are thermal-like with inverse temperature exactly $2\pi$, and generically there are no pure states at all. This behavior is very similar to Rindler quantum field theory, where all states are mixed with an inverse temperature $2\pi$ as one approaches the Rindler horizon to reproduce the original Minkowski pure states. 

Lastly, a state like \eqref{osc state general} already takes into account the presence of exponential operators with the correct net charge, and so the nonvanishing expectation values in it are still those for which the total charge vanishes. We already wrote the zero-mode part of the state as depending on one continuous parameter for the noncompact case and on two integral parameters for the compact case because the slope-mode sector gets eliminated or discretized in exactly the same way as described in Section \ref{zero net charge}. We must also note that there is some ambiguity in how one defines the zero-mode part of the density matrix. The algebra of operators is graded by the total charge, and we chose $\rho(\tau_0,\sigma_0)$ to act nontrivially only on the charge zero sector. We can also define the zero-mode part to be given by $|p\rangle\langle p'|$ or $|m,m_{\text{s}}\rangle\langle m',m_{\text{s}}'|$ instead, for which the state acts nontrivially only on the charge $p-p'$ or $(m'-m,m_{\text{s}}'-m_{\text{s}})$ sector. We stick with the charge zero definition so that the expectation value $\langle X\rangle = \mathrm{Tr}[\rho X]$ is nonvanishing.

It remains to verify the above claims by actually computing the expectation values in the state \eqref{osc state general}, where we keep the parameters $\tau_0$ and $\sigma_0$ arbitrary for now.

\subsection{Infinitely Excited Oscillator State in the Spectrum}

In order to compute $\langle X(\tau,\sigma)\rangle$ in the full state $\rho(\tau_0,\sigma_0)$, we first calculate the expectation values $\langle\alpha_{\pm\ell}\rangle_{\ell}$ of one oscillator in the state $\rho^{\text{osc}}_{\ell}(\tau_0,\sigma_0)$ given in \eqref{osc state general}. These fixed mode number expectation values are
\begin{align}
\notag \langle\alpha_{\pm\ell}\rangle_{\ell} & = \mathrm{Tr}\left[\rho^{\text{osc}}_{\ell}(\tau_0,\sigma_0)\alpha_{\pm\ell}\right] 
\\ \notag & = \frac{1}{\mathcal{N}_{\ell}}\sum_{N_{\ell}=0}^{\infty}\langle N_{\ell}|e^{\frac{\alpha_{\ell}}{\ell}f_{\ell}^{-k}(-\tau_0,\sigma_0)^*}e^{-\frac{\pi^2}{2L}\alpha_{-\ell}\alpha_{\ell}}\alpha_{\pm\ell}e^{-\frac{\pi^2}{2L}\alpha_{-\ell}\alpha_{\ell}}e^{\frac{\alpha_{-\ell}}{\ell}f_{\ell}^k(\tau_0,\sigma_0)}|N_{\ell}\rangle
\\ \notag & = \frac{e^{-\frac{\pi^2\ell}{2L}}}{\mathcal{N}_{\ell}}\genfrac{\{}{\}}{0pt}{0}{\hspace{-2pt}f_{\ell}^k\hspace{-1pt}(\hspace{-1pt}\tau_0,\hspace{-1pt}\sigma_0\hspace{-1pt})}{f_{\ell}^{-k}(-\tau_0,\sigma_0)^*}\hspace{-3pt}\sum_{N_{\ell}=0}^{\infty}\hspace{-3pt}\frac{e^{-\frac{\pi^2\ell}{L}\hspace{-1pt}N_{\ell}}}{N_{\ell}!}\hspace{-2pt}\sum_{m=0}^{\infty}\hspace{-3pt}\frac{(N_{\ell}\hspace{-2pt}+\hspace{-2pt}m\hspace{-2pt}+\hspace{-2pt}1)!}{m!(m+1)!}\hspace{-3pt}\left(\hspace{-2pt}\frac{e^{-\frac{\pi^2\ell}{L}}}{\ell}\hspace{-1pt}f_{\ell}^{-k}\hspace{-1pt}(\hspace{-1pt}-\tau_0,\hspace{-1pt}\sigma_0\hspace{-1pt})^* \hspace{-2pt}f_{\ell}^k\hspace{-1pt}(\hspace{-1pt}\tau_0,\hspace{-1pt}\sigma_0\hspace{-1pt})\hspace{-3pt}\right)^{\hspace{-2pt}m}
%\\ \notag & = \frac{e^{-\frac{\pi^2\ell}{2L}}e^{\frac{e^{-\frac{\pi^2\ell}{L}}}{\ell}\hspace{-1pt}f_{\ell}^{-k}\hspace{-1pt}(\hspace{-1pt}-\tau_0,\hspace{-1pt}\sigma_0\hspace{-1pt})^* \hspace{-2pt}f_{\ell}^k\hspace{-1pt}(\hspace{-1pt}\tau_0,\hspace{-1pt}\sigma_0\hspace{-1pt})}}{\mathcal{N}_{\ell}}\genfrac{\{}{\}}{0pt}{0}{\hspace{-2pt}f_{\ell}^k\hspace{-1pt}(\hspace{-1pt}\tau_0,\hspace{-1pt}\sigma_0\hspace{-1pt})}{f_{\ell}^{-k}(-\tau_0,\sigma_0)^*}\hspace{-3pt}\sum_{N_{\ell}=0}^{\infty}e^{-\frac{\pi^2\ell}{L}N_{\ell}}L_{N_{\ell}}^{(1)}\left(-\frac{e^{-\frac{\pi^2\ell}{L}}}{\ell}\hspace{-1pt}f_{\ell}^{-k}\hspace{-1pt}(\hspace{-1pt}-\tau_0,\hspace{-1pt}\sigma_0\hspace{-1pt})^* \hspace{-2pt}f_{\ell}^k\hspace{-1pt}(\hspace{-1pt}\tau_0,\hspace{-1pt}\sigma_0\hspace{-1pt})\right)
\\ \notag & = \frac{e^{-\frac{\pi^2\ell}{2L}}}{\mathcal{N}_{\ell}\big(1-e^{-\frac{\pi^2\ell}{L}}\big)^2}\genfrac{\{}{\}}{0pt}{0}{f_{\ell}^k(\tau_0,\sigma_0)}{f_{\ell}^{-k}(-\tau_0,\sigma_0)^*}\exp\left[\frac{e^{-\frac{\pi^2\ell}{L}}}{\ell(1-e^{-\frac{\pi^2\ell}{L}})}\hspace{-1pt}f_{\ell}^{-k}\hspace{-1pt}(\hspace{-1pt}-\tau_0,\hspace{-1pt}\sigma_0\hspace{-1pt})^* \hspace{-2pt}f_{\ell}^k\hspace{-1pt}(\hspace{-1pt}\tau_0,\hspace{-1pt}\sigma_0\hspace{-1pt})\right]
\\ & = \frac{e^{-\frac{\pi^2\ell}{2L}}}{1-e^{-\frac{\pi^2\ell}{L}}}\genfrac{\{}{\}}{0pt}{0}{f_{\ell}^k(\tau_0,\sigma_0)}{f_{\ell}^{-k}(-\tau_0,\sigma_0)^*}.
\end{align}
The corresponding free boson oscillator expectation values are then
\begin{align}
\text{(N):} \ \langle \hspace{-1pt}X_{\text{osc}}\hspace{-1pt}(\hspace{-1pt}\tau,\hspace{-1pt}\sigma\hspace{-1pt})\hspace{-1pt}\rangle & \hspace{-2pt}=\hspace{-2pt} i\sqrt{2}\hspace{-1pt}\sum_{\ell=1}^{\infty}\hspace{-3pt}\frac{e^{\hspace{-1pt}-\hspace{-1pt}\frac{\pi^2\hspace{-1pt}\ell\hspace{-1pt}}{2L}}\hspace{-2pt}[\hspace{-1pt}e^{\hspace{-1pt}-\hspace{-1pt}\frac{\pi\ell\tau\hspace{-1pt}}{2L}}\hspace{-3pt}f_{\ell}^{\hspace{-1pt}-\hspace{-1pt}k_1\hspace{-2pt}-\hspace{-1pt}k_2}\hspace{-2pt}(\hspace{-1pt}\tau_0,\hspace{-2pt}\sigma_0\hspace{-1pt}) \hspace{-3pt}-\hspace{-3pt} e^{\hspace{-1pt}\frac{\pi\ell\tau\hspace{-1pt}}{2L}}\hspace{-3pt}f_{\ell}^{k_1\hspace{-2pt}+\hspace{-1pt}k_2}\hspace{-2pt}(\hspace{-2pt}-\hspace{-1pt}\tau_0,\hspace{-2pt}\sigma_0\hspace{-1pt})^{\hspace{-1pt}*}]}{\ell(1-e^{-\frac{\pi^2\ell}{L}})}\hspace{-2pt}\cos\hspace{-3pt}\left(\hspace{-3pt}\frac{\pi\ell(\hspace{-1pt}\sigma\hspace{-3pt}+\hspace{-3pt}L\hspace{-1pt})}{2L}\hspace{-3pt}\right)
\\ \text{(D):} \ \langle \hspace{-1pt}X_{\text{osc}}\hspace{-1pt}(\hspace{-1pt}\tau,\hspace{-1pt}\sigma\hspace{-1pt})\hspace{-1pt}\rangle & \hspace{-2pt}=\hspace{-2pt} \sqrt{2}\hspace{-1pt}\sum_{\ell=1}^{\infty}\hspace{-3pt}\frac{e^{\hspace{-1pt}-\hspace{-1pt}\frac{\pi^2\hspace{-1pt}\ell\hspace{-1pt}}{2L}}\hspace{-2pt}[\hspace{-1pt}e^{\hspace{-1pt}-\hspace{-1pt}\frac{\pi\ell\tau\hspace{-1pt}}{2L}}\hspace{-3pt}f_{\ell}^{\hspace{-1pt}-\hspace{-1pt}k_1\hspace{-2pt}-\hspace{-1pt}k_2}\hspace{-2pt}(\hspace{-1pt}\tau_0,\hspace{-2pt}\sigma_0\hspace{-1pt}) \hspace{-3pt}+\hspace{-3pt} e^{\hspace{-1pt}\frac{\pi\ell\tau\hspace{-1pt}}{2L}}\hspace{-3pt}f_{\ell}^{k_1\hspace{-2pt}+\hspace{-1pt}k_2}\hspace{-2pt}(\hspace{-2pt}-\hspace{-1pt}\tau_0,\hspace{-2pt}\sigma_0\hspace{-1pt})^{\hspace{-1pt}*}]}{\ell(1-e^{-\frac{\pi^2\ell}{L}})}\hspace{-2pt}\sin\hspace{-3pt}\left(\hspace{-3pt}\frac{\pi\ell(\hspace{-1pt}\sigma\hspace{-3pt}+\hspace{-3pt}L\hspace{-1pt})}{2L}\hspace{-3pt}\right)
\\ \hspace{-30pt}\text{compact:} \ \langle\hspace{-1pt} X_{\text{osc}}\hspace{-1pt}(\hspace{-1pt}\tau,\hspace{-1pt}\sigma\hspace{-1pt})\hspace{-1pt}\rangle & \hspace{-2pt}=\hspace{-2pt} \sqrt{\hspace{-1pt}2}\hspace{-1pt}\sum_{\ell=1}^{\infty}\hspace{-3pt}\frac{e^{\hspace{-2pt}-\hspace{-1pt}\frac{\pi^2\hspace{-1pt}\ell\hspace{-1pt}}{2L}}\hspace{-2pt}[\hspace{-1pt}e^{\hspace{-2pt}-\hspace{-1pt}\frac{\pi\ell\tau\hspace{-1pt}}{2L}}\hspace{-3pt}f_{\ell}^{\hspace{-1pt}-\hspace{-1pt}n_1\hspace{-2pt}-\hspace{-1pt}n_2,\hspace{-2pt}-\hspace{-1pt}w_1\hspace{-2pt}-\hspace{-1pt}w_2}\hspace{-2pt}(\hspace{-1pt}\tau_0,\hspace{-2pt}\sigma_0\hspace{-1pt}) \hspace{-3pt}+\hspace{-3pt} e^{\hspace{-2pt}\frac{\pi\ell\tau\hspace{-1pt}}{2L}}\hspace{-3pt}f_{\ell}^{\hspace{-1pt}n_1\hspace{-2pt}+\hspace{-1pt}n_2,\hspace{-1pt}w_1\hspace{-2pt}+\hspace{-1pt}w_2}\hspace{-2pt}(\hspace{-2pt}-\hspace{-1pt}\tau_0,\hspace{-2pt}\sigma_0\hspace{-1pt})^{\hspace{-1pt}*}\hspace{-1pt}]}{\ell(1-e^{-\frac{\pi^2\ell}{L}})}\hspace{-2pt}\sin\hspace{-3pt}\left(\hspace{-3pt}\frac{\pi\ell(\hspace{-1pt}\sigma\hspace{-3pt}+\hspace{-3pt}L\hspace{-1pt})}{2L}\hspace{-3pt}\right)\hspace{-2pt}.
\end{align}
With the expressions given above for the functions $f_{\ell}^k(\tau_0,\sigma_0)$, these sums can be performed in closed form involving $q$-hypergeometric functions, but this result is not illuminating. Fortunately, we need only the leading terms which do not vanish in the shrinking limit, so we may perform an asymptotic expansion in $\frac{1}{L}$. Doing so eliminates all factors of $e^{-\frac{\pi^2\ell}{L}}$, reducing the above expressions to
\begin{align}
\text{(N):} \ \langle \hspace{-1pt}X_{\text{osc}}\hspace{-1pt}(\hspace{-1pt}\tau,\hspace{-1pt}\sigma\hspace{-1pt})\hspace{-1pt}\rangle & \stackrel{L\rightarrow\infty}{\longrightarrow} -\frac{2i(\hspace{-1pt}k_1\hspace{-3pt}+\hspace{-3pt}k_2\hspace{-1pt})\hspace{-1pt}L}{\pi^2}\hspace{-2pt}\sum_{\ell=1}^{\infty}\hspace{-3pt}\frac{(\hspace{-1pt}e^{\hspace{-1pt}-\hspace{-1pt}\frac{\hspace{-2pt}\pi\ell(\hspace{-1pt}\tau\hspace{-1pt}-\hspace{-1pt}\tau_0\hspace{-1pt})\hspace{-2pt}}{2L}}\hspace{-6pt}+\hspace{-3pt}e^{\frac{\pi\ell(\hspace{-1pt}\tau\hspace{-1pt}-\hspace{-1pt}\tau_0\hspace{-1pt})\hspace{-2pt}}{2L}}\hspace{-1pt})}{\ell^2}\hspace{-2pt}\cos\hspace{-3pt}\left(\hspace{-3.5pt}\frac{\pi\ell(\hspace{-1pt}\sigma\hspace{-3pt}+\hspace{-3pt}L\hspace{-1pt})}{2L}\hspace{-3.5pt}\right)\hspace{-3pt}\cos\hspace{-3pt}\left(\hspace{-3.5pt}\frac{\pi\ell(\hspace{-1pt}\sigma_{\hspace{-1pt}0}\hspace{-3pt}+\hspace{-3pt}L\hspace{-1pt})}{2L}\hspace{-3.5pt}\right)
\\ \text{(D):} \ \langle \hspace{-1pt}X_{\text{osc}}\hspace{-1pt}(\hspace{-1pt}\tau,\hspace{-1pt}\sigma\hspace{-1pt})\hspace{-1pt}\rangle & \stackrel{L\rightarrow\infty}{\longrightarrow} -\frac{2i(\hspace{-1pt}k_1\hspace{-3pt}+\hspace{-3pt}k_2\hspace{-1pt})\hspace{-1pt}L}{\pi^2}\hspace{-2pt}\sum_{\ell=1}^{\infty}\hspace{-3pt}\frac{(\hspace{-1pt}e^{\hspace{-1pt}-\hspace{-1pt}\frac{\hspace{-2pt}\pi\ell(\hspace{-1pt}\tau\hspace{-1pt}-\hspace{-1pt}\tau_0\hspace{-1pt})\hspace{-2pt}}{2L}}\hspace{-6pt}+\hspace{-3pt}e^{\frac{\pi\ell(\hspace{-1pt}\tau\hspace{-1pt}-\hspace{-1pt}\tau_0\hspace{-1pt})\hspace{-2pt}}{2L}}\hspace{-1pt})}{\ell^2}\hspace{-2pt}\sin\hspace{-3pt}\left(\hspace{-3.5pt}\frac{\pi\ell(\hspace{-1pt}\sigma\hspace{-3pt}+\hspace{-3pt}L\hspace{-1pt})}{2L}\hspace{-3.5pt}\right)\hspace{-3pt}\sin\hspace{-3pt}\left(\hspace{-3.5pt}\frac{\pi\ell(\hspace{-1pt}\sigma_{\hspace{-1pt}0}\hspace{-3pt}+\hspace{-3pt}L\hspace{-1pt})}{2L}\hspace{-3.5pt}\right)
\\ \notag \hspace{-30pt}\text{compact:} \ \langle \hspace{-1pt}X_{\text{osc}}\hspace{-1pt}(\hspace{-1pt}\tau,\hspace{-1pt}\sigma)\hspace{-1pt}\rangle & \stackrel{L\rightarrow\infty}{\longrightarrow} -\frac{2iL}{\pi^2}\hspace{-2pt}\sum_{\ell=1}^{\infty}\hspace{-3pt}\frac{e^{\hspace{-1pt}-\hspace{-1pt}\frac{\pi\ell(\hspace{-1pt}\tau\hspace{-1pt}-\hspace{-1pt}\tau_0\hspace{-1pt})\hspace{-1pt}}{2L}}\hspace{-2pt}[\hspace{-1pt}\frac{n_1\hspace{-2pt}+\hspace{-1pt}n_2}{R}\hspace{-3pt}\sin\hspace{-1pt}(\hspace{-1pt}\frac{\pi\ell(\hspace{-1pt}\sigma_{\hspace{-1pt}0}\hspace{-1pt}+\hspace{-1pt}L\hspace{-1pt})}{2L}\hspace{-2pt}) \hspace{-3pt}+\hspace{-3pt} i(\hspace{-1pt}w_1\hspace{-4pt}+\hspace{-3pt}w_2\hspace{-1pt})\hspace{-1pt}R\hspace{-1pt}\cos\hspace{-1pt}(\hspace{-1pt}\frac{\pi\ell(\hspace{-1pt}\sigma_{\hspace{-1pt}0}\hspace{-1pt}+\hspace{-1pt}L\hspace{-1pt})}{2L}\hspace{-2pt})\hspace{-1pt}]}{\ell^2}\hspace{-2pt}\sin\hspace{-3.5pt}\left(\hspace{-3.5pt}\frac{\pi\ell(\hspace{-1pt}\sigma\hspace{-3pt}+\hspace{-3pt}L\hspace{-1pt})}{2L}\hspace{-3.5pt}\right) 
\\ & \hspace{-40pt} - \frac{2iL}{\pi^2}\sum_{\ell=1}^{\infty}\frac{e^{\hspace{-1pt}\frac{\pi\ell(\hspace{-1pt}\tau\hspace{-1pt}-\hspace{-1pt}\tau_0\hspace{-1pt})\hspace{-1pt}}{2L}}\hspace{-2pt}[\hspace{-1pt}\frac{n_1\hspace{-2pt}+\hspace{-1pt}n_2}{R}\hspace{-2pt}\sin(\hspace{-1pt}\frac{\pi\ell(\hspace{-1pt}\sigma_{\hspace{-1pt}0}\hspace{-1pt}+\hspace{-1pt}L\hspace{-1pt})}{2L}\hspace{-1pt})\hspace{-2pt}-\hspace{-2pt}i(w_1\hspace{-3pt}+\hspace{-2pt}w_2)\hspace{-1pt}R\hspace{-1pt}\cos(\hspace{-1pt}\frac{\pi\ell(\hspace{-1pt}\sigma_{\hspace{-1pt}0}\hspace{-1pt}+\hspace{-1pt}L\hspace{-1pt})}{2L}\hspace{-2pt})]}{\ell^2}\hspace{-2pt}\sin\hspace{-3pt}\left(\hspace{-3pt}\frac{\pi\ell(\hspace{-1pt}\sigma\hspace{-3pt}+\hspace{-3pt}L\hspace{-1pt})}{2L}\hspace{-3pt}\right), 
\end{align}
which are now of dilogarithmic form. The sums that we need here are
\begin{align}
\notag \hspace{10pt}&\hspace{-10pt}\sum_{\ell=1}^{\infty}\frac{1}{\ell^2}\cosh\left(\frac{\pi\ell\tau_{12}}{2L}\right)\cos\left(\frac{\pi\ell(\sigma_1+L)}{2L}\right)\cos\left(\frac{\pi\ell(\sigma_2+L)}{2L}\right)
\\ & = \frac{\pi^2}{8}\hspace{-3pt}\left[\hspace{-1pt}\frac{1}{3} \hspace{-3pt}-\hspace{-3pt} \theta(|\sigma_{12}|\hspace{-3pt}-\hspace{-3pt}|\mathrm{Im}\s\tau_{12}|)\frac{|\sigma_{12}|}{L} \hspace{-3pt}+\hspace{-3pt} \theta(|\mathrm{Im}\s\tau_{12}|\hspace{-3pt}-\hspace{-3pt}|\sigma_{12}|)\mathrm{sgn}(\mathrm{Im}\s\tau_{12})\frac{i\tau_{12}}{L} \hspace{-3pt}+\hspace{-3pt} \frac{\sigma_1^2\hspace{-3pt}+\hspace{-3pt}\sigma_2^2\hspace{-3pt}-\hspace{-3pt}\tau_{12}^2}{2L^2}\hspace{-1pt}\right]
\\ \notag \hspace{10pt}&\hspace{-10pt}\sum_{\ell=1}^{\infty}\frac{1}{\ell^2}\cosh\left(\frac{\pi\ell\tau_{12}}{2L}\right)\sin\left(\frac{\pi\ell(\sigma_1+L)}{2L}\right)\sin\left(\frac{\pi\ell(\sigma_2+L)}{2L}\right)
\\ & = \frac{\pi^2}{8}\hspace{-3pt}\left[1 - \theta(|\sigma_{12}|\hspace{-2pt}-\hspace{-2pt}|\mathrm{Im}\s\tau_{12}|)\frac{|\sigma_{12}|}{L} + \theta(|\mathrm{Im}\s\tau_{12}|\hspace{-2pt}-\hspace{-2pt}|\sigma_{12}|)\mathrm{sgn}(\mathrm{Im}\s\tau_{12})\frac{i\tau_{12}}{L} - \frac{\sigma_1\sigma_2}{L^2}\right]
\\ \notag \hspace{10pt}&\hspace{-10pt}\sum_{\ell=1}^{\infty}\frac{1}{\ell^2}\sinh\left(\frac{\pi\ell\tau_{12}}{2L}\right)\cos\left(\frac{\pi\ell(\sigma_1+L)}{2L}\right)\sin\left(\frac{\pi\ell(\sigma_2+L)}{2L}\right)
\\ & = \frac{\pi^2 i}{8}\hspace{-3pt}\left[\hspace{-2pt}-\theta(|\mathrm{Im}\s\tau_{12}|\hspace{-3pt}-\hspace{-3pt}|\sigma_{12}|)\mathrm{sgn}(\mathrm{Im}\s\tau_{12}\hspace{-1pt})\hspace{-1pt}\frac{\sigma_{12}}{L} \hspace{-3pt}+\hspace{-3pt} \theta(|\sigma_{12}|\hspace{-3pt}-\hspace{-3pt}|\mathrm{Im}\s\tau_{12}|)\mathrm{sgn}(\hspace{-1pt}\sigma_{12}\hspace{-1pt})\hspace{-1pt}\frac{i\tau_{12}}{L} \hspace{-3pt}+\hspace{-3pt} \frac{2i\sigma_2\tau_{12}}{L^2}\hspace{-2pt}\right]\hspace{-2pt}.
\end{align}
The final term in each expression above vanishes in the shrinking limit. Adding these results to the zero-momentum-mode contributions \eqref{noncompact osc vac} and \eqref{compact osc vac}, we conclude that the free boson expectation values in the shrinking limit of the infinitely excited oscillator state $\rho(\tau_0,\sigma_0)$ are
\begin{multline}\label{noncompact N full}
\left\langle X(\tau,\sigma)\right\rangle_{\rho(\tau_0,\sigma_0)} = \langle x\rangle - \frac{i(k_1-k_2)}{2}\sigma - \frac{i(k_1+k_2)L}{6}
\\ + \frac{i(k_1+k_2)}{2}\theta\Big(|\sigma-\sigma_0|-|\mathrm{Im}(\tau-\tau_0)|\Big)|\sigma-\sigma_0|
\\ + \frac{k_1+k_2}{2}\theta\Big(|\mathrm{Im}(\tau-\tau_0)|-|\sigma-\sigma_0|\Big)\mathrm{sgn}\big(\mathrm{Im}(\tau-\tau_0)\big)(\tau-\tau_0)
\end{multline}
for the noncompact Neumann-class angular quantization, 
\begin{multline}\label{noncompact D full}
\left\langle X(\tau,\sigma)\right\rangle_{\rho(\tau_0,\sigma_0)} = \langle x\rangle - \frac{i(k_1-k_2)}{2}\sigma - \frac{i(k_1+k_2)L}{2}
\\ + \frac{i(k_1+k_2)}{2}\theta\Big(|\sigma-\sigma_0|-|\mathrm{Im}(\tau-\tau_0)|\Big)|\sigma-\sigma_0|
\\ + \frac{k_1+k_2}{2}\theta\Big(|\mathrm{Im}(\tau-\tau_0)|-|\sigma-\sigma_0|\Big)\mathrm{sgn}\big(\mathrm{Im}(\tau-\tau_0)\big)(\tau-\tau_0)
\end{multline}
for the noncompact Dirichlet-class angular quantization and finally
\begin{multline}\label{compact full}
\left\langle X(\tau,\sigma)\right\rangle_{\rho(\tau_0,\sigma_0)} = \langle x\rangle - \frac{i(n_1-n_2)}{2R}\sigma + \frac{(w_1 - w_2)R}{2}\tau - \frac{i(n_1+n_2)L}{2R}
\\ +\theta\Big(|\sigma\hspace{-2pt}-\hspace{-2pt}\sigma_0|-|\mathrm{Im}(\tau\hspace{-2pt}-\hspace{-2pt}\tau_0)|\Big)\left[\frac{i(n_1\hspace{-2pt}+\hspace{-2pt}n_2)}{2R}|\sigma\hspace{-2pt}-\hspace{-2pt}\sigma_0| - \frac{(w_1\hspace{-2pt}+\hspace{-2pt}w_2)R}{2}\mathrm{sgn}(\sigma\hspace{-2pt}-\hspace{-2pt}\sigma_0)(\tau\hspace{-2pt}-\hspace{-2pt}\tau_0)\right]
\\ + \theta\Big(|\mathrm{Im}(\tau\hspace{-2pt}-\hspace{-2pt}\tau_0)|-|\sigma\hspace{-2pt}-\hspace{-2pt}\sigma_0|\Big)\mathrm{sgn}\big(\mathrm{Im}(\tau\hspace{-2pt}-\hspace{-2pt}\tau_0)\big)\hspace{-3pt}\left[\frac{n_1\hspace{-2pt}+\hspace{-2pt}n_2}{2R}(\tau\hspace{-2pt}-\hspace{-2pt}\tau_0) - \frac{i(w_1\hspace{-3pt}+\hspace{-2pt}w_2)R}{2}(\sigma\hspace{-3pt}-\hspace{-2pt}\sigma_0)\right]
\end{multline}
for the compact boson angular quantization. At last we may easily verify that these expectation values do actually obey the requisite asymptotic conditions. Restricting to Euclidean time ($\tau,\tau_0 \in \mathds{R}$), the derivative expectation values in the shrinking limit are
\begin{align}
\text{noncompact (N):} \ \ \left\langle \partial_{\sigma}X(\tau,\sigma)\right\rangle_{\rho(\tau_0,\sigma_0)} & = -\frac{i(k_1\hspace{-2pt}-\hspace{-2pt}k_2)}{2} + \frac{i(k_1\hspace{-2pt}+\hspace{-2pt}k_2)}{2}\mathrm{sgn}(\sigma\hspace{-2pt}-\hspace{-2pt}\sigma_0)
\\ \left\langle \partial_{\tau}X(\tau,\sigma)\right\rangle_{\rho(\tau_0,\sigma_0)} & = 0
\\ \text{noncompact (D):} \ \ \left\langle \partial_{\sigma}X(\tau,\sigma)\right\rangle_{\rho(\tau_0,\sigma_0)} & = -\frac{i(k_1\hspace{-2pt}-\hspace{-2pt}k_2)}{2} + \frac{i(k_1\hspace{-2pt}+\hspace{-2pt}k_2)}{2}\mathrm{sgn}(\sigma\hspace{-2pt}-\hspace{-2pt}\sigma_0)
\\ \left\langle \partial_{\tau}X(\tau,\sigma)\right\rangle_{\rho(\tau_0,\sigma_0)} & = 0
\\ \text{compact:} \ \ \left\langle \partial_{\sigma}X(\tau,\sigma)\right\rangle_{\rho(\tau_0,\sigma_0)} & = -\frac{i(n_1\hspace{-2pt}-\hspace{-2pt}n_2)}{2R} + \frac{i(n_1\hspace{-2pt}+\hspace{-2pt}n_2)}{2R}\mathrm{sgn}(\sigma\hspace{-2pt}-\hspace{-2pt}\sigma_0)
\\ \left\langle \partial_{\tau}X(\tau,\sigma)\right\rangle_{\rho(\tau_0,\sigma_0)} & = \frac{(w_1\hspace{-2pt}-\hspace{-2pt}w_2)R}{2} - \frac{(w_1\hspace{-2pt}+\hspace{-2pt}w_2)R}{2}\mathrm{sgn}(\sigma\hspace{-2pt}-\hspace{-2pt}\sigma_0),
\end{align}
where we do not write the delta function terms for brevity since they play no role here. Therefore, we immediately see that the noncompact boson expectation values \eqref{noncompact N full} and \eqref{noncompact D full} satisfy the asymptotic conditions \eqref{noncompact asymptotics 1}-\eqref{noncompact asymptotics 4} and that the compact boson expectation value \eqref{compact full} satisfies the asymptotic conditions \eqref{compact asymptotics 1}-\eqref{compact asymptotics 4} for arbitrary $\tau_0$ and $\sigma_0$, showing that the infinitely excited oscillator state $\rho(\tau_0,\sigma_0)$ is indeed in the final spectrum of angular quantization. Moreover, the full expectation values \eqref{noncompact N full} and \eqref{noncompact D full} are equal (up to an additive constant), which almost completes the proof that the Neumann- and Dirichlet-class boundary conditions are equivalent in angular quantization.

\subsection{The Other Oscillator States}

Having constructed one oscillator state in the Minkowski spectrum, we must now deduce how to obtain all remaining oscillator states. The most natural guess is that a Fock-like structure still prevails but now built on top of a different state than the vacuum; the most important difference is that the base state is already infinitely excited compared to the original vacuum, so one should act on it with all combinations of creation \emph{and} annihilation operators. One would then expect that no finite combinations of such oscillators could alter the asymptotic conditions, which is what we shall establish in this subsection.

Note that the creation and annihilation operators map states to states via conjugation of the density operator. Starting from the $\rho(\tau_0,\sigma_0)$ defined by $\rho^{\text{osc}}_{\ell}(\tau_0,\sigma_0)$ in \eqref{osc state general}, we define a new oscillator state $\rho_{\{N_{\ell}\}}(\tau_0,\sigma_0)$ by using
\begin{equation}
\rho^{\text{osc}}_{\ell,N_{\ell}}(\tau_0,\sigma_0) \equiv \frac{1}{c(N_{\ell})}\alpha_{-\ell}^{N_{\ell}}\rho^{\text{osc}}_{\ell}(\tau_0,\sigma_0)\alpha_{\ell}^{N_{\ell}}
\end{equation}
in place of $\rho^{\text{osc}}_{\ell}(\tau_0,\sigma_0)$. As usual only finitely many of the `occupation' numbers $N_{\ell}$ can be nonzero, and $c(N_{\ell})$ are normalization constants. In the definition of $\rho^{\text{osc}}_{\ell,N_{\ell}}$, we now let $N_{\ell} \in \mathds{Z}$, where $\alpha_{-\ell}$ to a power $N_{\ell} < 0$ is defined to be $\alpha_{\ell}$ to the power $-N_{\ell}$, i.e.
\begin{equation}
\alpha_{-\ell}^{N_{\ell}} \equiv \alpha_{-\mathrm{sgn}(N_{\ell})\ell}^{|N_{\ell}|}.
\end{equation}
It is clear that, with this definition, all such states $\rho_{\{N_{\ell}\}}$ are independent but they are not orthogonal. The normalization constants $c(N_{\ell})$ ensuring $\mathrm{Tr}[\rho^{\text{osc}}_{\ell,N_{\ell}}] = 1$ must still be computed; in our conventions, $c(0) = 1$ because the state $\rho^{\text{osc}}_{\ell}$ is already normalized. From
\begin{equation}
\mathrm{Tr}[\rho^{\text{osc}}_{\ell}] = \frac{1}{c(N_{\ell})\mathcal{N}_{\ell}}\sum_{N'_{\ell}=0}^{\infty}\langle N'_{\ell}|e^{\frac{\alpha_{\ell}}{\ell}f_{\ell}^{-k}(-\tau_0,\sigma_0)^*}e^{-\frac{\pi^2}{2L}\alpha_{-\ell}\alpha_{\ell}}\alpha_{\ell}^{N_{\ell}}\alpha_{-\ell}^{N_{\ell}}e^{-\frac{\pi^2}{2L}\alpha_{-\ell}\alpha_{\ell}}e^{\frac{\alpha_{-\ell}}{\ell}f_{\ell}^k(\tau_0,\sigma_0)}|N'_{\ell}\rangle,
\end{equation}
we compute
\begin{align}
\notag\hspace{-10pt}\text{$N_{\ell} \geqslant 0$:} \ \ c(N_{\ell}) % & = \hspace{-2pt}\frac{1}{\mathcal{N}_{\ell}}\hspace{-3pt}\sum_{N'_{\ell}=0}^{\infty}\hspace{-3pt}\langle N'_{\ell}|e^{\frac{\alpha_{\ell}}{\ell}f_{\ell}^{-k}\hspace{-1pt}(\hspace{-1pt}-\hspace{-1pt}\tau_0,\sigma_0\hspace{-1pt})^*}\hspace{-2pt}e^{\hspace{-1pt}-\hspace{-1pt}\frac{\pi^2}{2L}\alpha_{-\ell}\alpha_{\ell}}\hspace{-1pt}\alpha_{\ell}^{N_{\ell}}\hspace{-1pt}\alpha_{-\ell}^{N_{\ell}}e^{\hspace{-1pt}-\hspace{-1pt}\frac{\pi^2}{2L}\alpha_{-\ell}\alpha_{\ell}}e^{\hspace{-1pt}\frac{\alpha_{-\ell}}{\ell}f_{\ell}^k\hspace{-1pt}(\hspace{-1pt}\tau_0,\sigma_0\hspace{-1pt})}|N'_{\ell}\rangle
%\\ & = \frac{1}{\mathcal{N}_{\ell}}\sum_{N'_{\ell}=0}^{\infty}\frac{\ell^{N_{\ell}}e^{-\frac{\pi^2\ell}{L}N'_{\ell}}}{N'_{\ell}!}\sum_{m=0}^{\infty}\frac{(N_{\ell}+N'_{\ell}+m)!}{m!^2}\left(\frac{e^{-\frac{\pi^2\ell}{L}}}{\ell}f_{\ell}^k f_{\ell}^{-k*}\right)^m
& = \frac{1}{\mathcal{N}_{\ell}}\sum_{N'_{\ell}=0}^{\infty}\frac{\ell^{N_{\ell}}(N_{\ell}\hspace{-3pt}+\hspace{-3pt}N'_{\ell})!e^{-\frac{\pi^2\ell}{L}N'_{\ell}}}{N'_{\ell}!}e^{\frac{e^{-\frac{\pi^2\ell\hspace{-2pt}}{L}}\hspace{-2pt}}{\ell}f_{\ell}^{-k*}f_{\ell}^k }L_{N_{\ell}+N'_{\ell}}\hspace{-3pt}\left(\hspace{-3pt}-\frac{e^{-\frac{\pi^2\ell}{L}}\hspace{-2pt}}{\ell}f_{\ell}^{-k*}f_{\ell}^k \hspace{-3pt}\right)
\\ & = \frac{\ell^{N_{\ell}}N_{\ell}!}{(1-e^{-\frac{\pi^2\ell}{L}})^{N_{\ell}}}L_{N_{\ell}}\left(-\frac{e^{-\frac{\pi^2\ell}{L}}}{\ell(1-e^{-\frac{\pi^2\ell}{L}})}f_{\ell}^{-k*}f_{\ell}^k\right)
\\ \notag\hspace{-10pt}\text{$N_{\ell} < 0$:} \ \ c(N_{\ell}) % & = \hspace{-2pt}\frac{1}{\mathcal{N}_{\ell}}\hspace{-3pt}\sum_{N'_{\ell}=0}^{\infty}\hspace{-3pt}\langle N'_{\ell}|e^{\frac{\alpha_{\ell}}{\ell}f_{\ell}^{-k}\hspace{-1pt}(\hspace{-1pt}-\hspace{-1pt}\tau_0,\sigma_0\hspace{-1pt})^*}\hspace{-2pt}e^{\hspace{-1pt}-\hspace{-1pt}\frac{\pi^2}{2L}\alpha_{-\ell}\alpha_{\ell}}\hspace{-1pt}\alpha_{-\ell}^{|\hspace{-1pt}N_{\ell}\hspace{-1pt}|}\hspace{-1pt}\alpha_{\ell}^{|\hspace{-1pt}N_{\ell}\hspace{-1pt}|}\hspace{-1pt}e^{\hspace{-1pt}-\hspace{-1pt}\frac{\pi^2}{2L}\alpha_{-\ell}\alpha_{\ell}}e^{\hspace{-1pt}\frac{\alpha_{-\ell}}{\ell}f_{\ell}^k\hspace{-1pt}(\hspace{-1pt}\tau_0,\sigma_0\hspace{-1pt})}|N'_{\ell}\rangle
%\\ & = \frac{(-1)^{|N_{\ell}|}\ell^{|N_{\ell}|}}{\mathcal{N}_{\ell}}\sum_{N'_{\ell}=0}^{\infty}\frac{e^{-\frac{\pi^2\ell}{L}N'_{\ell}}}{N'_{\ell}!}\sum_{n=0}^{|N_{\ell}|}\frac{(-1)^n |N_{\ell}|!^2}{n!^2(|N_{\ell}|\hspace{-3pt}-\hspace{-3pt}n)!}\sum_{m=0}^{\infty}\frac{(N'_{\ell}\hspace{-3pt}+\hspace{-3pt}m\hspace{-3pt}+\hspace{-3pt}n)!}{m!^2}\hspace{-3pt}\left(\hspace{-3pt}\frac{e^{-\frac{\pi^2\ell}{L}}}{\ell}f_{\ell}^{-k*}f_{\ell}^k \hspace{-3pt}\right)^{\hspace{-3pt}m} 
& = \frac{\ell^{|\hspace{-1pt}N_{\ell}\hspace{-1pt}|}\hspace{-2pt}}{\mathcal{N}_{\ell}}e^{\hspace{-1pt}\frac{e^{-\frac{\pi^2\ell\hspace{-2pt}}{L}}\hspace{-2pt}}{\ell}\hspace{-2pt}f_{\ell}^{-k*}\hspace{-2pt}f_{\ell}^k}\hspace{-4pt}\sum_{N'_{\ell}=0}^{\infty}\hspace{-3pt}\frac{e^{\hspace{-1pt}-\hspace{-1pt}\frac{\pi^2\ell\hspace{-1pt}}{L}\hspace{-2pt}N'_{\ell}}}{N'_{\ell}!}\hspace{-2pt}\sum_{n=0}^{|N_{\ell}|}\hspace{-2pt}\frac{(\hspace{-1pt}-1\hspace{-1pt})^{n+N_{\ell}} |N_{\ell}|!^2\hspace{-2pt}}{n!^2(|N_{\ell}|\hspace{-3pt}-\hspace{-3pt}n)!}\hspace{-1pt}(\hspace{-1pt}N'_{\ell}\hspace{-3pt}+\hspace{-3pt}n\hspace{-1pt})!L_{N'_{\ell}+n}\hspace{-4pt}\left(\hspace{-3pt}-\frac{e^{-\frac{\pi^2\ell}{L}}}{\ell}f_{\ell}^{-k*}f_{\ell}^k \hspace{-3pt}\right)
%\\ & = (-1)^{|N_{\ell}|}\ell^{|N_{\ell}|}|N_{\ell}|!\sum_{n=0}^{|N_{\ell}|}\frac{(-1)^n |N_{\ell}|!}{n!(|N_{\ell}|\hspace{-3pt}-\hspace{-3pt}n)!}\frac{1}{(1-e^{-\frac{\pi^2\ell}{L}})^{n}}L_{n}\left(\hspace{-3pt}-\frac{e^{-\frac{\pi^2\ell}{L}}}{\ell(1-e^{-\frac{\pi^2\ell}{L}})}f_{\ell}^{-k*}f_{\ell}^k \hspace{-3pt}\right)
\\ & = \ell^{|N_{\ell}|}|N_{\ell}|!\left(\frac{e^{-\frac{\pi^2\ell}{L}}}{1-e^{-\frac{\pi^2\ell}{L}}}\right)^{|N_{\ell}|} L_{|N_{\ell}|}\left(\hspace{-3pt}-\frac{1}{\ell(1-e^{-\frac{\pi^2\ell}{L}})}f_{\ell}^{-k*}f_{\ell}^k \hspace{-3pt}\right),
\end{align}
where it is understood that $f_{\ell}^k$ always means $f_{\ell}^k(\tau_0,\sigma_0)$ and that $f_{\ell}^{-k*}$ always means $f_{\ell}^{-k}(-\tau_0,\sigma_0)^*$. Hence, these normalization constants are given by the single expression
\begin{equation}\label{general oscillator normalization}
c(N_{\ell}) = \frac{\ell^{|N_{\ell}|}|N_{\ell}|!}{(1-e^{-\frac{\pi^2\ell}{L}})^{|N_{\ell}|}}e^{-\frac{\pi^2\ell}{L}|N_{\ell}|\theta(-N_{\ell})}L_{|N_{\ell}|}\left(-\frac{e^{-\frac{\pi^2\ell}{L}\theta(N_{\ell})}}{\ell(1-e^{-\frac{\pi^2\ell}{L}})}f_{\ell}^{-k*}f_{\ell}^k\right).
\end{equation}
Note that $c(N_{\ell})$ and $c(-N_{\ell})$ become equal in the shrinking limit $L\rightarrow\infty$.

Now let us show that $\rho_{\{N_{\ell}\}}(\tau_0,\sigma_0)$ also belongs to the Minkowksi CFT state space by verifying the asymptotic conditions of $X(\tau,\sigma)$ in it. To do so, we must first compute
\begin{multline}
\left\langle\alpha_{\pm\ell}\right\rangle_{\rho_{\{N_{\ell}\}}} = \frac{1}{c(N_{\ell})}\mathrm{Tr}\left[\alpha_{\pm\ell}\left(\prod_{\ell'=1}^{\infty}\alpha_{-\ell'}^{N_{\ell'}}\right)\bigotimes_{\ell''=1}^{\infty}\rho^{\text{osc}}_{\ell''}(\tau_0,\sigma_0)\left(\prod_{\ell'''=1}^{\infty}\alpha_{\ell'''}^{N_{\ell'''}}\right)\right]
\\ = \frac{1}{c(N_{\ell})\mathcal{N}_{\ell}}\sum_{N'_{\ell}=0}^{\infty}\langle N'_{\ell}|e^{\frac{\alpha_{\ell}}{\ell}f_{\ell}^{-k}(-\tau_0,\sigma_0)^*}e^{-\frac{\pi^2}{2L}\alpha_{-\ell}\alpha_{\ell}}\alpha_{\ell}^{N_{\ell}}\alpha_{\pm\ell}\alpha_{-\ell}^{N_{\ell}}e^{-\frac{\pi^2}{2L}\alpha_{-\ell}\alpha_{\ell}}e^{\frac{\alpha_{-\ell}}{\ell}f_{\ell}^k(\tau_0,\sigma_0)}|N'_{\ell}\rangle,
\end{multline}
where all the other states $|\{N'_{\ell'}\}\rangle$ with $\ell'\neq \ell$ do not contribute since they have already been normalized. For $N_{\ell} \geqslant 0$, we have
\begin{multline}
\frac{1}{\mathcal{N}_{\ell}}\sum_{N'_{\ell}=0}^{\infty}\frac{1}{\ell^{N'_{\ell}}N'_{\ell}!}\langle 0|\alpha_{\ell}^{N'_{\ell}}e^{\frac{\alpha_{\ell}}{\ell}f_{\ell}^{-k}(-\tau_0,\sigma_0)^*}e^{-\frac{\pi^2}{2L}\alpha_{-\ell}\alpha_{\ell}}\alpha_{\ell}^{N_{\ell}}\alpha_{\pm\ell}\alpha_{-\ell}^{N_{\ell}}e^{-\frac{\pi^2}{2L}\alpha_{-\ell}\alpha_{\ell}}e^{\frac{\alpha_{-\ell}}{\ell}f_{\ell}^k(\tau_0,\sigma_0)}\alpha_{-\ell}^{N'_{\ell}}|0\rangle
\\ = \frac{\ell^{N_{\ell}}e^{-\frac{\pi^2\ell}{2L}}}{\mathcal{N}_{\ell}}\genfrac{\{}{\}}{0pt}{0}{f_{\ell}^k}{f_{\ell}^{-k*}}\sum_{N'_{\ell}=0}^{\infty}\frac{e^{-\frac{\pi^2\ell}{L}N'_{\ell}}}{N'_{\ell}!}\sum_{m=0}^{\infty}\frac{(N_{\ell}+N'_{\ell}+m+1)!}{m!(m+1)!}\left(\frac{e^{-\frac{\pi^2\ell}{L}}}{\ell}f_{\ell}^{-k*}f_{\ell}^k\right)^{m}
%\\ = \frac{\ell^{N_{\ell}}e^{-\frac{\pi^2\ell}{2L}}}{\mathcal{N}_{\ell}}\genfrac{\{}{\}}{0pt}{0}{f_{\ell}^k}{f_{\ell}^{-k*}}e^{\frac{e^{-\frac{\pi^2\ell}{L}}}{\ell}f_{\ell}^{-k*}f_{\ell}^k}\sum_{N'_{\ell}=0}^{\infty}\frac{(N_{\ell}+N'_{\ell})!}{N'_{\ell}!}e^{-\frac{\pi^2\ell}{L}N'_{\ell}}L_{N_{\ell}+N'_{\ell}}^{(1)}\left(-\frac{e^{-\frac{\pi^2\ell}{L}}}{\ell}f_{\ell}^{-k*}f_{\ell}^k\right)
\\ = \genfrac{\{}{\}}{0pt}{0}{f_{\ell}^k}{f_{\ell}^{-k*}}\frac{\ell^{N_{\ell}}N_{\ell}!e^{-\frac{\pi^2\ell}{2L}}}{(1-e^{-\frac{\pi^2\ell}{L}})^{N_{\ell}+1}}L_{N_{\ell}}^{(1)}\left(-\frac{e^{-\frac{\pi^2\ell}{L}}}{\ell(1-e^{-\frac{\pi^2\ell}{L}})}f_{\ell}^{-k*}f_{\ell}^k\right).
\end{multline}
Likewise, for $N_{\ell} < 0$, we have
\begin{multline}
\frac{1}{\mathcal{N}_{\ell}}\sum_{N'_{\ell}=0}^{\infty}\frac{1}{\ell^{N'_{\ell}}N'_{\ell}!}\langle 0|\alpha_{\ell}^{N'_{\ell}}e^{\frac{\alpha_{\ell}}{\ell}f_{\ell}^{-k}(-\tau_0,\sigma_0)^*}e^{-\frac{\pi^2}{2L}\alpha_{-\ell}\alpha_{\ell}}\alpha_{-\ell}^{|N_{\ell}|}\alpha_{\pm\ell}\alpha_{\ell}^{|N_{\ell}|}e^{-\frac{\pi^2}{2L}\alpha_{-\ell}\alpha_{\ell}}e^{\frac{\alpha_{-\ell}}{\ell}f_{\ell}^k(\tau_0,\sigma_0)}\alpha_{-\ell}^{N'_{\ell}}|0\rangle
\\ = \frac{(-1)^{|N_{\ell}|}\ell^{|N_{\ell}|}e^{-\frac{\pi^2\ell}{2L}}}{\mathcal{N}_{\ell}}\genfrac{\{}{\}}{0pt}{0}{f_{\ell}^{k}}{f_{\ell}^{-k*}}\sum_{N'_{\ell}=0}^{\infty}\frac{e^{-\frac{\pi^2\ell}{L}N'_{\ell}}}{N'_{\ell}!}\sum_{n=0}^{|N_{\ell}|}\frac{(-1)^n |N_{\ell}|!^2}{n!^2(|N_{\ell}|-n)!}
\\ \times \left[\sum_{m=0}^{\infty}\frac{(N'_{\ell}\hspace{-3pt}+\hspace{-3pt}m\hspace{-3pt}+\hspace{-3pt}n\hspace{-3pt}+\hspace{-3pt}1)!}{m!(m+1)!}\left(\frac{e^{-\frac{\pi^2\ell}{L}}}{\ell}f_{\ell}^{-k*}f_{\ell}^k\right)^{m} - n\sum_{m=0}^{\infty}\frac{(N'_{\ell}\hspace{-3pt}+\hspace{-3pt}m\hspace{-3pt}+\hspace{-3pt}n)!}{m!(m+1)!}\left(\frac{e^{-\frac{\pi^2\ell}{L}}}{\ell}f_{\ell}^{-k*}f_{\ell}^k\right)^m\right] 
\\ = \frac{\ell^{|N_{\ell}|}|N_{\ell}|!e^{-\frac{\pi^2\ell}{2L}}}{1-e^{-\frac{\pi^2\ell}{L}}}\genfrac{\{}{\}}{0pt}{0}{f_{\ell}^{k}}{f_{\ell}^{-k*}}\left(\frac{e^{-\frac{\pi^2\ell}{L}}}{1-e^{-\frac{\pi^2\ell}{L}}}\right)^{|N_{\ell}|} L_{|N_{\ell}|}^{(1)}\left(-\frac{1}{\ell(1-e^{-\frac{\pi^2\ell}{L}})}f_{\ell}^{-k*}f_{\ell}^k\right).
\end{multline}
Dividing by the normalization constant \eqref{general oscillator normalization}, we have thus computed 
\begin{equation}
\left\langle\alpha_{\pm\ell}\right\rangle_{\rho_{\{N_{\ell}\}}} = \frac{e^{-\frac{\pi^2\ell}{2L}}}{1-e^{-\frac{\pi^2\ell}{L}}}\genfrac{\{}{\}}{0pt}{0}{f_{\ell}^k}{f_{\ell}^{-k*}}\frac{L_{|N_{\ell}|}^{(1)}\left(-\frac{e^{-\frac{\pi^2\ell}{L}\theta(N_{\ell})}}{\ell(1-e^{-\frac{\pi^2\ell}{L}})}f_{\ell}^{-k*}f_{\ell}^k\right)}{L_{|N_{\ell}|}\left(-\frac{e^{-\frac{\pi^2\ell}{L}\theta(N_{\ell})}}{\ell(1-e^{-\frac{\pi^2\ell}{L}})}f_{\ell}^{-k*}f_{\ell}^k\right)}.
\end{equation}
Noting that the large-$L$ limit of the Laguerre polynomial ratio above is
\begin{equation}
\frac{L_{|N_{\ell}|}^{(1)}\left(-\frac{e^{-\frac{\pi^2\ell}{L}\theta(N_{\ell})}}{\ell(1-e^{-\frac{\pi^2\ell}{L}})}f_{\ell}^{-k*}f_{\ell}^k\right)}{L_{|N_{\ell}|}\left(-\frac{e^{-\frac{\pi^2\ell}{L}\theta(N_{\ell})}}{\ell(1-e^{-\frac{\pi^2\ell}{L}})}f_{\ell}^{-k*}f_{\ell}^k\right)} \ \stackrel{L\rightarrow\infty}{\longrightarrow} \ 1,
\end{equation}
we conclude that the single-oscillator expectation value in the shrinking limit is
\begin{equation}
\left\langle\alpha_{\pm\ell}\right\rangle_{\rho_{\{N_{\ell}\}}(\tau_0,\sigma_0)} \ \stackrel{L\rightarrow\infty}{\longrightarrow} \ \frac{e^{-\frac{\pi^2\ell}{2L}}}{1-e^{-\frac{\pi^2\ell}{L}}}\genfrac{\{}{\}}{0pt}{0}{f_{\ell}^k}{f_{\ell}^{-k*}} = \left\langle\alpha_{\pm\ell}\right\rangle_{\rho(\tau_0,\sigma_0)}.
\end{equation}
Therefore, all such finite oscillator excitations also satisfy the correct asymptotic conditions. The basis of nontrivial oscillator excitations in the shrinking limit are then formed in the same way as before, namely by scaling $\ell = \frac{qL}{\pi}$ with $q$ finite. The resulting state is written $\rho_{\{\hat{q}_1,\dotsc,\hat{q}_i,q_{i+1},\dotsc,q_{i+j}\}}(\tau_0,\sigma_0)$ and consists of $j$ particles and $i$ holes on top of the basic state $\rho(\tau_0,\sigma_0)$, where a hole of the corresponding momentum is indicated by a circumflex. Note that acting with all possible combinations of creation and annihilation operators automatically constructs states with any possible values of $\tau_0$ and $\sigma_0$ as well as states corresponding to arbitrary descendants of $\mathcal{O}_{k}(\tau_0,\sigma_0)$, thus generating all correlators with arbitrary insertions of local operators via the previous OPE argument.

\subsection{Exponential Operator Normalization}

For future calculations, it is useful to compute the expectation values $\langle\alpha_{\pm\ell}^m\rangle_{\rho_{\{N_{\ell}\}}}$ and $\langle\alpha_{-\ell}^m\alpha_{\ell}^{m'}\rangle_{\rho_{\{N_{\ell}\}}}$ for any non-negative integers $m$ and $m'$. 
For the former, we find
\begin{equation}
\langle\alpha_{\pm\ell}^m\rangle_{\rho_{\{N_{\ell}\}}} = \left(\frac{e^{-\frac{\pi^2\ell}{2L}}}{1-e^{-\frac{\pi^2\ell}{L}}}\right)^m\genfrac{\{}{\}}{0pt}{0}{(f_{\ell}^k)^m}{(f_{\ell}^{-k*})^m}\frac{L_{|N_{\ell}|}^{(m)}\left(-\frac{e^{-\frac{\pi^2\ell}{L}\theta(N_{\ell})}}{\ell(1-e^{-\frac{\pi^2\ell}{L}})}f_{\ell}^{-k*}f_{\ell}^k\right)}{L_{|N_{\ell}|}\left(-\frac{e^{-\frac{\pi^2\ell}{L}\theta(N_{\ell})}}{\ell(1-e^{-\frac{\pi^2\ell}{L}})}f_{\ell}^{-k*}f_{\ell}^k\right)}.
\end{equation}
Note that, in the shrinking limit, this expectation value factorizes as
\begin{equation}
\langle\alpha_{\pm\ell}^m\rangle_{\rho_{\{N_{\ell}\}}} \ \stackrel{L\rightarrow\infty}{\longrightarrow} \ \Big(\langle\alpha_{\pm\ell}\rangle_{\rho_{\{N_{\ell}\}}}\Big)^m.
\end{equation}
For the latter, define $m_+ \equiv \mathrm{max}(m,m')$ and $m_- \equiv \mathrm{min}(m,m')$. Then, the expectation value for $N_{\ell} \geqslant 0$ is
\begin{multline}
\langle\alpha_{-\ell}^m\alpha_{\ell}^{m'}\rangle_{\rho_{\{N_{\ell}\}}} %= \sum_{j=0}^{m_-}\frac{(-1)^j \ell^j m_+!m_-!}{j!(m_+-j)!(m_--j)!}\langle\alpha_{\ell}^{m'-j}\alpha_{-\ell}^{m-j}\rangle_{\rho_{\{N_{\ell}\}}}
= \frac{(-1)^{m_-}\ell^{m_-}m_+!m_-!}{N_{\ell}!(1-e^{-\frac{\pi^2\ell}{L}})^{m_+-m_-}}\genfrac{\{}{\}}{0pt}{0}{(e^{-\frac{\pi^2\ell}{2L}}f_{\ell}^{-k*})^{m_+-m_-}}{(e^{-\frac{\pi^2\ell}{2L}}f_{\ell}^k)^{m_+-m_-}}
\\ \times \sum_{j=0}^{m_-}\frac{(-1)^j(N_{\ell}+j)!}{j!(j+m_+-m_-)!(m_--j)!}\frac{1}{(1-e^{-\frac{\pi^2\ell}{L}})^{j}}\frac{L_{N_{\ell}+j}^{(m_+-m_-)}\left(-\frac{e^{-\frac{\pi^2\ell}{L}}}{\ell(1-e^{-\frac{\pi^2\ell}{L}})}f_{\ell}^{-k*}f_{\ell}^k\right)}{L_{N_{\ell}}\left(-\frac{e^{-\frac{\pi^2\ell}{L}}}{\ell(1-e^{-\frac{\pi^2\ell}{L}})}f_{\ell}^{-k*}f_{\ell}^k\right)},
\end{multline}
where the top line applies for $m \geqslant m'$ and the bottom line applies for $m < m'$, and for $N_{\ell} < 0$ this expectation value is
\begin{multline}
\langle\alpha_{-\ell}^m\alpha_{\ell}^{m'}\rangle_{\rho_{\{N_{\ell}\}}}
%\\ = \frac{1}{c(N_{\ell})\mathcal{N}_{\ell}}\sum_{N'_{\ell}=0}^{\infty}\frac{1}{\ell^{N'_{\ell}}N'_{\ell}!}\langle 0|\alpha_{\ell}^{N'_{\ell}}e^{\frac{\alpha_{\ell}}{\ell}f_{\ell}^{-k}(-\tau_0,\sigma_0)^*}\alpha_{-\ell}^{|N_{\ell}|}\alpha_{-\ell}^{m}\alpha_{\ell}^{m'}\alpha_{\ell}^{|N_{\ell}|}e^{-\frac{\pi^2}{L}\alpha_{-\ell}\alpha_{\ell}}e^{\frac{\alpha_{-\ell}}{\ell}f_{\ell}^k(\tau_0,\sigma_0)}\alpha_{-\ell}^{N'_{\ell}}|0\rangle
%\\ = (-1)^{|N_{\ell}|+m_-}\frac{\ell^{|N_{\ell}|+m_-}}{c(N_{\ell})\mathcal{N}_{\ell}}\genfrac{\{}{\}}{0pt}{0}{(e^{-\frac{\pi^2\ell}{2L}}f_{\ell}^{-k*})^{m_+-m_-}}{(e^{-\frac{\pi^2\ell}{2L}}f_{\ell}^k)^{m_+-m_-}}e^{\frac{e^{-\frac{\pi^2\ell}{L}}}{\ell}f_{\ell}^{-k*}f_{\ell}^k}\sum_{j=0}^{|N_{\ell}|+m_-}\frac{(-1)^j(|N_{\ell}|+m_+)!(|N_{\ell}|+m_-)!}{j!(j+m_+-m_-)!(|N_{\ell}|+m_--j)!}
%\\ \times \sum_{N'_{\ell}=0}^{\infty}\frac{(N'_{\ell}+j)!}{N'_{\ell}!}e^{-\frac{\pi^2\ell}{L}N'_{\ell}}L_{N'_{\ell}+j}^{(m_+-m_-)}\left(-\frac{e^{-\frac{\pi^2\ell}{L}}}{\ell}f_{\ell}^{-k*}f_{\ell}^k\right)
%\\ = \frac{(-1)^{|N_{\ell}|+m_-}\ell^{|N_{\ell}|+m_-}}{c(N_{\ell})(1-e^{-\frac{\pi^2\ell}{L}})^{m_+-m_-}}\genfrac{\{}{\}}{0pt}{0}{(e^{-\frac{\pi^2\ell}{2L}}f_{\ell}^{-k*})^{m_+-m_-}}{(e^{-\frac{\pi^2\ell}{2L}}f_{\ell}^k)^{m_+-m_-}}\sum_{j=0}^{|N_{\ell}|+m_-}\frac{(-1)^j(|N_{\ell}|+m_+)!(|N_{\ell}|+m_-)!}{(j+m_+-m_-)!(|N_{\ell}|+m_--j)!}
%\\ \times \frac{1}{(1-e^{-\frac{\pi^2\ell}{L}})^{j}}L_{j}^{(m_+-m_-)}\left(-\frac{e^{-\frac{\pi^2\ell}{L}}}{\ell(1-e^{-\frac{\pi^2\ell}{L}})}f_{\ell}^{-k*}f_{\ell}^k\right)
= \frac{\ell^{m_-}(|N_{\ell}|+m_-)!}{|N_{\ell}|!(1-e^{-\frac{\pi^2\ell}{L}})^{m_+-m_-}}\genfrac{\{}{\}}{0pt}{0}{(e^{-\frac{\pi^2\ell}{2L}}f_{\ell}^{-k*})^{m_+-m_-}}{(e^{-\frac{\pi^2\ell}{2L}}f_{\ell}^k)^{m_+-m_-}}
\\ \times \left(\frac{e^{-\frac{\pi^2\ell}{L}}}{1-e^{-\frac{\pi^2\ell}{L}}}\right)^{m_-} \frac{L_{|N_{\ell}|+m_-}^{(m_+-m_-)}\left(-\frac{1}{\ell(1-e^{-\frac{\pi^2\ell}{L}})}f_{\ell}^{-k*}f_{\ell}^k\right)}{L_{|N_{\ell}|}\left(-\frac{1}{\ell(1-e^{-\frac{\pi^2\ell}{L}})}f_{\ell}^{-k*}f_{\ell}^k\right)}.
\end{multline}
We now obtain the normalization of the exponential operator in the generating state \eqref{osc state general}. Setting $N_{\ell} = 0$ above, the expectation value of $\alpha_{-\ell}^m\alpha_{\ell}^{m'}$ is
\begin{equation}
\langle\alpha_{-\ell}^m\alpha_{\ell}^{m'}\rangle_{\rho} % = \ell^{m_-}m_-!\genfrac{\{}{\}}{0pt}{0}{(\hspace{-1pt}f_{\ell}^{-k*}\hspace{-1pt})^{m_+\hspace{-1pt}-m_-}}{(f_{\ell}^k)^{m_+-m_-}}\frac{(\hspace{-1pt}e^{-\frac{\pi^2\ell}{2L}}\hspace{-1pt})^{m_+\hspace{-1pt}+m_-}}{(1\hspace{-1pt}-\hspace{-1pt}e^{-\frac{\pi^2\ell}{L}})^{m_+}} L_{m_-}^{(m_+\hspace{-1pt}-m_-)}\hspace{-4pt}\left(\hspace{-3pt}-\frac{1}{\ell(1\hspace{-2pt}-\hspace{-2pt}e^{-\frac{\pi^2\ell}{L}})}f_{\ell}^{-k*}f_{\ell}^k\hspace{-3pt}\right)
= \ell^{m'}m'!(f_{\ell}^{-k*})^{m-m'}\frac{e^{-\frac{\pi^2\ell}{2L}(m+m')}}{(1-e^{-\frac{\pi^2\ell}{L}})^m}L_{m'}^{(m-m')}\left(-\frac{f_{\ell}^{-k*}f_{\ell}^k}{\ell(1-e^{-\frac{\pi^2\ell}{L}})}\right),
\end{equation}
which works for both $m\geqslant m'$ and $m < m'$. We are evaluating the expectation value of $\mathcal{O}_{k''}(\tau_2,\sigma_2)\mathcal{O}_{k'}(\tau_1,\sigma_1)$ in the state $\rho(\tau_0,\sigma_0)$. The zero-mode part of this expectation value provides the familiar $\delta(k'+k'')$ for the noncompact case and $\delta_{n'+n'',0}\delta_{w'+w'',0}$ for the compact case. Letting $f_{j,\ell}^{k'}$ be short for $f_{\ell}^{k'}(\tau_j,\sigma_j)$, for instance, the pure oscillator part of the expectation value is then the product over $\ell$ of
\begin{multline}
\left\langle e^{\frac{\alpha_{-\ell}}{\ell}f_{2,\ell}^{-k'}}e^{\frac{\alpha_{\ell}}{\ell}f_{2,\ell}^{k'*}}e^{\frac{\alpha_{-\ell}}{\ell}f_{1,\ell}^{k'}}e^{\frac{\alpha_{\ell}}{\ell}f_{1,\ell}^{-k'*}}\right\rangle_{\rho^{\text{osc}}_{\ell}} = e^{\frac{1}{\ell}f_{2,\ell}^{k'*}f_{1,\ell}^{k'}}\left\langle e^{\frac{\alpha_{-\ell}}{\ell}(f_{1,\ell}^{k'}+f_{2,\ell}^{-k'})}e^{\frac{\alpha_{\ell}}{\ell}(f_{1,\ell}^{-k'*}+f_{2,\ell}^{k'*})}\right\rangle_{\rho^{\text{osc}}_{\ell}}
\\ = e^{\frac{1}{\ell}f_{2,\ell}^{k'*}f_{1,\ell}^{k'}}\hspace{-2pt}\exp\hspace{-4pt}\left[\hspace{-2pt}\frac{e^{-\frac{\pi^2\ell}{2L}}\hspace{-2pt}[f_{0,\ell}^{-k*}(f_{1,\ell}^{k'}\hspace{-3pt}+\hspace{-3pt}f_{2,\ell}^{-k'}\hspace{-1pt})\hspace{-3pt}+\hspace{-3pt}(f_{1,\ell}^{-k'*}\hspace{-4pt}+\hspace{-3pt}f_{2,\ell}^{k'*})\hspace{-1pt}f_{0,\ell}^k]\hspace{-3pt}+\hspace{-3pt}e^{-\frac{\pi^2\ell}{L}}\hspace{-2pt}(f_{1,\ell}^{-k'*}\hspace{-4pt}+\hspace{-3pt}f_{2,\ell}^{k'*})(f_{1,\ell}^{k'}\hspace{-3pt}+\hspace{-3pt}f_{2,\ell}^{-k'}\hspace{-1pt})}{\ell(1-e^{-\frac{\pi^2\ell}{L}})}\hspace{-2pt}\right]\hspace{-3pt},
\end{multline}
where the functions $f_{\ell}^{k'}(\tau_j,\sigma_j)$ are given explicitly in \eqref{f noncompact N}, \eqref{f noncompact D} and \eqref{f compact}. For the first exponential $e^{\frac{1}{\ell}f_{2,\ell}^{k'*}f_{1,\ell}^{k'}}$, the infinite product over $\ell$ is easily performed in the three cases `N' (noncompact Neumann-class), `D' (noncompact Dirichlet-class) and `c' (compact) via the exact summations
\begin{align}
\sum_{\ell=1}^{\infty}\frac{1}{\ell}f_{2,\ell}^{k'*}f_{1,\ell}^{k'} % & \stackrel{\text{N}}{=} 2k'^2\sum_{\ell=1}^{\infty}\frac{1}{\ell}e^{\frac{\pi\ell\tau_{12}}{2L}}\cos\left(\frac{\pi\ell(\sigma_1+L)}{2L}\right)\cos\left(\frac{\pi\ell(\sigma_2+L)}{2L}\right)
& \ \stackrel{\text{N}}{=} \ \frac{k'^2}{2}\ln\left[\frac{(\frac{L}{\pi})^2 e^{i\varphi_{12}}}{(\sigma_{12}^2 + |\tau_{12}|^2)^2 - (2\sigma_{12}\mathrm{Im}\tau_{12})^2}\right]
\\ \sum_{\ell=1}^{\infty}\frac{1}{\ell}f_{2,\ell}^{k'*}f_{1,\ell}^{k'} % & \stackrel{\text{D}}{=} 2k'^2\sum_{\ell=1}^{\infty}\frac{1}{\ell}e^{\frac{\pi\ell\tau_{12}}{2L}}\sin\left(\frac{\pi\ell(\sigma_1+L)}{2L}\right)\sin\left(\frac{\pi\ell(\sigma_2+L)}{2L}\right)
& \ \stackrel{\text{D}}{=} \ \frac{k'^2}{2}\ln\left[\frac{(\frac{4L}{\pi})^2 e^{i\varphi_{12}}}{(\sigma_{12}^2 + |\tau_{12}|^2)^2 - (2\sigma_{12}\mathrm{Im}\tau_{12})^2}\right]
\\ \sum_{\ell=1}^{\infty}\frac{1}{\ell}f_{2,\ell}^{k'*}f_{1,\ell}^{k'} % & \stackrel{\text{c}}{=} 2\sum_{\ell=1}^{\infty}\frac{1}{\ell}e^{\frac{\pi\ell\tau_{12}}{2L}}\left[\frac{n'}{R}\sin\left(\frac{\pi\ell(\sigma_1+L)}{2L}\right) + iw'R\cos\left(\frac{\pi\ell(\sigma_1+L)}{2L}\right)\right]\left[\frac{n'}{R}\sin\left(\frac{\pi\ell(\sigma_2+L)}{2L}\right) - iw'R\cos\left(\frac{\pi\ell(\sigma_2+L)}{2L}\right)\right]
\notag & \ \stackrel{\text{c}}{=} \ \frac{1}{2}\left(\frac{n'^2}{R^2} + w'^2 R^2\right)\ln\left[\frac{(\frac{L}{\pi})^2 e^{i\varphi_{12}}}{(\sigma_{12}^2+|\tau_{12}|^2)^2-(2\sigma_{12}\mathrm{Im}\s\tau_{12})^2}\right] + \frac{n'^2}{R^2}\ln 4
\\ & \hspace{20pt} - n'w'\left[\ln\left(\frac{\sigma_{12}^2+|\tau_{12}|^2-2\sigma_{12}\mathrm{Im}\s\tau_{12}}{\sigma_{12}^2+|\tau_{12}|^2+2\sigma_{12}\mathrm{Im}\s\tau_{12}}\right) + i\phi_{12}\right],
\end{align}
where we have defined the angles
\begin{align}
\varphi_{12} & \equiv \tan^{-1}\hspace{-3pt}\left(\hspace{-2pt}\frac{\mathrm{Re}\s\tau_{12}}{\sigma_{12}\hspace{-2pt}+\hspace{-2pt}\mathrm{Im}\s\tau_{12}}\hspace{-2pt}\right) - \tan^{-1}\hspace{-3pt}\left(\hspace{-2pt}\frac{\mathrm{Re}\s\tau_{12}}{\sigma_{12}\hspace{-2pt}-\hspace{-2pt}\mathrm{Im}\s\tau_{12}}\hspace{-2pt}\right) + \pi\theta(|\mathrm{Im}\s\tau_{12}|\hspace{-2pt}-\hspace{-2pt}|\sigma_{12}|)\mathrm{sgn}(\mathrm{Im}\s\tau_{12})
\\ \phi_{12} & \equiv \tan^{-1}\hspace{-3pt}\left(\hspace{-2pt}\frac{\mathrm{Re}\s\tau_{12}}{\sigma_{12}\hspace{-2pt}+\hspace{-2pt}\mathrm{Im}\s\tau_{12}}\hspace{-2pt}\right) + \tan^{-1}\hspace{-3pt}\left(\hspace{-2pt}\frac{\mathrm{Re}\s\tau_{12}}{\sigma_{12}\hspace{-2pt}-\hspace{-2pt}\mathrm{Im}\s\tau_{12}}\hspace{-2pt}\right) + \pi\theta(|\sigma_{12}|\hspace{-2pt}-\hspace{-2pt}|\mathrm{Im}\s\tau_{12}|)\mathrm{sgn}(\sigma_{12}).
\end{align}
The infinite product over the lengthier exponential seems much more daunting. Fortunately, as before, we care only about the shrinking limit and so we may perform an asymptotic expansion in $\frac{1}{L}$. The leading contributions in this expansion are then due to replacing $\frac{e^{-\frac{\pi^2\ell}{aL}}}{1-e^{-\frac{\pi^2\ell}{L}}}$ in the exponential with $\frac{L}{\pi^2\ell}$. The resulting sums are then easy to perform. It is readily verified that in all three cases the shrinking limits of these sums are
\begin{equation}
L\sum_{\ell=1}^{\infty}\frac{(f_{1,\ell}^{-k'*}+f_{2,\ell}^{k'*})(f_{1,\ell}^{k'}+f_{2,\ell}^{-k'})}{\ell^2} \ \stackrel{L\rightarrow\infty}{\longrightarrow} \ 0
\end{equation}
and
\begin{equation}
L\sum_{\ell=1}^{\infty}\frac{f_{0,\ell}^{-k*}(f_{1,\ell}^{k'}+f_{2,\ell}^{-k'})+(f_{1,\ell}^{-k'*}+f_{2,\ell}^{k'*})f_{0,\ell}^k}{\ell^2} \ \stackrel{L\rightarrow\infty}{\longrightarrow} \ 0.
\end{equation}
The vanishing of the leading terms simply arises because each such contribution is independent of all positions, and the terms in $f_{1,\ell}^{k'}+f_{2,\ell}^{-k'}$ or $f_{1,\ell}^{-k'*}+f_{2,\ell}^{k'*}$ have opposite signs. The vanishing of the subleading terms is then because the nontrivial position dependence in all the sums is quadratic as a result of the linear pieces canceling. Therefore, including the zero-mode contributions and the numerical prefactors, the exponential operator two-point function for the Neumann-class noncompact boson is
\begin{multline}
\left\langle\mathcal{O}_{k''}(\tau_2,\sigma_2)\mathcal{O}_{k'}(\tau_1,\sigma_1)\right\rangle_{\rho(\tau_0,\sigma_0)}
\\ \stackrel{L\rightarrow\infty}{\longrightarrow} 2\pi\delta(k'+k'')e^{\frac{(k_1-k_2)k'}{2}\sigma_{12}}\left[\frac{e^{i\varphi_{12}}}{(\sigma_{12}^2+|\tau_{12}|^2)^2-(2\sigma_{12}\mathrm{Im}\s\tau_{12})^2}\right]^{\frac{k'^2}{2}},
\end{multline}
for the Dirichlet-class noncompact boson is
\begin{multline}
\left\langle\mathcal{O}_{k''}(\tau_2,\sigma_2)\mathcal{O}_{k'}(\tau_1,\sigma_1)\right\rangle_{\rho(\tau_0,\sigma_0)}
\\ \stackrel{L\rightarrow\infty}{\longrightarrow} 2\pi\delta(k'+k'')e^{\frac{(k_1-k_2)k'}{2}\sigma_{12}}\left[\frac{e^{i\varphi_{12}}}{(\sigma_{12}^2+|\tau_{12}|^2)^2-(2\sigma_{12}\mathrm{Im}\s\tau_{12})^2}\right]^{\frac{k'^2}{2}}
\end{multline}
and finally for the compact boson is
\begin{multline}
\left\langle\mathcal{O}_{n'',w''}(\tau_2,\sigma_2)\mathcal{O}_{n',w'}(\tau_1,\sigma_1)\right\rangle_{\rho(\tau_0,\sigma_0)}
\\ \stackrel{L\rightarrow\infty}{\longrightarrow} 2\pi \delta_{n'+n'',0}\delta_{w'+w'',0}e^{[\frac{(n_1-n_2)n'}{2R^2}+\frac{(w_1-w_2)w'R^2}{2}]\sigma_{12}}e^{\frac{i[(n_1-n_2)w'+n'(w_1-w_2)]}{2}\tau_{12}}
\\ \times \left[\frac{e^{i\varphi_{12}}}{(\sigma_{12}^2+|\tau_{12}|^2)^2-(2\sigma_{12}\mathrm{Im}\s\tau_{12})^2}\right]^{\frac{1}{2}(\frac{n'^2}{R^2}+w'^2 R^2)}\left(-\frac{\sigma_{12}^2+|\tau_{12}|^2+2\sigma_{12}\mathrm{Im}\s\tau_{12}}{\sigma_{12}^2+|\tau_{12}|^2-2\sigma_{12}\mathrm{Im}\s\tau_{12}}e^{-i\phi_{12}}\right)^{n'w'}.
\end{multline}
These results are independent of $(\tau_0,\sigma_0)$ and are manifestly finite, proving that the exponentials are indeed (Dirac-)normalizable operators in the infinitely excited Minkowski CFT states generated by $\rho(\tau_0,\sigma_0)$. The elimination or quantization of the slope-mode sector then follows from the same reasoning as in Section \ref{zero net charge}.

\subsection{Equivalence of Neumann and Dirichlet}

We are finally in the position to provide the remaining detail proving that the Neumann-class and Dirichlet-class regulators are equivalent in angular quantization. As in Section \ref{zero net charge}, we simply need to provide the appropriate isometry between the states and operators, where the only difference lies in the zero-momentum-mode sector. To wit, due to the underlying shift symmetry of the theory, we must be careful to ensure that the constant-mode $x$ is defined in the same way in the two approaches. 

Since $x$ appears explicitly in the Hamiltonian as the term $-\frac{i(k_1+k_2)}{2\pi}x$, constant shifts in $x$ change the energy. For this subsection, denote the constant-modes in the Neumann-class and Dirichlet-class approaches as $x_{\text{N}}$ and $x_{\text{D}}$, respectively. We shall deduce the map between $x_{\text{N}}$ and $x_{\text{D}}$ by comparing the expectation values of the Hamiltonian in both cases. Using the previous result for $\langle \alpha^m_{-\ell}\alpha_{\ell}^{m'}\rangle_{\rho}$ with $m = m' = 1$, the expectation value of the oscillator part of the Hamiltonian is
\begin{align}
\frac{\pi}{2L}\sum_{\ell=1}^{\infty}\langle\alpha_{-\ell}\alpha_{\ell}\rangle_{\rho} & = \frac{\pi}{2L}\sum_{\ell=1}^{\infty}\frac{\ell e^{-\frac{\pi^2\ell}{L}}}{1-e^{-\frac{\pi^2\ell}{L}}}\left[1+\frac{f_{\ell}^{-k*}f_{\ell}^k}{\ell(1-e^{-\frac{\pi^2\ell}{L}})}\right]
\\ & \stackrel{L\rightarrow\infty}{\longrightarrow} \frac{e^{-\frac{\pi^2}{L}}}{2\pi(1-e^{-\frac{\pi^2}{L}})} + \frac{L}{2\pi^3}\sum_{\ell=1}^{\infty}\frac{1}{\ell^2}f_{\ell}^{-k*}f_{\ell}^k.
\end{align}
The remaining sums for the noncompact Neumann-class and Dirichlet-class cases are
\begin{align}
\sum_{\ell=1}^{\infty}\frac{1}{\ell^2}f_{\ell}^{-k*}f_{\ell}^k & \stackrel{\text{N}}{=} -\frac{\pi^2 k^2}{4}\left(\frac{1}{3}+\frac{\sigma_0^2}{L^2}\right)
\\ \sum_{\ell=1}^{\infty}\frac{1}{\ell^2}f_{\ell}^{-k*}f_{\ell}^k & \stackrel{\text{D}}{=} -\frac{\pi^2 k^2}{4}\left(1-\frac{\sigma_0^2}{L^2}\right),
%\\ \sum_{\ell=1}^{\infty}\frac{1}{\ell^2}f_{\ell}^{-k*}f_{\ell}^k & \stackrel{\text{c}}{=} -\frac{\pi^2}{6}\left(\frac{n^2}{R^2}+w^2 R^2\right) - \frac{\pi^2}{4}\left(\frac{n^2}{R^2} - w^2 R^2\right)\left(\frac{1}{3} - \frac{\sigma_0^2}{L^2}\right),
\end{align}
where we recall that $k = -(k_1+k_2)$. Thus, the expectation values of the full Hamiltonians are
\begin{align}
\langle H_{\text{R}}^{\text{N}}(L)\rangle_{\rho} & \ \stackrel{L\rightarrow\infty}{\longrightarrow} \ - \frac{i(k_1+k_2)}{2\pi}\langle x_{\text{N}}\rangle + \frac{(k_1-k_2)^2}{8\pi}L + \frac{(k_1+k_2)^2}{24\pi}L + \frac{L}{2\pi^3} - \frac{1}{4\pi} 
\\ \langle H_{\text{R}}^{\text{D}}(L)\rangle_{\rho} & \ \stackrel{L\rightarrow\infty}{\longrightarrow} \ -\frac{i(k_1+k_2)}{2\pi}\langle x_{\text{D}}\rangle + \frac{(k_1-k_2)^2}{8\pi}L - \frac{(k_1+k_2)^2}{8\pi}L + \frac{L}{2\pi^3} - \frac{1}{4\pi}.
% \\ \notag \text{compact:} \ \ \langle H_{\text{R}}\rangle & \ \stackrel{L\rightarrow\infty}{\longrightarrow} \ -\frac{i(n_1+n_2)}{2\pi R}\langle x\rangle + \frac{i(w_1-w_2)R}{2}\langle p\rangle - \frac{i(w_1+w_2)R}{2}\langle p_{\text{s}}\rangle
% \\ & \hspace{20pt} + \frac{(n_1-n_2)^2}{8\pi R^2}L - \frac{(n_1+n_2)^2}{8\pi R^2}L + \frac{(w_1-w_2)^2 R^2}{8\pi}L + \frac{L}{2\pi^3} - \frac{1}{4\pi} 
\end{align}
From these expressions, we see that the appropriate map between the zero-momentum-modes is
\begin{equation}\label{zero-mode isometry}
x_{\text{N}} \longmapsto x_{\text{D}} - \frac{i(k_1+k_2)}{3}L.
\end{equation}
Under this map, we immediately see from \eqref{noncompact N full} and \eqref{noncompact D full} that
\begin{align}
\notag \left\langle X(\tau,\sigma)\right\rangle_{\rho(\tau_0,\sigma_0)}^{\text{N}} & \longmapsto \left\langle X(\tau,\sigma)\right\rangle_{\rho(\tau_0,\sigma_0)}^{\text{D}},
\end{align}
showing that the free boson expectation values now agree exactly including the constant term. Under the isometry defined by \eqref{zero-mode isometry} and $a_{\text{N}}(q) \mapsto \mathrm{sgn}(q)a_{\text{D}}(q)$, we have shown that the actions of all states on the operator algebra manifestly agree in the two cases. Therefore, the Neumann-class and Dirichlet-class regulators are exactly equivalent in the shrinking limit of angular quantization and hence define identical Minkowski CFTs.

\section{Prelude to String Theory}\label{prelude}

In Section \ref{nonzero net charge}, we computed the expectation value $\langle X(\tau,\sigma)\rangle$ of the free boson in the base state $\rho$ in angular quantization for the noncompact case and for the compact case in \eqref{noncompact N full} and \eqref{compact full}, respectively, from which it trivially followed that the Euclidean asymptotic conditions are satisfied. Those computations were performed for arbitrary complex $\tau$, and hence they may be applied to the physically relevant Lorentzian theory for which $\tau = it$ with $t \in \mathds{R}$. In the context of string theory, $\langle X^{\mu}(t,\sigma)\rangle$ represents the embedding of a string worldsheet in target spacetime. These worldsheets belong to a new class of fundamental string which is defined from angular quantization. These strings are characterized by having asymptotic boundaries and are hence neither open nor closed, which have finite boundaries and no boundaries, respectively. We dub this new class ``stretched strings'', and in this section we provide snapshot descriptions of the different types of stretched strings. We focus on the basic explicit states $\rho(\tau_0,\sigma_0)$ which will describe classical string worldsheets of single stretched strings in isolation. The full quantum-mechanical treatment of stretched strings will later be presented in \cite{future}.

From \eqref{noncompact N full} and \eqref{compact full}, the Lorentzian noncompact free boson expectation value in the state $\rho(t_0,\sigma_0)$ with asymptotic charges $k_1$ and $k_2$ is
\begin{multline}\label{noncompact Lorentzian expectation}
\langle X(t,\sigma)\rangle_{\rho(t_0,\sigma_0)} = \langle x\rangle - \frac{i(k_1-k_2)}{2}\sigma + \frac{i(k_1+k_2)}{2}\theta\Big(|\sigma-\sigma_0|-|t-t_0|\Big)|\sigma-\sigma_0|
\\ + \frac{i(k_1+k_2)}{2}\theta\Big(|t-t_0| - |\sigma-\sigma_0|\Big)|t-t_0|,
\end{multline}
and the Lorentzian compact free boson expectation value in the state $\rho(t_0,\sigma_0)$ is
\begin{multline}\label{compact Lorentzian expectation}
\langle X(t,\sigma)\rangle_{\rho(t_0,\sigma_0)} = \langle x\rangle - \frac{i(n_1-n_2)}{2R}\sigma + \frac{i(w_1-w_2)R}{2}t
\\ + \theta\Big(|\sigma-\sigma_0|-|t-t_0|\Big)\left[\frac{i(n_1+n_2)}{2R}|\sigma-\sigma_0| - \frac{i(w_1+w_2)R}{2}\mathrm{sgn}(\sigma-\sigma_0)(t-t_0)\right]
\\ + \theta\Big(|t-t_0|-|\sigma-\sigma_0|\Big)\left[\frac{i(n_1+n_2)}{2R}|t-t_0| - \frac{i(w_1+w_2)R}{2}\mathrm{sgn}(t-t_0)(\sigma-\sigma_0)\right].
\end{multline}
In angular quantization, the free boson is defined on a complexified contour essentially defined by its mode expansion. Here, we interpret $X$ as a target space coordinate and plot constant-time snapshots in the coordinate $iX$. In the actual string theory applications of interest, the field $X$ is either a time-like boson or a linear dilaton (more generally, Liouville field), for which the corresponding string profiles will indeed be real-valued. Furthermore, we shall take $iX$ as noncompact, as even the compact boson is generally noncompact when continued to the imaginary direction. As such, we need only plot \eqref{compact Lorentzian expectation}, as it subsumes \eqref{noncompact Lorentzian expectation}, and we set $\langle x\rangle = 0$ since we already understand the shift symmetry. We shall plot the snapshots as a function of the worldsheet time; it is understood that in any slicing of an actual string worldsheet, a gauge choice is made to fix the relation between $t$ and $X^0$. Finally, we consider the pure momentum and pure winding cases separately in what follows. The most general stretched string worldsheet exhibits a combination of the two effects.

\subsection{Pure Momentum Charges}

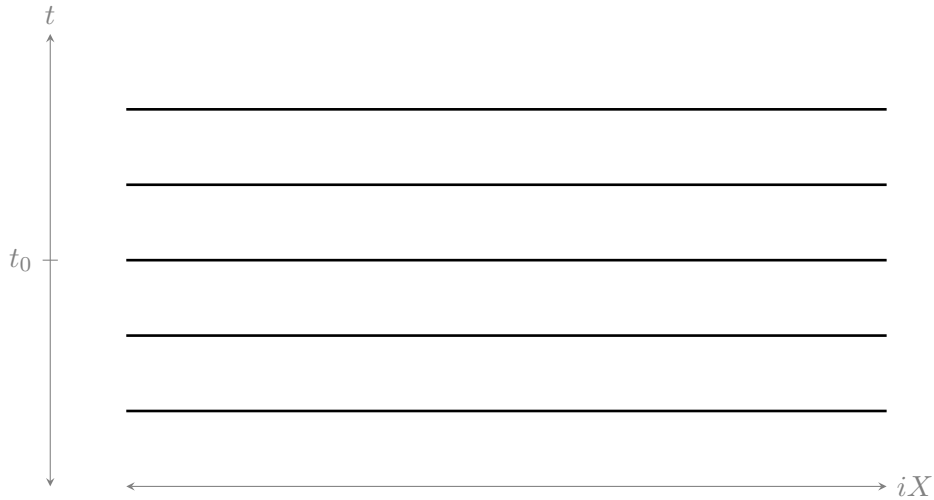
\begin{figure}
\centering
\begin{tikzpicture}
\draw[gray,stealth-stealth] (0,0) -- (0,6) node[above]{$t$};
\draw[gray] (0.1,3) -- (-0.1,3) node[left]{$t_0$};
\draw[gray,stealth-stealth] (1,0) -- (11,0) node[right]{$iX$};
\draw[line width=1pt] (1,3) -- (11,3);
\draw[line width=1pt] (1,2) -- (11,2);
\draw[line width=1pt] (1,1) -- (11,1);
\draw[line width=1pt] (1,4) -- (11,4);
\draw[line width=1pt] (1,5) -- (11,5);
\end{tikzpicture}
\caption{The pure-momentum stretched string in the generic case $n_1 \neq n_2$. All snapshots are identical, consisting of a space-filling string.}
\label{pure momentum generic}
\end{figure}

First consider $w_1 = w_2 = 0$ so that only asymptotic momentum charges are present. There are two distinct cases. The generic case is $n_1 \neq n_2$, for which the linear term $-\frac{i(n_1-n_2)}{2R}\sigma$ in \eqref{compact Lorentzian expectation} automatically creates a stretched string from one end of space to the other, as shown in Figure \ref{pure momentum generic}.
Of course, which value of the worldsheet coordinate $\sigma$ gives rise to which value of $iX$ does change over time (as long as $n_1 \neq -n_2$), but the point is that all values of $iX$ are always reached. Since a fundamental string has no width, the coordinate $\sigma$ itself is not physical. 

\begin{figure}
\centering
\begin{tikzpicture}
\draw[gray,stealth-stealth] (0,0) -- (0,6) node[above]{$t$};
\draw[gray] (0.1,3) -- (-0.1,3) node[left]{$t_0$};
\draw[gray,stealth-stealth] (1,0) -- (11,0) node[right]{$iX$};
\foreach \y in {-2,...,2}{
\draw[line width=1pt] (1,{2.95+\y}) -- ({8-3*abs(\y},{2.95+\y}) to[out=0,in=0] ({8-3*abs(\y},{3.05+\y}) -- (1,{3.05+\y});
}
\draw[-stealth] (9,3.5) node[above,scale=0.75]{$\sigma=\sigma_0$} to[out=-90,in=0] (8.2,3);
\draw[decorate,decoration={brace,mirror}] (5,1.7) -- node[midway,below,scale=0.9,yshift=-0.05cm]{$\frac{|n|}{R}\Delta t$} (8,1.7);
\end{tikzpicture}
\caption{The pure-momentum stretched string in the special case $n_1 = n_2$. Each snapshot is a half-space-filling string doubled back on itself. Over time, the folding point ($\sigma = \sigma_0$) evenly travels into space, reaches a maximum (at $t = t_0$) and then recedes back to infinity.}
\label{pure momentum special}
\end{figure}
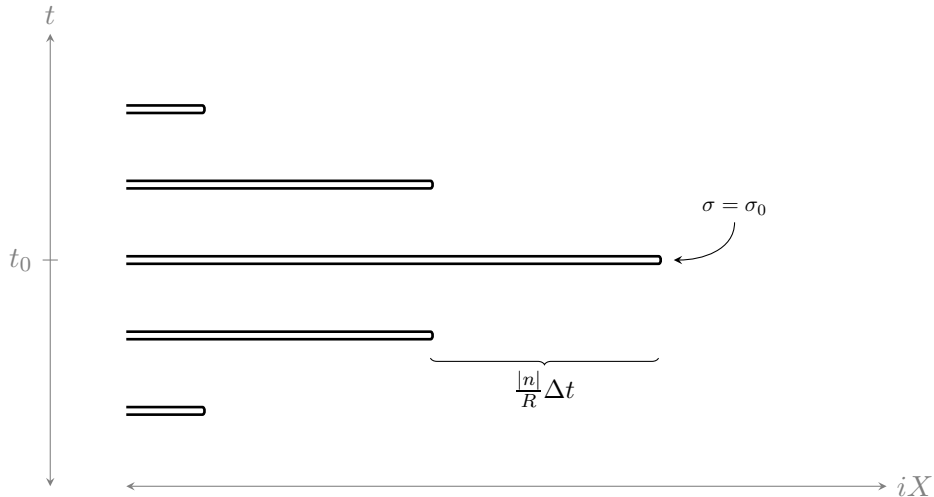

The far more interesting scenario is the special case $n_1 = n_2$ for which the linear term $-\frac{i(n_1-n_2)}{2R}\sigma$ is absent. Snapshots of this stretched string are plotted in Figure \ref{pure momentum special}.
Unlike the generic case, for $n_1 = n_2$ both asymptotic ends of the stretched string emerge from the same spatial infinity. While the string itself is infinitely long, there is a distinguished point where the string folds and doubles back on itself. This folding point travels into the bulk of space at a constant speed, reaches a maximum and then recedes again at the same speed, which is controlled by the asymptotic momentum $\frac{|n|}{R}$. This heuristic picture agrees with that of the ``long strings'' in the $c = 1$ string theory discovered by Maldacena in \cite{Maldacena05}. In that work, however, the ``long string'' was obtained less directly as a scaling limit of an infinitely-energetic open string on a receding FZZT-brane. For more appearances of long and folded strings, see for instance \cite{Balthazar18} and \cite{Giveon20}. We propose that the proper string-theoretic construction of such long string states is via the angular quantization procedure which defines stretched strings more generally. We shall have more to say about this connection when we present the full BRST quantization in \cite{future}.

\subsection{Pure Winding Charges}

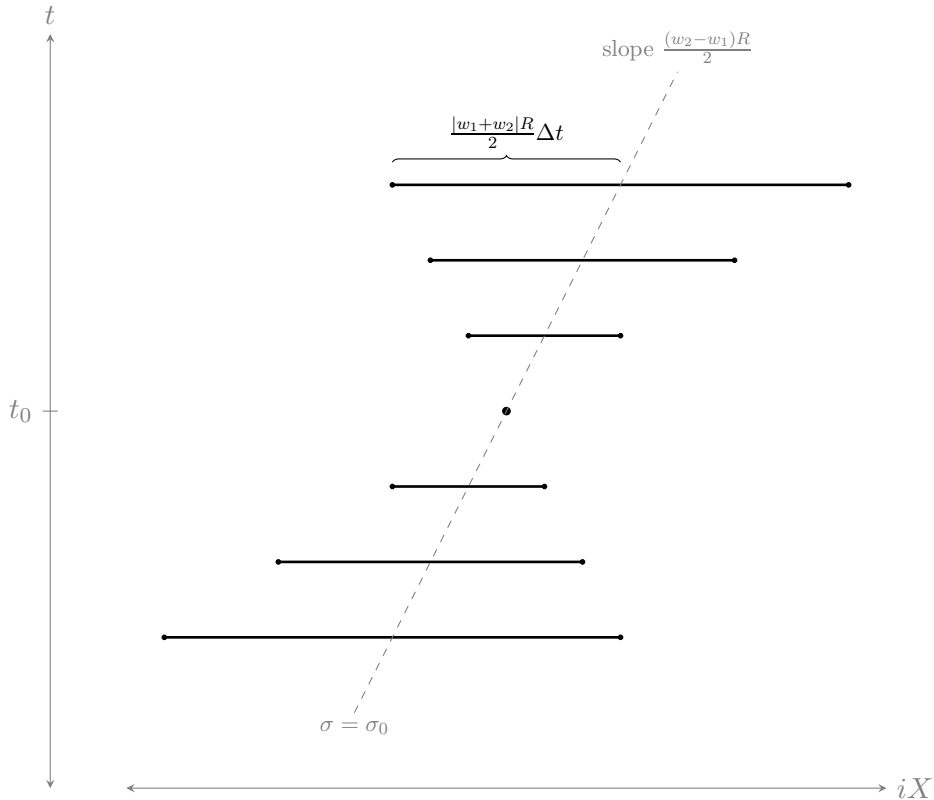
\begin{figure}
\centering
\begin{tikzpicture}
\draw[gray,stealth-stealth] (0,-1) -- (0,9) node[above]{$t$};
\draw[gray] (0.1,4) -- (-0.1,4) node[left]{$t_0$};
\draw[gray,stealth-stealth] (1,-1) -- (11,-1) node[right]{$iX$};
\draw[fill] (6,4) circle (0.05);
\foreach \y in {-3,...,3}{
\draw[fill] ({6+0.5*\y - 1*\y},{4+\y}) circle (0.03);
\draw[fill] ({6+0.5*\y + 1*\y},{4+\y}) circle (0.03);
\draw[line width=1pt] ({6+0.5*\y - 1*\y},{4+\y}) -- ({6+0.5*\y + 1*\y},{4+\y});
}
\draw[gray,dashed] ({6-2},{4-4}) node[below,scale=0.8]{$\sigma=\sigma_0$} -- ({6+2.25},{4+4.5}) node[above,scale=0.8]{slope $\frac{(w_2-w_1)R}{2}$};
\draw[decorate,decoration=brace] (4.5,7.3) -- node[midway,above,scale=0.8,yshift=0.1cm]{$\frac{|w_1+w_2|R}{2}\Delta t$} (7.5,7.3);
\end{tikzpicture}
\caption{The pure-winding stretched string. Such a string has finite length at all times, which contracts to a point (at $t = t_0$) before expanding again all at a constant rate.}
\label{pure winding} 
\end{figure}

Now we consider $n_1 = n_2 = 0$ so that only asymptotic winding charges are present. From \eqref{compact Lorentzian expectation}, we note that the point $\sigma = \sigma_0$ simply moves at a constant speed set by $w_1 - w_2$. At $t = t_0$, the entire string is at the same location as $\sigma = \sigma_0$. As time elapses, all points with $|\sigma-\sigma_0| > |t-t_0|$ are at the same distance from $\sigma = \sigma_0$, set by $w_1 + w_2$. The remaining points with $|\sigma - \sigma_0| < |t-t_0|$ simply fill in the interval between $\sigma = \sigma_0$ and this maximum distance. The resulting stretched string is shown in Figure \ref{pure winding}.
Since in applications the compact boson will be that of Euclidean time, the result of Figure \ref{pure winding} is to be interpreted that there exists a one-to-one map between $t$ and $X^0$ (for instance given by the line $\sigma = \sigma_0$) so that there indeed exists a well-defined global gauge relating the worldsheet and target time coordinates. To wit, in the usual case of interest the angular quantization endpoint operators will have opposite Euclidean time winding, in which Figure \ref{pure winding} already collapses to a single line.

\section{Summary}

In this work, we detailed the states of the 2d free boson on the infinite line with specified asymptotic charges at spatial infinity, corresponding to a superselection sector in the full Minkowski CFT. These states are constructed by taking the shrinking limit of angular quantization with exponential endpoint operators, which in particular retains only those states obeying the appropriate asymptotic conditions. In the special case that the net asymptotic charge vanishes, we found a simple vector space structure of pure states inherited from the finite-regulator Fock space construction. In the general case that the net asymptotic charge does not vanish, we do not find a basis of pure states and instead must work with mixed states originating from the Euclidean cylinder having the correct periodicity. While working with the purely algebraic definition of states may be more cumbersome, we found a similar structure as the previous case --- a one-parameter continuous zero-mode (noncompact theory) or a two-parameter discrete zero-mode (compact theory) combined with an oscillator sector which is generated from a basic state by acting with all combinations of creation and annihilation operators. One novelty of this construction is that the resulting ``Fock space'' is bi-directional. We also showed that the Neumann-class and Dirichlet-class regulators for angular quantization, which are seemingly quite different in the finite-regulator theory, are in fact entirely equivalent after the shrinking limit is taken.

As a basic motivating example, we also briefly discussed the ramifications of defining CFTs in Minkowski space via angular quantization on string theory. This leads to a new type of fundamental string which we called ``stretched strings'', which will be the purview of a future paper. We provided snapshots of the stretched string worldsheets that are obtained from a free boson sector of the worldsheet CFT. While the general construction of strings defined by angular quantization is new, some examples of stretched strings have been discussed in disguise over the years. One of the most famous such examples is Maldacena's ``long strings'' in $c=1$ string theory, and one special case of the stretched strings heuristically matches this example. This briefest of connections is meant as an appetizer for the next paper in the installment, in which we plan to detail the angular quantization of ``long strings'' in the actual framework of string theory.

\section*{Acknowledgments}

This work is supported in part by DOE grant DE-SC0007870.

\appendix
\section{Identities}\label{appendix}

The following identities are useful for the calculations involved in this paper.
\begin{align}
\alpha_{-\ell}^{m'} \alpha_{\ell}^m & = \sum_{j=0}^m\frac{(-1)^j \ell^j m'!m!}{j!(m'-j)!(m-j)!}\alpha_{\ell}^{m-j}\alpha_{-\ell}^{m'-j}
\\ {}_1F_1(a;b|z) & = e^z {}_1F_1(b-a;b|\!-\!z)
\\ {}_1F_1(-N;\alpha+1|x) & = \frac{N!\alpha!}{(N+\alpha)!}L_N^{(\alpha)}(x)
\\ \sum_{N=0}^{\infty}t^N L_{N}^{(\alpha)}(x) & = \frac{e^{-\frac{tx}{1-t}}}{(1-t)^{\alpha+1}}
\\ \frac{(-x)^m}{m!}L_n^{(m-n)}(x) & = \frac{(-x)^n}{n!}L_m^{(n-m)}(x)
\end{align}
Moreover, the following identities are found in \cite{Prudnikov} ---
\begin{align}
\sum_{m=0}^n \frac{t^m}{(n-m)!(m+\alpha)!}L_m^{(\alpha)}(x) & = \frac{(1+t)^n}{(n+\alpha)!}L_n^{(\alpha)}\left(\frac{xt}{1+t}\right)
\\ \sum_{m=0}^{\infty}\frac{(n+m)!}{n!m!}t^m L_{n+m}^{(\alpha)}(x) & = \frac{e^{-\frac{xt}{1-t}}}{(1-t)^{n+\alpha+1}}L_n^{(\alpha)}\left(\frac{x}{1-t}\right)
\\ \sum_{k=0}^{\infty}\frac{1}{k!}t^k L_n^{(k-n)}(x) & = \frac{1}{n!}e^t (t-x)^n,
\end{align}
respectively as identity 4.4.1.7, identity 5.11.2.8 and identity 5.11.4.2.

\end{document}